\documentclass[letter,longauth]{aa}
\usepackage[utf8]{inputenc}
\usepackage{natbib,xcolor}
\usepackage[varg]{txfonts}
\usepackage{hyperref}

\makeatletter
\DeclareRobustCommand*{\fieldName}[1]{%
  \begingroup\@fieldName\scantokens{\texttt{\small {#1}}\noexpand}\endgroup}
\begingroup\lccode`\~=`\_\relax
   \lowercase{\endgroup\def\@fieldName{\catcode`\_=\active \let~\_}}
\makeatother

\newcommand\secref[1]{Sect.~\ref{#1}}

\newcommand\figref[1]{Fig.~\ref{#1}}

\newcommand\tabref[1]{Table~\ref{#1}}
\newcommand\appref[1]{Appendix~\ref{#1}}

\bibpunct{(}{)}{;}{a}{}{,} 

\newcommand\gaia{\textit{Gaia}\xspace}
\newcommand\gdrthree{\gaia~DR3\xspace}

\newcommand\bpminrp{\ensuremath{G_\mathrm{BP}-G_\mathrm{RP}}}
\newcommand\ebpminrp{\ensuremath{E(G_\mathrm{BP}-G_\mathrm{RP})}}

\def\bprp{\bpminrp}

\newcommand{\columnImage}[1]{\includegraphics[width=\columnwidth]{#1}}

\def\arcsec{\,$''$}
\def\ltsim{\ifmmode\stackrel{<}{_{\sim}}\else$\stackrel{<}{_{\sim}}$\fi}

\providecommand{\kms}{\,km\,s$^{-1}$}

\providecommand{\muas}{\,$\mu$as}

\providecommand{\Msun}{\ensuremath{\,{M}_{\odot}}\xspace}
\providecommand{\Rsun}{\ensuremath{\,{R}_{\odot}}\xspace}

\def\teff{$T_{\rm eff}$}
\def\logg{$\log g$}
\def\feh{[Fe/H]}
\def\mh{[M/H]}
\def\a0{$A_{\rm 0}$}

\def\aabun{[$\alpha$/Fe]}

\def\grvs{$G_{\rm RVS}$}

\newcommand{\orcit}[1]{\protect\href{https://orcid.org/#1}{\protect\includegraphics[width=8pt]{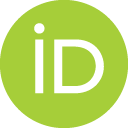}}}

\newcommand\hip{\textsc{Hipparcos}\xspace}
\newcommand\tyc{\textit{Tycho}}
\newcommand\tyctwo{\textit{Tycho}-2}

\newcommand\gdr[1]{\gaia~DR#1}

\title{Discovery of a dormant 33 solar-mass black hole in pre-release Gaia astrometry\thanks{\tabref{tab:epoch_astrometry} and \tabref{tab:epoch_rvs} with \gaia epoch data are available in electronic form at the CDS via
anonymous ftp to cdsarc.u-strasbg.fr (130.79.128.5) or via http://cdsweb.u-strasbg.fr/cgi-bin/qcat?J/A+A/, see \appref{sec:epoch_data}.}}

\author{
{\it Gaia} Collaboration: P.        ~Panuzzo                       \orcit{0000-0002-0016-8271}\inst{\ref{inst:0001}}\thanks{Corresponding author: P. Panuzzo, \email{pasquale.panuzzo@observatoiredeparis.psl.eu}}
\and T.        ~Mazeh                         \orcit{0000-0002-3569-3391}\inst{\ref{inst:0002}}
\and F.        ~Arenou                        \orcit{0000-0003-2837-3899}\inst{\ref{inst:0001}}
\and B.        ~Holl                          \orcit{0000-0001-6220-3266}\inst{\ref{inst:0004},\ref{inst:0005}}
\and E.        ~Caffau                        \orcit{0000-0001-6011-6134}\inst{\ref{inst:0001}}
\and A.        ~Jorissen                      \orcit{0000-0002-1883-4578}\inst{\ref{inst:0007}}
\and C.        ~Babusiaux                     \orcit{0000-0002-7631-348X}\inst{\ref{inst:0008}}
\and P.        ~Gavras                        \orcit{0000-0002-4383-4836}\inst{\ref{inst:0009}}
\and J.        ~Sahlmann                      \orcit{0000-0001-9525-3673}\inst{\ref{inst:0009}}
\and U.        ~Bastian                       \orcit{0000-0002-8667-1715}\inst{\ref{inst:0011}}
\and \L{}.     ~Wyrzykowski                   \orcit{0000-0002-9658-6151}\inst{\ref{inst:0012}}
\and L.        ~Eyer                          \orcit{0000-0002-0182-8040}\inst{\ref{inst:0004}}
\and N.        ~Leclerc                       \orcit{0009-0001-5569-6098}\inst{\ref{inst:0001}}
\and N.        ~Bauchet                       \orcit{0000-0002-2307-8973}\inst{\ref{inst:0001}}
\and A.        ~Bombrun                       \inst{\ref{inst:0016}}
\and N.        ~Mowlavi                       \orcit{0000-0003-1578-6993}\inst{\ref{inst:0004}}
\and G.M.      ~Seabroke                      \orcit{0000-0003-4072-9536}\inst{\ref{inst:0018}}
\and D.        ~Teyssier                      \orcit{0000-0002-6261-5292}\inst{\ref{inst:0019}}
\and E.        ~Balbinot                      \orcit{0000-0002-1322-3153}\inst{\ref{inst:0020},\ref{inst:0021}}
\and A.        ~Helmi                         \orcit{0000-0003-3937-7641}\inst{\ref{inst:0020}}
\and A.G.A.    ~Brown                         \orcit{0000-0002-7419-9679}\inst{\ref{inst:0021}}
\and A.        ~Vallenari                     \orcit{0000-0003-0014-519X}\inst{\ref{inst:0024}}
\and T.        ~Prusti                        \orcit{0000-0003-3120-7867}\inst{\ref{inst:0025}}
\and J.H.J.    ~de Bruijne                    \orcit{0000-0001-6459-8599}\inst{\ref{inst:0025}}
\and A.        ~Barbier                       \orcit{0009-0004-0983-931X}\inst{\ref{inst:0027}}
\and M.        ~Biermann                      \orcit{0000-0002-5791-9056}\inst{\ref{inst:0011}}
\and O.L.      ~Creevey                       \orcit{0000-0003-1853-6631}\inst{\ref{inst:0029}}
\and C.        ~Ducourant                     \orcit{0000-0003-4843-8979}\inst{\ref{inst:0030}}
\and D.W.      ~Evans                         \orcit{0000-0002-6685-5998}\inst{\ref{inst:0031}}
\and R.        ~Guerra                        \orcit{0000-0002-9850-8982}\inst{\ref{inst:0032}}
\and A.        ~Hutton                        \inst{\ref{inst:0033}}
\and C.        ~Jordi                         \orcit{0000-0001-5495-9602}\inst{\ref{inst:0034},\ref{inst:0035}}
\and S.A.      ~Klioner                       \orcit{0000-0003-4682-7831}\inst{\ref{inst:0036}}
\and U.        ~Lammers                       \orcit{0000-0001-8309-3801}\inst{\ref{inst:0032}}
\and L.        ~Lindegren                     \orcit{0000-0002-5443-3026}\inst{\ref{inst:0038}}
\and X.        ~Luri                          \inst{\ref{inst:0034},\ref{inst:0040},\ref{inst:0035}}
\and F.        ~Mignard                       \inst{\ref{inst:0029}}
\and C.        ~Nicolas                       \inst{\ref{inst:0027}}
\and S.        ~Randich                       \orcit{0000-0003-2438-0899}\inst{\ref{inst:0044}}
\and P.        ~Sartoretti                    \orcit{0000-0002-6574-7565}\inst{\ref{inst:0001}}
\and R.        ~Smiljanic                     \orcit{0000-0003-0942-7855}\inst{\ref{inst:0046}}
\and P.        ~Tanga                         \orcit{0000-0002-2718-997X}\inst{\ref{inst:0029}}
\and N.A.      ~Walton                        \orcit{0000-0003-3983-8778}\inst{\ref{inst:0031}}
\and C.        ~Aerts                         \orcit{0000-0003-1822-7126}\inst{\ref{inst:0049},\ref{inst:0050},\ref{inst:0051}}
\and C.A.L.    ~Bailer-Jones                  \inst{\ref{inst:0051}}
\and M.        ~Cropper                       \orcit{0000-0003-4571-9468}\inst{\ref{inst:0018}}
\and R.        ~Drimmel                       \orcit{0000-0002-1777-5502}\inst{\ref{inst:0054}}
\and F.        ~Jansen                        \inst{\ref{inst:0055}}
\and D.        ~Katz                          \orcit{0000-0001-7986-3164}\inst{\ref{inst:0001}}
\and M.G.      ~Lattanzi                      \orcit{0000-0003-0429-7748}\inst{\ref{inst:0054},\ref{inst:0058}}
\and C.        ~Soubiran                      \orcit{0000-0003-3304-8134}\inst{\ref{inst:0030}}
\and F.        ~Th\'{e}venin                  \orcit{0000-0002-5032-2476}\inst{\ref{inst:0029}}
\and F.        ~van Leeuwen                   \orcit{0000-0003-1781-4441}\inst{\ref{inst:0031}}
\and R.        ~Andrae                        \orcit{0000-0001-8006-6365}\inst{\ref{inst:0051}}
\and M.        ~Audard                        \orcit{0000-0003-4721-034X}\inst{\ref{inst:0004}}
\and J.        ~Bakker                        \inst{\ref{inst:0032}}
\and R.        ~Blomme                        \orcit{0000-0002-2526-346X}\inst{\ref{inst:0065}}
\and J.        ~Casta\~{n}eda                 \orcit{0000-0001-7820-946X}\inst{\ref{inst:0066},\ref{inst:0034},\ref{inst:0035}}
\and F.        ~De Angeli                     \orcit{0000-0003-1879-0488}\inst{\ref{inst:0031}}
\and C.        ~Fabricius                     \inst{\ref{inst:0035},\ref{inst:0034},\ref{inst:0040}}
\and M.        ~Fouesneau                     \orcit{0000-0001-9256-5516}\inst{\ref{inst:0051}}
\and Y.        ~Fr\'{e}mat                    \orcit{0000-0002-4645-6017}\inst{\ref{inst:0065}}
\and L.        ~Galluccio                     \orcit{0000-0002-8541-0476}\inst{\ref{inst:0029}}
\and A.        ~Guerrier                      \orcit{0009-0009-8218-5670}\inst{\ref{inst:0027}}
\and U.        ~Heiter                        \orcit{0000-0001-6825-1066}\inst{\ref{inst:0077}}
\and E.        ~Masana                        \inst{\ref{inst:0035},\ref{inst:0034},\ref{inst:0040}}
\and R.        ~Messineo                      \inst{\ref{inst:0081}}
\and K.        ~Nienartowicz                  \orcit{0000-0001-5415-0547}\inst{\ref{inst:0082},\ref{inst:0005}}
\and F.        ~Pailler                       \orcit{0000-0002-4834-481X}\inst{\ref{inst:0027}}
\and F.        ~Riclet                        \inst{\ref{inst:0027}}
\and W.        ~Roux                          \orcit{0000-0002-7816-1950}\inst{\ref{inst:0027}}
\and R.        ~Sordo                         \orcit{0000-0003-4979-0659}\inst{\ref{inst:0024}}
\and G.        ~Gracia-Abril                  \inst{\ref{inst:0088},\ref{inst:0011}}
\and J.        ~Portell                       \orcit{0000-0002-8886-8925}\inst{\ref{inst:0034},\ref{inst:0040},\ref{inst:0035}}
\and M.        ~Altmann                       \orcit{0000-0002-0530-0913}\inst{\ref{inst:0011},\ref{inst:0094}}
\and K.        ~Benson                        \inst{\ref{inst:0018}}
\and J.        ~Berthier                      \orcit{0000-0003-1846-6485}\inst{\ref{inst:0096}}
\and P.W.      ~Burgess                       \orcit{0009-0002-6668-4559}\inst{\ref{inst:0031}}
\and D.        ~Busonero                      \orcit{0000-0002-3903-7076}\inst{\ref{inst:0054}}
\and G.        ~Busso                         \orcit{0000-0003-0937-9849}\inst{\ref{inst:0031}}
\and C.        ~Cacciari                      \orcit{0000-0001-5174-3179}\inst{\ref{inst:0100}}
\and H.        ~C\'{a}novas                   \orcit{0000-0001-7668-8022}\inst{\ref{inst:0019}}
\and J.M.      ~Carrasco                      \orcit{0000-0002-3029-5853}\inst{\ref{inst:0035},\ref{inst:0034},\ref{inst:0040}}
\and B.        ~Carry                         \orcit{0000-0001-5242-3089}\inst{\ref{inst:0029}}
\and A.        ~Cellino                       \orcit{0000-0002-6645-334X}\inst{\ref{inst:0054}}
\and N.        ~Cheek                         \inst{\ref{inst:0107}}
\and G.        ~Clementini                    \orcit{0000-0001-9206-9723}\inst{\ref{inst:0100}}
\and Y.        ~Damerdji                      \orcit{0000-0002-3107-4024}\inst{\ref{inst:0109},\ref{inst:0110}}
\and M.        ~Davidson                      \orcit{0000-0001-9271-4411}\inst{\ref{inst:0111}}
\and P.        ~de Teodoro                    \inst{\ref{inst:0032}}
\and L.        ~Delchambre                    \orcit{0000-0003-2559-408X}\inst{\ref{inst:0109}}
\and A.        ~Dell'Oro                      \orcit{0000-0003-1561-9685}\inst{\ref{inst:0044}}
\and E.        ~Fraile Garcia                 \orcit{0000-0001-7742-9663}\inst{\ref{inst:0009}}
\and D.        ~Garabato                      \orcit{0000-0002-7133-6623}\inst{\ref{inst:0116}}
\and P.        ~Garc\'{i}a-Lario              \orcit{0000-0003-4039-8212}\inst{\ref{inst:0032}}
\and R.        ~Haigron                       \inst{\ref{inst:0001}}
\and N.C.      ~Hambly                        \orcit{0000-0002-9901-9064}\inst{\ref{inst:0111}}
\and D.L.      ~Harrison                      \orcit{0000-0001-8687-6588}\inst{\ref{inst:0031},\ref{inst:0121}}
\and D.        ~Hatzidimitriou                \orcit{0000-0002-5415-0464}\inst{\ref{inst:0122}}
\and J.        ~Hern\'{a}ndez                 \orcit{0000-0002-0361-4994}\inst{\ref{inst:0032}}
\and D.        ~Hestroffer                    \orcit{0000-0003-0472-9459}\inst{\ref{inst:0096}}
\and S.T.      ~Hodgkin                       \orcit{0000-0002-5470-3962}\inst{\ref{inst:0031}}
\and S.        ~Jamal                         \orcit{0000-0002-3929-6668}\inst{\ref{inst:0051}}
\and G.        ~Jevardat de Fombelle          \orcit{0000-0001-6166-8221}\inst{\ref{inst:0004}}
\and S.        ~Jordan                        \orcit{0000-0001-6316-6831}\inst{\ref{inst:0011}}
\and A.        ~Krone-Martins                 \orcit{0000-0002-2308-6623}\inst{\ref{inst:0129},\ref{inst:0130}}
\and A.C.      ~Lanzafame                     \orcit{0000-0002-2697-3607}\inst{\ref{inst:0131},\ref{inst:0132}}
\and W.        ~L\"{ o}ffler                  \orcit{0009-0003-1319-5601}\inst{\ref{inst:0011}}
\and A.        ~Lorca                         \orcit{0000-0002-7985-250X}\inst{\ref{inst:0033}}
\and O.        ~Marchal                       \orcit{ 0000-0001-7461-892}\inst{\ref{inst:0135}}
\and P.M.      ~Marrese                       \orcit{0000-0002-8162-3810}\inst{\ref{inst:0136},\ref{inst:0137}}
\and A.        ~Moitinho                      \orcit{0000-0003-0822-5995}\inst{\ref{inst:0130}}
\and K.        ~Muinonen                      \orcit{0000-0001-8058-2642}\inst{\ref{inst:0139},\ref{inst:0140}}
\and M.        ~Nu\~{n}ez Campos              \inst{\ref{inst:0033}}
\and I.        ~Oreshina-Slezak               \inst{\ref{inst:0029}}
\and P.        ~Osborne                       \orcit{0000-0003-4482-3538}\inst{\ref{inst:0031}}
\and E.        ~Pancino                       \orcit{0000-0003-0788-5879}\inst{\ref{inst:0044},\ref{inst:0137}}
\and T.        ~Pauwels                       \inst{\ref{inst:0065}}
\and A.        ~Recio-Blanco                  \orcit{0000-0002-6550-7377}\inst{\ref{inst:0029}}
\and M.        ~Riello                        \orcit{0000-0002-3134-0935}\inst{\ref{inst:0031}}
\and L.        ~Rimoldini                     \orcit{0000-0002-0306-585X}\inst{\ref{inst:0005}}
\and A.C.      ~Robin                         \orcit{0000-0001-8654-9499}\inst{\ref{inst:0150}}
\and T.        ~Roegiers                      \orcit{0000-0002-1231-4440}\inst{\ref{inst:0151}}
\and L.M.      ~Sarro                         \orcit{0000-0002-5622-5191}\inst{\ref{inst:0152}}
\and M.        ~Schultheis                    \orcit{0000-0002-6590-1657}\inst{\ref{inst:0029}}
\and M.        ~Smith                         \inst{\ref{inst:0018}}
\and A.        ~Sozzetti                      \orcit{0000-0002-7504-365X}\inst{\ref{inst:0054}}
\and E.        ~Utrilla                       \inst{\ref{inst:0033}}
\and M.        ~van Leeuwen                   \orcit{0000-0001-9698-2392}\inst{\ref{inst:0031}}
\and K.        ~Weingrill                     \orcit{0000-0002-8163-2493}\inst{\ref{inst:0158}}
\and U.        ~Abbas                         \orcit{0000-0002-5076-766X}\inst{\ref{inst:0054}}
\and P.        ~\'{A}brah\'{a}m               \orcit{0000-0001-6015-646X}\inst{\ref{inst:0160},\ref{inst:0161},\ref{inst:0162}}
\and A.        ~Abreu Aramburu                \orcit{0000-0003-3959-0856}\inst{\ref{inst:0163}}
\and S.        ~Ahmed                         \orcit{0000-0001-8304-5586}\inst{\ref{inst:0031}}
\and G.        ~Altavilla                     \orcit{0000-0002-9934-1352}\inst{\ref{inst:0136},\ref{inst:0137}}
\and M.A.      ~\'{A}lvarez                   \orcit{0000-0002-6786-2620}\inst{\ref{inst:0116}}
\and F.        ~Anders                        \orcit{0000-0003-4524-9363}\inst{\ref{inst:0034},\ref{inst:0040},\ref{inst:0035}}
\and R.I.      ~Anderson                      \orcit{0000-0001-8089-4419}\inst{\ref{inst:0171}}
\and E.        ~Anglada Varela                \orcit{0000-0001-7563-0689}\inst{\ref{inst:0033}}
\and T.        ~Antoja                        \orcit{0000-0003-2595-5148}\inst{\ref{inst:0034},\ref{inst:0040},\ref{inst:0035}}
\and S.        ~Baig                          \orcit{0009-0004-8992-680X}\inst{\ref{inst:0176}}
\and D.        ~Baines                        \orcit{0000-0002-6923-3756}\inst{\ref{inst:0177}}
\and S.G.      ~Baker                         \orcit{0000-0002-6436-1257}\inst{\ref{inst:0018}}
\and L.        ~Balaguer-N\'{u}\~{n}ez        \orcit{0000-0001-9789-7069}\inst{\ref{inst:0035},\ref{inst:0034},\ref{inst:0040}}
\and Z.        ~Balog                         \orcit{0000-0003-1748-2926}\inst{\ref{inst:0011},\ref{inst:0051}}
\and C.        ~Barache                       \inst{\ref{inst:0094}}
\and M.        ~Barros                        \orcit{0000-0002-9728-9618}\inst{\ref{inst:0185}}
\and M.A.      ~Barstow                       \orcit{0000-0002-7116-3259}\inst{\ref{inst:0186}}
\and S.        ~Bartolom\'{e}                 \orcit{0000-0002-6290-6030}\inst{\ref{inst:0035},\ref{inst:0034},\ref{inst:0040}}
\and D.        ~Bashi                         \orcit{0000-0002-9035-2645}\inst{\ref{inst:0002},\ref{inst:0191}}
\and J.-L.     ~Bassilana                     \inst{\ref{inst:0192}}
\and N.        ~Baudeau                       \inst{\ref{inst:0193}}
\and U.        ~Becciani                      \orcit{0000-0002-4389-8688}\inst{\ref{inst:0131}}
\and L.R.      ~Bedin                         \orcit{0000-0003-4080-6466}\inst{\ref{inst:0024}}
\and I.        ~Bellas-Velidis                \inst{\ref{inst:0196}}
\and M.        ~Bellazzini                    \orcit{0000-0001-8200-810X}\inst{\ref{inst:0100}}
\and W.        ~Beordo                        \orcit{0000-0002-5094-1306}\inst{\ref{inst:0054},\ref{inst:0058}}
\and M.        ~Bernet                        \orcit{0000-0001-7503-1010}\inst{\ref{inst:0034},\ref{inst:0040},\ref{inst:0035}}
\and C.        ~Bertolotto                    \inst{\ref{inst:0081}}
\and S.        ~Bertone                       \orcit{0000-0001-9885-8440}\inst{\ref{inst:0054},\ref{inst:0205}}
\and L.        ~Bianchi                       \orcit{0000-0002-7999-4372}\inst{\ref{inst:0206}}
\and A.        ~Binnenfeld                    \orcit{0000-0002-9319-3838}\inst{\ref{inst:0207}}
\and S.        ~Blanco-Cuaresma               \orcit{0000-0002-1584-0171}\inst{\ref{inst:0208},\ref{inst:0209}}
\and J.        ~Bland-Hawthorn                \orcit{0000-0001-7516-4016}\inst{\ref{inst:0210},\ref{inst:0211}}
\and A.        ~Blazere                       \inst{\ref{inst:0212}}
\and T.        ~Boch                          \orcit{0000-0001-5818-2781}\inst{\ref{inst:0135}}
\and D.        ~Bossini                       \orcit{0000-0002-9480-8400}\inst{\ref{inst:0214},\ref{inst:0024}}
\and S.        ~Bouquillon                    \inst{\ref{inst:0094},\ref{inst:0217}}
\and A.        ~Bragaglia                     \orcit{0000-0002-0338-7883}\inst{\ref{inst:0100}}
\and J.        ~Braine                        \orcit{0000-0003-1740-1284}\inst{\ref{inst:0030}}
\and E.        ~Bratsolis                     \inst{\ref{inst:0220}}
\and E.        ~Breedt                        \orcit{0000-0001-6180-3438}\inst{\ref{inst:0031}}
\and A.        ~Bressan                       \orcit{0000-0002-7922-8440}\inst{\ref{inst:0222}}
\and N.        ~Brouillet                     \orcit{0000-0002-3274-7024}\inst{\ref{inst:0030}}
\and E.        ~Brugaletta                    \orcit{0000-0003-2598-6737}\inst{\ref{inst:0131}}
\and B.        ~Bucciarelli                   \orcit{0000-0002-5303-0268}\inst{\ref{inst:0054},\ref{inst:0058}}
\and A.G.      ~Butkevich                     \orcit{0000-0002-4098-3588}\inst{\ref{inst:0054}}
\and R.        ~Buzzi                         \orcit{0000-0001-9389-5701}\inst{\ref{inst:0054}}
\and A.        ~Camut                         \inst{\ref{inst:0229}}
\and R.        ~Cancelliere                   \orcit{0000-0002-9120-3799}\inst{\ref{inst:0230}}
\and T.        ~Cantat-Gaudin                 \orcit{0000-0001-8726-2588}\inst{\ref{inst:0051}}
\and D.        ~Capilla Guilarte              \inst{\ref{inst:0031}}
\and R.        ~Carballo                      \orcit{0000-0001-7412-2498}\inst{\ref{inst:0233}}
\and T.        ~Carlucci                      \inst{\ref{inst:0094}}
\and M.I.      ~Carnerero                     \orcit{0000-0001-5843-5515}\inst{\ref{inst:0054}}
\and J.        ~Carretero                     \orcit{0000-0002-3130-0204}\inst{\ref{inst:0236},\ref{inst:0237},\ref{inst:0238}}
\and S.        ~Carton                        \inst{\ref{inst:0192}}
\and L.        ~Casamiquela                   \orcit{0000-0001-5238-8674}\inst{\ref{inst:0001}}
\and A.        ~Casey                         \inst{\ref{inst:0241}}
\and M.        ~Castellani                    \orcit{0000-0002-7650-7428}\inst{\ref{inst:0136}}
\and A.        ~Castro-Ginard                 \orcit{0000-0002-9419-3725}\inst{\ref{inst:0021}}
\and L.        ~Ceraj                         \orcit{0000-0003-1651-7111}\inst{\ref{inst:0244}}
\and V.        ~Cesare                        \orcit{0000-0003-1119-4237}\inst{\ref{inst:0131}}
\and P.        ~Charlot                       \orcit{0000-0002-9142-716X}\inst{\ref{inst:0030}}
\and C.        ~Chaudet                       \orcit{0000-0003-2535-0202}\inst{\ref{inst:0005}}
\and L.        ~Chemin                        \orcit{0000-0002-3834-7937}\inst{\ref{inst:0248}}
\and A.        ~Chiavassa                     \orcit{0000-0003-3891-7554}\inst{\ref{inst:0029}}
\and N.        ~Chornay                       \orcit{0000-0002-8767-3907}\inst{\ref{inst:0005}}
\and D.        ~Chosson                       \orcit{0009-0009-2092-5514}\inst{\ref{inst:0251}}
\and W.J.      ~Cooper                        \orcit{0000-0003-3501-8967}\inst{\ref{inst:0176},\ref{inst:0054}}
\and T.        ~Cornez                        \inst{\ref{inst:0192}}
\and S.        ~Cowell                        \inst{\ref{inst:0031}}
\and M.        ~Crosta                        \orcit{0000-0003-4369-3786}\inst{\ref{inst:0054},\ref{inst:0257}}
\and C.        ~Crowley                       \orcit{0000-0002-9391-9360}\inst{\ref{inst:0016}}
\and M.        ~Cruz Reyes                    \orcit{0000-0003-2443-173X}\inst{\ref{inst:0171}}
\and C.        ~Dafonte                       \orcit{0000-0003-4693-7555}\inst{\ref{inst:0116}}
\and M.        ~Dal Ponte                     \orcit{0000-0003-1056-2747}\inst{\ref{inst:0024}}
\and M.        ~David                         \orcit{0000-0002-4172-3112}\inst{\ref{inst:0262}}
\and P.        ~de Laverny                    \orcit{0000-0002-2817-4104}\inst{\ref{inst:0029}}
\and F.        ~De Luise                      \orcit{0000-0002-6570-8208}\inst{\ref{inst:0264}}
\and R.        ~De March                      \orcit{0000-0003-0567-842X}\inst{\ref{inst:0081}}
\and A.        ~de Torres                     \inst{\ref{inst:0016}}
\and E.F.      ~del Peloso                    \inst{\ref{inst:0011}}
\and M.        ~Delbo                         \orcit{0000-0002-8963-2404}\inst{\ref{inst:0029}}
\and A.        ~Delgado                       \inst{\ref{inst:0009}}
\and J.-B.     ~Delisle                       \orcit{0000-0001-5844-9888}\inst{\ref{inst:0004}}
\and C.        ~Demouchy                      \inst{\ref{inst:0272}}
\and E.        ~Denis                         \inst{\ref{inst:0029},\ref{inst:0274}}
\and T.E.      ~Dharmawardena                 \orcit{0000-0002-9583-5216}\inst{\ref{inst:0275},\ref{inst:0276}}
\and F.        ~Di Giacomo                    \orcit{0000-0002-9180-1019}\inst{\ref{inst:0264}}
\and C.        ~Diener                        \inst{\ref{inst:0031}}
\and E.        ~Distefano                     \orcit{0000-0002-2448-2513}\inst{\ref{inst:0131}}
\and C.        ~Dolding                       \inst{\ref{inst:0018}}
\and K.        ~Dsilva                        \orcit{0000-0002-1476-9772}\inst{\ref{inst:0007}}
\and H.        ~Enke                          \orcit{0000-0002-2366-8316}\inst{\ref{inst:0158}}
\and C.        ~Fabre                         \inst{\ref{inst:0212}}
\and M.        ~Fabrizio                      \orcit{0000-0001-5829-111X}\inst{\ref{inst:0136},\ref{inst:0137}}
\and S.        ~Faigler                       \orcit{0000-0002-8368-5724}\inst{\ref{inst:0002}}
\and M.        ~Fatovi\'{c}                   \orcit{0000-0003-1911-4326}\inst{\ref{inst:0244}}
\and G.        ~Fedorets                      \orcit{0000-0002-8418-4809}\inst{\ref{inst:0288},\ref{inst:0139},\ref{inst:0290}}
\and J.        ~Fern\'{a}ndez-Hern\'{a}ndez   \inst{\ref{inst:0009}}
\and P.        ~Fernique                      \orcit{0000-0002-3304-2923}\inst{\ref{inst:0135}}
\and F.        ~Figueras                      \orcit{0000-0002-3393-0007}\inst{\ref{inst:0034},\ref{inst:0040},\ref{inst:0035}}
\and C.        ~Fouron                        \inst{\ref{inst:0193}}
\and F.        ~Fragkoudi                     \orcit{0000-0002-0897-3013}\inst{\ref{inst:0297}}
\and M.        ~Gai                           \orcit{0000-0001-9008-134X}\inst{\ref{inst:0054}}
\and M.        ~Galinier                      \orcit{0000-0001-7920-0133}\inst{\ref{inst:0029}}
\and A.        ~Garcia-Serrano                \inst{\ref{inst:0035},\ref{inst:0034},\ref{inst:0040}}
\and M.        ~Garc\'{i}a-Torres             \orcit{0000-0002-6867-7080}\inst{\ref{inst:0303}}
\and A.        ~Garofalo                      \orcit{0000-0002-5907-0375}\inst{\ref{inst:0100}}
\and E.        ~Gerlach                       \orcit{0000-0002-9533-2168}\inst{\ref{inst:0036}}
\and R.        ~Geyer                         \orcit{0000-0001-6967-8707}\inst{\ref{inst:0036}}
\and P.        ~Giacobbe                      \orcit{0000-0001-7034-7024}\inst{\ref{inst:0054}}
\and G.        ~Gilmore                       \orcit{0000-0003-4632-0213}\inst{\ref{inst:0031},\ref{inst:0309}}
\and S.        ~Girona                        \orcit{0000-0002-1975-1918}\inst{\ref{inst:0310}}
\and G.        ~Giuffrida                     \orcit{0000-0002-8979-4614}\inst{\ref{inst:0136}}
\and A.        ~Gomboc                        \orcit{0000-0002-0908-914X}\inst{\ref{inst:0312}}
\and A.        ~Gomez                         \orcit{0000-0002-3796-3690}\inst{\ref{inst:0116}}
\and I.        ~Gonz\'{a}lez-Santamar\'{i}a   \orcit{0000-0002-8537-9384}\inst{\ref{inst:0116}}
\and E.        ~Gosset                        \inst{\ref{inst:0109},\ref{inst:0316}}
\and M.        ~Granvik                       \orcit{0000-0002-5624-1888}\inst{\ref{inst:0139},\ref{inst:0318}}
\and V.        ~Gregori Barrera               \inst{\ref{inst:0035},\ref{inst:0034},\ref{inst:0040}}
\and R.        ~Guti\'{e}rrez-S\'{a}nchez     \orcit{0009-0003-1500-4733}\inst{\ref{inst:0019}}
\and M.        ~Haywood                       \orcit{0000-0003-0434-0400}\inst{\ref{inst:0001}}
\and A.        ~Helmer                        \inst{\ref{inst:0192}}
\and S.L.      ~Hidalgo                       \orcit{0000-0002-0002-9298}\inst{\ref{inst:0325},\ref{inst:0326}}
\and T.        ~Hilger                        \orcit{0000-0003-1646-0063}\inst{\ref{inst:0036}}
\and D.        ~Hobbs                         \orcit{0000-0002-2696-1366}\inst{\ref{inst:0038}}
\and C.        ~Hottier                       \orcit{0000-0002-3498-3944}\inst{\ref{inst:0001}}
\and H.E.      ~Huckle                        \inst{\ref{inst:0018}}
\and \'{O}.    ~Jim\'{e}nez-Arranz            \orcit{0000-0001-7434-5165}\inst{\ref{inst:0034},\ref{inst:0040},\ref{inst:0035}}
\and J.        ~Juaristi Campillo             \inst{\ref{inst:0011}}
\and Z.        ~Kaczmarek                     \inst{\ref{inst:0011}}
\and P.        ~Kervella                      \orcit{0000-0003-0626-1749}\inst{\ref{inst:0251}}
\and S.        ~Khanna                        \orcit{0000-0002-2604-4277}\inst{\ref{inst:0054},\ref{inst:0020}}
\and M.        ~Kontizas                      \orcit{0000-0001-7177-0158}\inst{\ref{inst:0122}}
\and G.        ~Kordopatis                    \orcit{0000-0002-9035-3920}\inst{\ref{inst:0029}}
\and A.J.      ~Korn                          \orcit{0000-0002-3881-6756}\inst{\ref{inst:0077}}
\and \'{A}     ~K\'{o}sp\'{a}l                \orcit{0000-0001-7157-6275}\inst{\ref{inst:0160},\ref{inst:0051},\ref{inst:0161}}
\and Z.        ~Kostrzewa-Rutkowska           \inst{\ref{inst:0163},\ref{inst:0021}}
\and K.        ~Kruszy\'{n}ska                \orcit{0000-0002-2729-5369}\inst{\ref{inst:0347}}
\and M.        ~Kun                           \orcit{0000-0002-7538-5166}\inst{\ref{inst:0160}}
\and S.        ~Lambert                       \orcit{0000-0001-6759-5502}\inst{\ref{inst:0094}}
\and A.F.      ~Lanza                         \orcit{0000-0001-5928-7251}\inst{\ref{inst:0131}}
\and Y.        ~Lebreton                      \orcit{0000-0002-4834-2144}\inst{\ref{inst:0251},\ref{inst:0352}}
\and T.        ~Lebzelter                     \orcit{0000-0002-0702-7551}\inst{\ref{inst:0162}}
\and S.        ~Leccia                        \orcit{0000-0001-5685-6930}\inst{\ref{inst:0354}}
\and G.        ~Lecoutre                      \inst{\ref{inst:0150}}
\and S.        ~Liao                          \orcit{0000-0002-9346-0211}\inst{\ref{inst:0356},\ref{inst:0054},\ref{inst:0358}}
\and L.        ~Liberato                      \orcit{0000-0003-3433-6269}\inst{\ref{inst:0029},\ref{inst:0360}}
\and E.        ~Licata                        \orcit{0000-0002-5203-0135}\inst{\ref{inst:0054}}
\and E.        ~Livanou                       \orcit{0000-0003-0628-2347}\inst{\ref{inst:0122}}
\and A.        ~Lobel                         \orcit{0000-0001-5030-019X}\inst{\ref{inst:0065}}
\and J.        ~L\'{o}pez-Miralles            \orcit{0000-0002-6187-2713}\inst{\ref{inst:0033}}
\and C.        ~Loup                          \inst{\ref{inst:0135}}
\and M.        ~Madar\'{a}sz                  \orcit{0009-0005-4037-5506}\inst{\ref{inst:0160}}
\and L.        ~Mahy                          \orcit{0000-0003-0688-7987}\inst{\ref{inst:0065}}
\and R.G.      ~Mann                          \orcit{0000-0002-0194-325X}\inst{\ref{inst:0111}}
\and M.        ~Manteiga                      \orcit{0000-0002-7711-5581}\inst{\ref{inst:0369}}
\and C.P.      ~Marcellino                    \orcit{0009-0004-1508-1218}\inst{\ref{inst:0131}}
\and J.M.      ~Marchant                      \orcit{0000-0002-3678-3145}\inst{\ref{inst:0371}}
\and M.        ~Marconi                       \orcit{0000-0002-1330-2927}\inst{\ref{inst:0354}}
\and D.        ~Mar\'{i}n Pina                \orcit{0000-0001-6482-1842}\inst{\ref{inst:0034},\ref{inst:0040},\ref{inst:0035}}
\and S.        ~Marinoni                      \orcit{0000-0001-7990-6849}\inst{\ref{inst:0136},\ref{inst:0137}}
\and D.J.      ~Marshall                      \orcit{0000-0003-3956-3524}\inst{\ref{inst:0378}}
\and J.        ~Mart\'{i}n Lozano             \orcit{0009-0001-2435-6680}\inst{\ref{inst:0107}}
\and L.        ~Martin Polo                   \inst{\ref{inst:0380}}
\and J.M.      ~Mart\'{i}n-Fleitas            \orcit{0000-0002-8594-569X}\inst{\ref{inst:0033}}
\and G.        ~Marton                        \orcit{0000-0002-1326-1686}\inst{\ref{inst:0160}}
\and D.        ~Mascarenhas                   \orcit{0000-0003-1574-4304}\inst{\ref{inst:0192}}
\and A.        ~Masip                         \orcit{0000-0003-1419-0020}\inst{\ref{inst:0035},\ref{inst:0034},\ref{inst:0040}}
\and A.        ~Mastrobuono-Battisti          \orcit{0000-0002-2386-9142}\inst{\ref{inst:0001}}
\and P.J.      ~McMillan                      \orcit{0000-0002-8861-2620}\inst{\ref{inst:0186}}
\and J.        ~Meichsner                     \orcit{0000-0002-9900-7864}\inst{\ref{inst:0036}}
\and J.        ~Merc                          \orcit{0000-0001-6355-2468}\inst{\ref{inst:0390}}
\and S.        ~Messina                       \orcit{0000-0002-2851-2468}\inst{\ref{inst:0131}}
\and N.R.      ~Millar                        \inst{\ref{inst:0031}}
\and A.        ~Mints                         \orcit{0000-0002-8440-1455}\inst{\ref{inst:0158}}
\and D.        ~Mohamed                       \inst{\ref{inst:0229}}
\and D.        ~Molina                        \orcit{0000-0003-4814-0275}\inst{\ref{inst:0040},\ref{inst:0034},\ref{inst:0035}}
\and R.        ~Molinaro                      \orcit{0000-0003-3055-6002}\inst{\ref{inst:0354}}
\and L.        ~Moln\'ar                      \orcit{0000-0002-8159-1599}\inst{\ref{inst:0160},\ref{inst:0161}}
\and M.        ~Mongui\'{o}                   \orcit{0000-0002-4519-6700}\inst{\ref{inst:0034},\ref{inst:0400}}
\and P.        ~Montegriffo                   \orcit{0000-0001-5013-5948}\inst{\ref{inst:0100}}
\and L.        ~Monti                         \orcit{0000-0002-2087-0535}\inst{\ref{inst:0100}}
\and A.        ~Mora                          \inst{\ref{inst:0033}}
\and R.        ~Morbidelli                    \orcit{0000-0001-7627-4946}\inst{\ref{inst:0054}}
\and D.        ~Morris                        \orcit{0000-0002-1952-6251}\inst{\ref{inst:0111}}
\and R.        ~Mudimadugula                  \inst{\ref{inst:0158}}
\and T.        ~Muraveva                      \orcit{0000-0002-0969-1915}\inst{\ref{inst:0100}}
\and I.        ~Musella                       \orcit{0000-0001-5909-6615}\inst{\ref{inst:0354}}
\and Z.        ~Nagy                          \orcit{0000-0002-3632-1194}\inst{\ref{inst:0160}}
\and N.        ~Nardetto                      \orcit{0000-0002-7399-0231}\inst{\ref{inst:0029}}
\and C.        ~Navarrete                     \orcit{0000-0002-4777-9934}\inst{\ref{inst:0029}}
\and S.        ~Oh                            \orcit{0000-0001-7790-5308}\inst{\ref{inst:0031},\ref{inst:0413}}
\and C.        ~Ordenovic                     \inst{\ref{inst:0029}}
\and O.        ~Orenstein                     \orcit{0009-0008-5052-6226}\inst{\ref{inst:0002}}
\and C.        ~Pagani                        \orcit{0000-0001-5477-4720}\inst{\ref{inst:0186}}
\and I.        ~Pagano                        \orcit{0000-0001-9573-4928}\inst{\ref{inst:0131}}
\and L.        ~Palaversa                     \orcit{0000-0003-3710-0331}\inst{\ref{inst:0244}}
\and P.A.      ~Palicio                       \orcit{0000-0002-7432-8709}\inst{\ref{inst:0029}}
\and L.        ~Pallas-Quintela               \orcit{0000-0001-9296-3100}\inst{\ref{inst:0116}}
\and M.        ~Pawlak                        \orcit{0000-0002-5632-9433}\inst{\ref{inst:0421},\ref{inst:0038}}
\and A.        ~Penttil\"{ a}                 \orcit{0000-0001-7403-1721}\inst{\ref{inst:0139}}
\and P.        ~Pesciullesi                   \inst{\ref{inst:0009}}
\and M.        ~Pinamonti                     \orcit{0000-0002-4445-1845}\inst{\ref{inst:0054}}
\and E.        ~Plachy                        \orcit{0000-0002-5481-3352}\inst{\ref{inst:0160},\ref{inst:0161}}
\and L.        ~Planquart                     \orcit{0000-0003-0392-6645}\inst{\ref{inst:0007}}
\and G.        ~Plum                          \inst{\ref{inst:0001}}
\and E.        ~Poggio                        \orcit{0000-0003-3793-8505}\inst{\ref{inst:0054},\ref{inst:0029}}
\and D.        ~Pourbaix$^\dagger$            \orcit{0000-0002-3020-1837}\inst{\ref{inst:0007},\ref{inst:0316}}
\and A.M.      ~Price-Whelan                  \orcit{0000-0003-0872-7098}\inst{\ref{inst:0434}}
\and L.        ~Pulone                        \orcit{0000-0002-5285-998X}\inst{\ref{inst:0136}}
\and V.        ~Rabin                         \inst{\ref{inst:0096}}
\and M.        ~Rainer                        \orcit{0000-0002-8786-2572}\inst{\ref{inst:0437},\ref{inst:0044}}
\and C.M.      ~Raiteri                       \orcit{0000-0003-1784-2784}\inst{\ref{inst:0054}}
\and P.        ~Ramos                         \orcit{0000-0002-5080-7027}\inst{\ref{inst:0440},\ref{inst:0034},\ref{inst:0035}}
\and M.        ~Ramos-Lerate                  \orcit{0009-0005-4677-8031}\inst{\ref{inst:0019}}
\and M.        ~Ratajczak                     \orcit{0000-0002-3218-2684}\inst{\ref{inst:0012}}
\and P.        ~Re Fiorentin                  \orcit{0000-0002-4995-0475}\inst{\ref{inst:0054}}
\and S.        ~Regibo                        \orcit{0000-0001-7227-9563}\inst{\ref{inst:0049}}
\and C.        ~Reyl\'{e}                     \orcit{0000-0003-2258-2403}\inst{\ref{inst:0150}}
\and V.        ~Ripepi                        \orcit{0000-0003-1801-426X}\inst{\ref{inst:0354}}
\and A.        ~Riva                          \orcit{0000-0002-6928-8589}\inst{\ref{inst:0054}}
\and H.-W.     ~Rix                           \orcit{0000-0003-4996-9069}\inst{\ref{inst:0051}}
\and G.        ~Rixon                         \orcit{0000-0003-4399-6568}\inst{\ref{inst:0031}}
\and G.        ~Robert                        \inst{\ref{inst:0192}}
\and N.        ~Robichon                      \orcit{0000-0003-4545-7517}\inst{\ref{inst:0001}}
\and C.        ~Robin                         \inst{\ref{inst:0192}}
\and M.        ~Romero-G\'{o}mez              \orcit{0000-0003-3936-1025}\inst{\ref{inst:0034},\ref{inst:0040},\ref{inst:0035}}
\and N.        ~Rowell                        \orcit{0000-0003-3809-1895}\inst{\ref{inst:0111}}
\and D.        ~Ruz Mieres                    \orcit{0000-0002-9455-157X}\inst{\ref{inst:0031}}
\and K.A.      ~Rybicki                       \orcit{0000-0002-9326-9329}\inst{\ref{inst:0460}}
\and G.        ~Sadowski                      \orcit{0000-0002-3411-1003}\inst{\ref{inst:0007}}
\and A.        ~Sagrist\`{a} Sell\'{e}s       \orcit{0000-0001-6191-2028}\inst{\ref{inst:0011}}
\and N.        ~Sanna                         \orcit{0000-0001-9275-9492}\inst{\ref{inst:0044}}
\and R.        ~Santove\~{n}a                 \orcit{0000-0002-9257-2131}\inst{\ref{inst:0116}}
\and M.        ~Sarasso                       \orcit{0000-0001-5121-0727}\inst{\ref{inst:0054}}
\and M.H.      ~Sarmiento                     \orcit{0000-0003-4252-5115}\inst{\ref{inst:0033}}
\and C.        ~Sarrate Riera                 \inst{\ref{inst:0066},\ref{inst:0034},\ref{inst:0035}}
\and E.        ~Sciacca                       \orcit{0000-0002-5574-2787}\inst{\ref{inst:0131}}
\and D.        ~S\'{e}gransan                 \orcit{0000-0003-2355-8034}\inst{\ref{inst:0004}}
\and M.        ~Semczuk                       \orcit{0000-0002-8191-8918}\inst{\ref{inst:0034},\ref{inst:0040},\ref{inst:0035}}
\and S.        ~Shahaf                        \orcit{0000-0001-9298-8068}\inst{\ref{inst:0460}}
\and A.        ~Siebert                       \orcit{0000-0001-8059-2840}\inst{\ref{inst:0135},\ref{inst:0477}}
\and E.        ~Slezak                        \inst{\ref{inst:0029}}
\and R.L.      ~Smart                         \orcit{0000-0002-4424-4766}\inst{\ref{inst:0054},\ref{inst:0176}}
\and O.N.      ~Snaith                        \orcit{0000-0003-1414-1296}\inst{\ref{inst:0001},\ref{inst:0482}}
\and E.        ~Solano                        \orcit{0000-0003-1885-5130}\inst{\ref{inst:0483}}
\and F.        ~Solitro                       \inst{\ref{inst:0081}}
\and D.        ~Souami                        \orcit{0000-0003-4058-0815}\inst{\ref{inst:0251},\ref{inst:0486}}
\and J.        ~Souchay                       \inst{\ref{inst:0094}}
\and E.        ~Spitoni                       \orcit{0000-0001-9715-5727}\inst{\ref{inst:0029},\ref{inst:0489}}
\and F.        ~Spoto                         \orcit{0000-0001-7319-5847}\inst{\ref{inst:0208}}
\and L.A.      ~Squillante                    \inst{\ref{inst:0081}}
\and I.A.      ~Steele                        \orcit{0000-0001-8397-5759}\inst{\ref{inst:0371}}
\and H.        ~Steidelm\"{ u}ller            \inst{\ref{inst:0036}}
\and J.        ~Surdej                        \orcit{0000-0002-7005-1976}\inst{\ref{inst:0109}}
\and L.        ~Szabados                      \orcit{0000-0002-2046-4131}\inst{\ref{inst:0160}}
\and F.        ~Taris                         \inst{\ref{inst:0094}}
\and M.B.      ~Taylor                        \orcit{0000-0002-4209-1479}\inst{\ref{inst:0497}}
\and R.        ~Teixeira                      \orcit{0000-0002-6806-6626}\inst{\ref{inst:0498}}
\and T.        ~Tepper-Garcia                 \orcit{0000-0002-1081-883X}\inst{\ref{inst:0210},\ref{inst:0211}}
\and W.        ~Thuillot                      \orcit{0000-0002-5203-6932}\inst{\ref{inst:0096}}
\and L.        ~Tolomei                       \orcit{0000-0002-3541-3230}\inst{\ref{inst:0081}}
\and N.        ~Tonello                       \orcit{0000-0003-0550-1667}\inst{\ref{inst:0310}}
\and F.        ~Torra                         \orcit{0000-0002-8429-299X}\inst{\ref{inst:0034},\ref{inst:0040},\ref{inst:0035}}
\and G.        ~Torralba Elipe                \orcit{0000-0001-8738-194X}\inst{\ref{inst:0116},\ref{inst:0508},\ref{inst:0509}}
\and M.        ~Trabucchi                     \orcit{0000-0002-1429-2388}\inst{\ref{inst:0214},\ref{inst:0004}}
\and E.        ~Trentin                       \orcit{0000-0002-3899-566X}\inst{\ref{inst:0512},\ref{inst:0158}}
\and M.        ~Tsantaki                      \orcit{0000-0002-0552-2313}\inst{\ref{inst:0044}}
\and C.        ~Turon                         \orcit{0000-0003-1236-5157}\inst{\ref{inst:0001}}
\and A.        ~Ulla                          \orcit{0000-0001-6424-5005}\inst{\ref{inst:0516},\ref{inst:0517}}
\and N.        ~Unger                         \orcit{0000-0003-3993-7127}\inst{\ref{inst:0004}}
\and I.        ~Valtchanov                    \orcit{0000-0001-9930-7886}\inst{\ref{inst:0019}}
\and O.        ~Vanel                         \orcit{0000-0002-7898-0454}\inst{\ref{inst:0001}}
\and A.        ~Vecchiato                     \orcit{0000-0003-1399-5556}\inst{\ref{inst:0054}}
\and D.        ~Vicente                       \orcit{0000-0002-1584-1182}\inst{\ref{inst:0310}}
\and E.        ~Villar                        \inst{\ref{inst:0066},\ref{inst:0034},\ref{inst:0035}}
\and M.        ~Weiler                        \inst{\ref{inst:0035},\ref{inst:0034},\ref{inst:0040}}
\and H.        ~Zhao                          \orcit{0000-0003-2645-6869}\inst{\ref{inst:0029},\ref{inst:0530}}
\and J.        ~Zorec                         \orcit{0000-0003-1257-6915}\inst{\ref{inst:0531}}
\and S.        ~Zucker                        \orcit{0000-0003-3173-3138}\inst{\ref{inst:0207},\ref{inst:0002}}
\and A.        ~\v{Z}upi\'{c}                 \orcit{0009-0002-0811-2931}\inst{\ref{inst:0116}}
\and T.        ~Zwitter                       \orcit{0000-0002-2325-8763}\inst{\ref{inst:0535}}
}
\institute{
     GEPI, Observatoire de Paris, Universit\'{e} PSL, CNRS, 5 Place Jules Janssen, 92190 Meudon, France\relax                                                                                                                                                                                                    \label{inst:0001}
\and School of Physics and Astronomy, Tel Aviv University, Tel Aviv 6997801, Israel\relax                                                                                                                                                                                                                        \label{inst:0002}
\and Department of Astronomy, University of Geneva, Chemin Pegasi 51, 1290 Versoix, Switzerland\relax                                                                                                                                                                                                            \label{inst:0004}
\and Department of Astronomy, University of Geneva, Chemin d'Ecogia 16, 1290 Versoix, Switzerland\relax                                                                                                                                                                                                          \label{inst:0005}
\and Institut d'Astronomie et d'Astrophysique, Universit\'{e} Libre de Bruxelles CP 226, Boulevard du Triomphe, 1050 Brussels, Belgium\relax                                                                                                                                                                     \label{inst:0007}
\and Univ. Grenoble Alpes, CNRS, IPAG, 38000 Grenoble, France\relax                                                                                                                                                                                                                                              \label{inst:0008}
\and RHEA for European Space Agency (ESA), Camino bajo del Castillo, s/n, Urbanizaci\'{o}n Villafranca del Castillo, Villanueva de la Ca\~{n}ada, 28692 Madrid, Spain\relax                                                                                                                                      \label{inst:0009}
\and Astronomisches Rechen-Institut, Zentrum f\"{ u}r Astronomie der Universit\"{ a}t Heidelberg, M\"{ o}nchhofstr. 12-14, 69120 Heidelberg, Germany\relax                                                                                                                                                       \label{inst:0011}
\and Astronomical Observatory, University of Warsaw,  Al. Ujazdowskie 4, 00-478 Warszawa, Poland\relax                                                                                                                                                                                                           \label{inst:0012}
\and HE Space Operations BV for European Space Agency (ESA), Camino bajo del Castillo, s/n, Urbanizaci\'{o}n Villafranca del Castillo, Villanueva de la Ca\~{n}ada, 28692 Madrid, Spain\relax                                                                                                                    \label{inst:0016}
\and Mullard Space Science Laboratory, University College London, Holmbury St Mary, Dorking, Surrey RH5 6NT, United Kingdom\relax                                                                                                                                                                                \label{inst:0018}
\and Telespazio UK S.L. for European Space Agency (ESA), Camino bajo del Castillo, s/n, Urbanizaci\'{o}n Villafranca del Castillo, Villanueva de la Ca\~{n}ada, 28692 Madrid, Spain\relax                                                                                                                        \label{inst:0019}
\and Kapteyn Astronomical Institute, University of Groningen, Landleven 12, 9747 AD Groningen, The Netherlands\relax                                                                                                                                                                                             \label{inst:0020}
\and Leiden Observatory, Leiden University, Einsteinweg 55, 2333 CC Leiden, The Netherlands\relax                                                                                                                                                                                                                \label{inst:0021}
\and INAF - Osservatorio astronomico di Padova, Vicolo Osservatorio 5, 35122 Padova, Italy\relax                                                                                                                                                                                                                 \label{inst:0024}
\and European Space Agency (ESA), European Space Research and Technology Centre (ESTEC), Keplerlaan 1, 2201AZ, Noordwijk, The Netherlands\relax                                                                                                                                                                  \label{inst:0025}
\and CNES Centre Spatial de Toulouse, 18 avenue Edouard Belin, 31401 Toulouse Cedex 9, France\relax                                                                                                                                                                                                              \label{inst:0027}
\and Universit\'{e} C\^{o}te d'Azur, Observatoire de la C\^{o}te d'Azur, CNRS, Laboratoire Lagrange, Bd de l'Observatoire, CS 34229, 06304 Nice Cedex 4, France\relax                                                                                                                                            \label{inst:0029}
\and Laboratoire d'astrophysique de Bordeaux, Univ. Bordeaux, CNRS, B18N, all{\'e}e Geoffroy Saint-Hilaire, 33615 Pessac, France\relax                                                                                                                                                                           \label{inst:0030}
\and Institute of Astronomy, University of Cambridge, Madingley Road, Cambridge CB3 0HA, United Kingdom\relax                                                                                                                                                                                                    \label{inst:0031}
\and European Space Agency (ESA), European Space Astronomy Centre (ESAC), Camino bajo del Castillo, s/n, Urbanizaci\'{o}n Villafranca del Castillo, Villanueva de la Ca\~{n}ada, 28692 Madrid, Spain\relax                                                                                                       \label{inst:0032}
\and Aurora Technology for European Space Agency (ESA), Camino bajo del Castillo, s/n, Urbanizaci\'{o}n Villafranca del Castillo, Villanueva de la Ca\~{n}ada, 28692 Madrid, Spain\relax                                                                                                                         \label{inst:0033}
\and Institut de Ci\`{e}ncies del Cosmos (ICCUB), Universitat  de  Barcelona  (UB), Mart\'{i} i  Franqu\`{e}s  1, 08028 Barcelona, Spain\relax                                                                                                                                                                   \label{inst:0034}
\and Institut d'Estudis Espacials de Catalunya (IEEC), c. Esteve Terradas 1, 08860 Castelldefels (Barcelona), Spain\relax                                                                                                                                                                                        \label{inst:0035}
\and Lohrmann Observatory, Technische Universit\"{ a}t Dresden, Mommsenstra{\ss}e 13, 01062 Dresden, Germany\relax                                                                                                                                                                                               \label{inst:0036}
\and Lund Observatory, Division of Astrophysics, Department of Physics, Lund University, Box 43, 22100 Lund, Sweden\relax                                                                                                                                                                                        \label{inst:0038}
\and Departament de F\'{i}sica Qu\`{a}ntica i Astrof\'{i}sica (FQA), Universitat de Barcelona (UB), c. Mart\'{i} i Franqu\`{e}s 1, 08028 Barcelona, Spain\relax                                                                                                                                                  \label{inst:0040}
\and INAF - Osservatorio Astrofisico di Arcetri, Largo Enrico Fermi 5, 50125 Firenze, Italy\relax                                                                                                                                                                                                                \label{inst:0044}
\and Nicolaus Copernicus Astronomical Center, Polish Academy of Sciences, ul. Bartycka 18, 00-716 Warsaw, Poland\relax                                                                                                                                                                                           \label{inst:0046}
\and Instituut voor Sterrenkunde, KU Leuven, Celestijnenlaan 200D, 3001 Leuven, Belgium\relax                                                                                                                                                                                                                    \label{inst:0049}
\and Department of Astrophysics/IMAPP, Radboud University, P.O.Box 9010, 6500 GL Nijmegen, The Netherlands\relax                                                                                                                                                                                                 \label{inst:0050}
\and Max Planck Institute for Astronomy, K\"{ o}nigstuhl 17, 69117 Heidelberg, Germany\relax                                                                                                                                                                                                                     \label{inst:0051}
\and INAF - Osservatorio Astrofisico di Torino, via Osservatorio 20, 10025 Pino Torinese (TO), Italy\relax                                                                                                                                                                                                       \label{inst:0054}
\and European Space Agency (ESA, retired), European Space Research and Technology Centre (ESTEC), Keplerlaan 1, 2201AZ, Noordwijk, The Netherlands\relax                                                                                                                                                         \label{inst:0055}
\and University of Turin, Department of Physics, Via Pietro Giuria 1, 10125 Torino, Italy\relax                                                                                                                                                                                                                  \label{inst:0058}
\and Royal Observatory of Belgium, Ringlaan 3, 1180 Brussels, Belgium\relax                                                                                                                                                                                                                                      \label{inst:0065}
\and DAPCOM Data Services, c. dels Vilabella, 5-7, 80500 Vic, Barcelona, Spain\relax                                                                                                                                                                                                                             \label{inst:0066}
\and Observational Astrophysics, Division of Astronomy and Space Physics, Department of Physics and Astronomy, Uppsala University, Box 516, 751 20 Uppsala, Sweden\relax                                                                                                                                         \label{inst:0077}
\and ALTEC S.p.a, Corso Marche, 79,10146 Torino, Italy\relax                                                                                                                                                                                                                                                     \label{inst:0081}
\and Sednai S\`{a}rl, Geneva, Switzerland\relax                                                                                                                                                                                                                                                                  \label{inst:0082}
\and Gaia DPAC Project Office, ESAC, Camino bajo del Castillo, s/n, Urbanizaci\'{o}n Villafranca del Castillo, Villanueva de la Ca\~{n}ada, 28692 Madrid, Spain\relax                                                                                                                                            \label{inst:0088}
\and SYRTE, Observatoire de Paris, Universit\'{e} PSL, CNRS, Sorbonne Universit\'{e}, LNE, 61 avenue de l'Observatoire 75014 Paris, France\relax                                                                                                                                                                 \label{inst:0094}
\and IMCCE, Observatoire de Paris, Universit\'{e} PSL, CNRS, Sorbonne Universit{\'e}, Univ. Lille, 77 av. Denfert-Rochereau, 75014 Paris, France\relax                                                                                                                                                           \label{inst:0096}
\and INAF - Osservatorio di Astrofisica e Scienza dello Spazio di Bologna, via Piero Gobetti 93/3, 40129 Bologna, Italy\relax                                                                                                                                                                                    \label{inst:0100}
\and Serco Gesti\'{o}n de Negocios for European Space Agency (ESA), Camino bajo del Castillo, s/n, Urbanizaci\'{o}n Villafranca del Castillo, Villanueva de la Ca\~{n}ada, 28692 Madrid, Spain\relax                                                                                                             \label{inst:0107}
\and Institut d'Astrophysique et de G\'{e}ophysique, Universit\'{e} de Li\`{e}ge, 19c, All\'{e}e du 6 Ao\^{u}t, B-4000 Li\`{e}ge, Belgium\relax                                                                                                                                                                  \label{inst:0109}
\and CRAAG - Centre de Recherche en Astronomie, Astrophysique et G\'{e}ophysique, Route de l'Observatoire Bp 63 Bouzareah 16340 Algiers, Algeria\relax                                                                                                                                                           \label{inst:0110}
\and Institute for Astronomy, University of Edinburgh, Royal Observatory, Blackford Hill, Edinburgh EH9 3HJ, United Kingdom\relax                                                                                                                                                                                \label{inst:0111}
\and CIGUS CITIC - Department of Computer Science and Information Technologies, University of A Coru\~{n}a, Campus de Elvi\~{n}a s/n, A Coru\~{n}a, 15071, Spain\relax                                                                                                                                           \label{inst:0116}
\and Kavli Institute for Cosmology Cambridge, Institute of Astronomy, Madingley Road, Cambridge, CB3 0HA\relax                                                                                                                                                                                                   \label{inst:0121}
\and Department of Astrophysics, Astronomy and Mechanics, National and Kapodistrian University of Athens, Panepistimiopolis, Zografos, 15783 Athens, Greece\relax                                                                                                                                                \label{inst:0122}
\and Donald Bren School of Information and Computer Sciences, University of California, Irvine, CA 92697, USA\relax                                                                                                                                                                                              \label{inst:0129}
\and CENTRA, Faculdade de Ci\^{e}ncias, Universidade de Lisboa, Edif. C8, Campo Grande, 1749-016 Lisboa, Portugal\relax                                                                                                                                                                                          \label{inst:0130}
\and INAF - Osservatorio Astrofisico di Catania, via S. Sofia 78, 95123 Catania, Italy\relax                                                                                                                                                                                                                     \label{inst:0131}
\and Dipartimento di Fisica e Astronomia ""Ettore Majorana"", Universit\`{a} di Catania, Via S. Sofia 64, 95123 Catania, Italy\relax                                                                                                                                                                             \label{inst:0132}
\and Universit\'{e} de Strasbourg, CNRS, Observatoire astronomique de Strasbourg, UMR 7550,  11 rue de l'Universit\'{e}, 67000 Strasbourg, France\relax                                                                                                                                                          \label{inst:0135}
\and INAF - Osservatorio Astronomico di Roma, Via Frascati 33, 00078 Monte Porzio Catone (Roma), Italy\relax                                                                                                                                                                                                     \label{inst:0136}
\and Space Science Data Center - ASI, Via del Politecnico SNC, 00133 Roma, Italy\relax                                                                                                                                                                                                                           \label{inst:0137}
\and Department of Physics, University of Helsinki, P.O. Box 64, 00014 Helsinki, Finland\relax                                                                                                                                                                                                                   \label{inst:0139}
\and Finnish Geospatial Research Institute FGI, Vuorimiehentie 5, 02150 Espoo, Finland\relax                                                                                                                                                                                                                     \label{inst:0140}
\and Institut UTINAM CNRS UMR6213, Universit\'{e} de Franche-Comt\'{e}, OSU THETA Franche-Comt\'{e} Bourgogne, Observatoire de Besan\c{c}on, BP1615, 25010 Besan\c{c}on Cedex, France\relax                                                                                                                      \label{inst:0150}
\and HE Space Operations BV for European Space Agency (ESA), Keplerlaan 1, 2201AZ, Noordwijk, The Netherlands\relax                                                                                                                                                                                              \label{inst:0151}
\and Dpto. de Inteligencia Artificial, UNED, c/ Juan del Rosal 16, 28040 Madrid, Spain\relax                                                                                                                                                                                                                     \label{inst:0152}
\and Leibniz Institute for Astrophysics Potsdam (AIP), An der Sternwarte 16, 14482 Potsdam, Germany\relax                                                                                                                                                                                                        \label{inst:0158}
\and Konkoly Observatory, HUN-REN Research Centre for Astronomy and Earth Sciences, MTA Centre of Excellence, H-1121, Konkoly Thege Mikl\'os \'ut 15-17, Budapest, Hungary\relax                                                                                                                                 \label{inst:0160}
\and Institute of Physics and Astronomy, ELTE E\"{ o}tv\"{ o}s Lor\'and University, H-1117, P\'azm\'any P\'eter s\'et\'any 1A, Budapest, Hungary\relax                                                                                                                                                           \label{inst:0161}
\and University of Vienna, Department of Astrophysics, T\"{ u}rkenschanzstra{\ss}e 17, A1180 Vienna, Austria\relax                                                                                                                                                                                               \label{inst:0162}
\and ATG Europe for European Space Agency (ESA), Camino bajo del Castillo, s/n, Urbanizaci\'{o}n Villafranca del Castillo, Villanueva de la Ca\~{n}ada, 28692 Madrid, Spain\relax                                                                                                                                \label{inst:0163}
\and Institute of Physics, Ecole Polytechnique F\'ed\'erale de Lausanne (EPFL), Observatoire de Sauverny, 1290 Versoix, Switzerland\relax                                                                                                                                                                        \label{inst:0171}
\and Centre for Astrophysics Research, University of Hertfordshire, College Lane, AL10 9AB, Hatfield, United Kingdom\relax                                                                                                                                                                                       \label{inst:0176}
\and Quasar Science Resources for European Space Agency (ESA), Camino bajo del Castillo, s/n, Urbanizaci\'{o}n Villafranca del Castillo, Villanueva de la Ca\~{n}ada, 28692 Madrid, Spain\relax                                                                                                                  \label{inst:0177}
\and LASIGE, Faculdade de Ci\^{e}ncias, Universidade de Lisboa, Edif. C6, Campo Grande, 1749-016 Lisboa, Portugal\relax                                                                                                                                                                                          \label{inst:0185}
\and School of Physics and Astronomy, University of Leicester, University Road, Leicester LE1 7RH, United Kingdom\relax                                                                                                                                                                                          \label{inst:0186}
\and Astrophysics Group, Cavendish Laboratory, University of Cambridge, JJ Thomson Avenue, Cambridge CB3 0HE, United Kingdom\relax                                                                                                                                                                               \label{inst:0191}
\and Thales Services for CNES Centre Spatial de Toulouse, 18 avenue Edouard Belin, 31401 Toulouse Cedex 9, France\relax                                                                                                                                                                                          \label{inst:0192}
\and Telespazio for CNES Centre Spatial de Toulouse, 18 avenue Edouard Belin, 31401 Toulouse Cedex 9, France\relax                                                                                                                                                                                               \label{inst:0193}
\and National Observatory of Athens, I. Metaxa and Vas. Pavlou, Palaia Penteli, 15236 Athens, Greece\relax                                                                                                                                                                                                       \label{inst:0196}
\and University of Maryland, College Park, MD, USA\relax                                                                                                                                                                                                                                                         \label{inst:0205}
\and EURIX S.r.l., Corso Vittorio Emanuele II 61, 10128, Torino, Italy\relax                                                                                                                                                                                                                                     \label{inst:0206}
\and Porter School of the Environment and Earth Sciences, Tel Aviv University, Tel Aviv 6997801, Israel\relax                                                                                                                                                                                                    \label{inst:0207}
\and Harvard-Smithsonian Center for Astrophysics, 60 Garden St., MS 15, Cambridge, MA 02138, USA\relax                                                                                                                                                                                                           \label{inst:0208}
\and Laboratoire de Recherche en Neuroimagerie, University Hospital (CHUV) and University of Lausanne (UNIL), Lausanne, Switzerland\relax                                                                                                                                                                        \label{inst:0209}
\and Sydney Institute for Astronomy, School of Physics, University of Sydney, NSW 2006, Australia\relax                                                                                                                                                                                                          \label{inst:0210}
\and ARC Centre of Excellence for All Sky Astrophysics in 3 Dimensions (ASTRO 3D), Australia\relax                                                                                                                                                                                                               \label{inst:0211}
\and ATOS for CNES Centre Spatial de Toulouse, 18 avenue Edouard Belin, 31401 Toulouse Cedex 9, France\relax                                                                                                                                                                                                     \label{inst:0212}
\and Department of Physics and Astronomy G. Galilei, University of Padova, Vicolo dell'Osservatorio 3, 35122, Padova, Italy\relax                                                                                                                                                                                \label{inst:0214}
\and LFCA/DAS,Universidad de Chile,CNRS,Casilla 36-D, Santiago, Chile\relax                                                                                                                                                                                                                                      \label{inst:0217}
\and University of West Attica, Ag. Spyridonos Str., Egaleo, 12243 Athens, Greece\relax                                                                                                                                                                                                                          \label{inst:0220}
\and SISSA - Scuola Internazionale Superiore di Studi Avanzati, via Bonomea 265, 34136 Trieste, Italy\relax                                                                                                                                                                                                      \label{inst:0222}
\and SII for CNES Centre Spatial de Toulouse, 18 avenue Edouard Belin, 31401 Toulouse Cedex 9, France\relax                                                                                                                                                                                                      \label{inst:0229}
\and University of Turin, Department of Computer Sciences, Corso Svizzera 185, 10149 Torino, Italy\relax                                                                                                                                                                                                         \label{inst:0230}
\and Dpto. de Matem\'{a}tica Aplicada y Ciencias de la Computaci\'{o}n, Univ. de Cantabria, ETS Ingenieros de Caminos, Canales y Puertos, Avda. de los Castros s/n, 39005 Santander, Spain\relax                                                                                                                 \label{inst:0233}
\and Institut de F\'{i}sica d'Altes Energies (IFAE), The Barcelona Institute of Science and Technology, Campus UAB, 08193 Bellaterra (Barcelona), Spain\relax                                                                                                                                                    \label{inst:0236}
\and Port d'Informaci\'{o} Cient\'{i}fica (PIC), Campus UAB, C. Albareda s/n, 08193 Bellaterra (Barcelona), Spain\relax                                                                                                                                                                                          \label{inst:0237}
\and Centro de Investigaciones Energ\'eticas, Medioambientales y Tecnol\'ogicas (CIEMAT), Avenida Complutense 40, E-28040 Madrid, Spain\relax                                                                                                                                                                    \label{inst:0238}
\and Monash University, 10 College Walk, Monash University, Clayton, VIC 3800, Australia\relax                                                                                                                                                                                                                   \label{inst:0241}
\and Ru{\dj}er Bo\v{s}kovi\'{c} Institute, Bijeni\v{c}ka cesta 54, 10000 Zagreb, Croatia\relax                                                                                                                                                                                                                   \label{inst:0244}
\and Universidad Andres Bello, Facultad de Ciencias Exactas, Departamento de Ciencias F\'{\i}sicas - Instituto de Astrof\'{\i}sica, Fernandez Concha 700, Las Condes, Santiago, Chile\relax                                                                                                                      \label{inst:0248}
\and LESIA, Observatoire de Paris, Universit\'{e} PSL, CNRS, Sorbonne Universit\'{e}, Universit\'{e} de Paris, 5 Place Jules Janssen, 92190 Meudon, France\relax                                                                                                                                                 \label{inst:0251}
\and University of Turin, Mathematical Department ""G.Peano"", Via Carlo Alberto 10, 10123 Torino, Italy\relax                                                                                                                                                                                                   \label{inst:0257}
\and University of Antwerp, Onderzoeksgroep Toegepaste Wiskunde, Middelheimlaan 1, 2020 Antwerp, Belgium\relax                                                                                                                                                                                                   \label{inst:0262}
\and INAF - Osservatorio Astronomico d'Abruzzo, Via Mentore Maggini, 64100 Teramo, Italy\relax                                                                                                                                                                                                                   \label{inst:0264}
\and APAVE EXPLOITATION for CNES Centre Spatial de Toulouse, 18 avenue Edouard Belin, 31401 Toulouse Cedex 9, France\relax                                                                                                                                                                                       \label{inst:0272}
\and Dpto. F\'\i sica Te\'orica y del Cosmos, Universidad de Granada, 18071 Granada, Spain\relax                                                                                                                                                                                                                 \label{inst:0274}
\and The Center for Cosmology and Particle Physics, New York University, 726 Broadway New York, NY 10003, USA\relax                                                                                                                                                                                              \label{inst:0275}
\and NASA Hubble Fellow\relax                                                                                                                                                                                                                                                                                    \label{inst:0276}
\and Finnish Centre for Astronomy with ESO, University of Turku, FI-20014 Turku, Finland\relax                                                                                                                                                                                                                   \label{inst:0288}
\and Astrophysics Research Centre, School of Mathematics and Physics, Queen's University Belfast, Belfast BT7 1NN, UK\relax                                                                                                                                                                                      \label{inst:0290}
\and Institute for Computational Cosmology, Department of Physics, Durham University, Durham DH1 3LE, UK\relax                                                                                                                                                                                                   \label{inst:0297}
\and Data Science and Big Data Lab, Pablo de Olavide University, 41013, Seville, Spain\relax                                                                                                                                                                                                                     \label{inst:0303}
\and Institute of Astrophysics, FORTH, Crete, Greece\relax                                                                                                                                                                                                                                                       \label{inst:0309}
\and Barcelona Supercomputing Center (BSC), Pla\c{c}a Eusebi G\"{ u}ell 1-3, 08034-Barcelona, Spain\relax                                                                                                                                                                                                        \label{inst:0310}
\and Center for Astrophysics and Cosmology, University of Nova Gorica, Vipavska 13, 5000 Nova Gorica, Slovenia\relax                                                                                                                                                                                             \label{inst:0312}
\and F.R.S.-FNRS, Rue d'Egmont 5, 1000 Brussels, Belgium\relax                                                                                                                                                                                                                                                   \label{inst:0316}
\and Asteroid Engineering Laboratory, Lule\aa{} University of Technology, Box 848, 981 28 Kiruna, Sweden\relax                                                                                                                                                                                                   \label{inst:0318}
\and IAC - Instituto de Astrofisica de Canarias, Via L\'{a}ctea s/n, 38200 La Laguna S.C., Tenerife, Spain\relax                                                                                                                                                                                                 \label{inst:0325}
\and Department of Astrophysics, University of La Laguna, Via L\'{a}ctea s/n, 38200 La Laguna S.C., Tenerife, Spain\relax                                                                                                                                                                                        \label{inst:0326}
\and Las Cumbres Observatory, 6740 Cortona Drive Suite 102, Goleta, CA 93117, USA\relax                                                                                                                                                                                                                          \label{inst:0347}
\and Universit\'{e} Rennes, CNRS, IPR (Institut de Physique de Rennes) - UMR 6251, 35000 Rennes, France\relax                                                                                                                                                                                                    \label{inst:0352}
\and INAF - Osservatorio Astronomico di Capodimonte, Via Moiariello 16, 80131, Napoli, Italy\relax                                                                                                                                                                                                               \label{inst:0354}
\and Shanghai Astronomical Observatory, Chinese Academy of Sciences, 80 Nandan Road, Shanghai 200030, People's Republic of China\relax                                                                                                                                                                           \label{inst:0356}
\and University of Chinese Academy of Sciences, No.19(A) Yuquan Road, Shijingshan District, Beijing 100049, People's Republic of China\relax                                                                                                                                                                     \label{inst:0358}
\and S\~{a}o Paulo State University, Grupo de Din\^{a}mica Orbital e Planetologia, CEP 12516-410, Guaratinguet\'{a}, SP, Brazil\relax                                                                                                                                                                            \label{inst:0360}
\and CIGUS CITIC, Department of Nautical Sciences and Marine Engineering, University of A Coru\~{n}a, Paseo de Ronda 51, 15071, A Coru\~{n}a, Spain\relax                                                                                                                                                        \label{inst:0369}
\and Astrophysics Research Institute, Liverpool John Moores University, 146 Brownlow Hill, Liverpool L3 5RF, United Kingdom\relax                                                                                                                                                                                \label{inst:0371}
\and IRAP, Universit\'{e} de Toulouse, CNRS, UPS, CNES, 9 Av. colonel Roche, BP 44346, 31028 Toulouse Cedex 4, France\relax                                                                                                                                                                                      \label{inst:0378}
\and Serco Gesti\'{o}n de Negocios for European Space Agency (ESA), Camino bajo del Castillo, s/n, Urbanizacion Villafranca del Castillo, Villanueva de la Ca\~{n}ada, 28692 Madrid, Spain\relax                                                                                                                 \label{inst:0380}
\and Astronomical Institute, Faculty of Mathematics and Physics, Charles University, V Hole\v{s}ovi\v{c}k{\'a}ch 2, 180 00 Prague, Czech Republic\relax                                                                                                                                                          \label{inst:0390}
\and Dribia Data Research S.L., Pg. de Gr\`acia, 55, 3r 4a, 08007 Barcelona\relax                                                                                                                                                                                                                                \label{inst:0400}
\and Pioneer Research Center for Climate and Earth Science, Institute for Basic Science, Daejeon 34126, Republic of Korea\relax                                                                                                                                                                                  \label{inst:0413}
\and Astronomical Observatory, Jagiellonian University, ul. Orla 171, 30-244 Krak\'{o}w, Poland\relax                                                                                                                                                                                                            \label{inst:0421}
\and Center for Computational Astrophysics, Flatiron Institute, 162 Fifth Ave, New York, NY 10010, USA\relax                                                                                                                                                                                                     \label{inst:0434}
\and INAF - Osservatorio Astronomico di Brera, via E. Bianchi, 46, 23807 Merate (LC), Italy\relax                                                                                                                                                                                                                \label{inst:0437}
\and National Astronomical Observatory of Japan, 2-21-1 Osawa, Mitaka, Tokyo 181-8588, Japan\relax                                                                                                                                                                                                               \label{inst:0440}
\and Department of Particle Physics and Astrophysics, Weizmann Institute of Science, Rehovot 7610001, Israel\relax                                                                                                                                                                                               \label{inst:0460}
\and Centre de Donn\'{e}es Astronomique de Strasbourg, Strasbourg, France\relax                                                                                                                                                                                                                                  \label{inst:0477}
\and Astrophysics Group, Department of Physics and Astronomy, University of Exeter, Exeter EX4 4QL, United Kingdom\relax                                                                                                                                                                                         \label{inst:0482}
\and Departamento de Astrof\'{i}sica, Centro de Astrobiolog\'{i}a (CSIC-INTA), ESA-ESAC. Camino Bajo del Castillo s/n. 28692 Villanueva de la Ca\~{n}ada, Madrid, Spain\relax                                                                                                                                    \label{inst:0483}
\and naXys, Department of Mathematics, University of Namur, Rue de Bruxelles 61, 5000 Namur, Belgium\relax                                                                                                                                                                                                       \label{inst:0486}
\and INAF. Osservatorio Astronomico di Trieste, via G.B. Tiepolo 11, 34131, Trieste, Italy\relax                                                                                                                                                                                                                 \label{inst:0489}
\and H H Wills Physics Laboratory, University of Bristol, Tyndall Avenue, Bristol BS8 1TL, United Kingdom\relax                                                                                                                                                                                                  \label{inst:0497}
\and Instituto de Astronomia, Geof\`{i}sica e Ci\^{e}ncias Atmosf\'{e}ricas, Universidade de S\~{a}o Paulo, Rua do Mat\~{a}o, 1226, Cidade Universitaria, 05508-900 S\~{a}o Paulo, SP, Brazil\relax                                                                                                              \label{inst:0498}
\and Escuela de Arquitectura y Polit\'{e}cnica - Universidad Europea de Valencia, Spain\relax                                                                                                                                                                                                                    \label{inst:0508}
\and Escuela Superior de Ingenier\'{i}a y Tecnolog\'{i}a - Universidad Internacional de la Rioja, Spain\relax                                                                                                                                                                                                    \label{inst:0509}
\and Institut f\"{ u}r Physik und Astronomie, Universit\"{ a}t Potsdam, Haus 28, Karl-Liebknecht-Str. 24/25, 14476 Golm (Potsdam), Germany\relax                                                                                                                                                                 \label{inst:0512}
\and Applied Physics Department, Universidade de Vigo, 36310 Vigo, Spain\relax                                                                                                                                                                                                                                   \label{inst:0516}
\and Instituto de F\'{\i}sica e Ciencias Aeroespaciais (IFCAE), Universidade de Vigo Campus de As Lagoas, 32004 Ourense, Spain\relax                                                                                                                                                                             \label{inst:0517}
\and Purple Mountain Observatory and Key Laboratory of Radio Astronomy, Chinese Academy of Sciences, 10 Yuanhua Road, Nanjing 210033, People's Republic of China\relax                                                                                                                                           \label{inst:0530}
\and Sorbonne Universit\'{e}, CNRS, UMR7095, Institut d'Astrophysique de Paris, 98bis bd. Arago, 75014 Paris, France\relax                                                                                                                                                                                       \label{inst:0531}
\and Faculty of Mathematics and Physics, University of Ljubljana, Jadranska ulica 19, 1000 Ljubljana, Slovenia\relax                                                                                                                                                                                             \label{inst:0535}
}

\date{Received ??? / Accepted ???}

\abstract{
Gravitational waves from black-hole (BH) merging events have revealed a population of extra-galactic BHs residing in short-period binaries with masses that are higher than expected based on most stellar evolution models -- and also higher than known stellar-origin black holes in our Galaxy. It has been proposed that those high-mass BHs are the remnants of massive metal-poor stars. 
}{
\gaia astrometry is expected to uncover many Galactic wide-binary systems containing dormant BHs, which may not have been detected before.
The study of this population will provide new information on the BH-mass distribution in binaries and shed light on their formation mechanisms and progenitors.
}{
As part of the validation efforts in preparation for the fourth \gaia data release (DR4), we analysed the preliminary astrometric binary solutions, obtained by the \gaia Non-Single Star pipeline, to verify their significance and to minimise false-detection rates in high-mass-function orbital solutions.
}{
The astrometric binary solution of one source, \gaia BH3, implies the presence of a $32.70\pm 0.82$ \Msun BH in a binary system with a period of 11.6 yr. 
\gaia radial velocities independently validate the astrometric orbit. Broad-band photometric and spectroscopic data show that the visible component is an old, very metal-poor giant of the Galactic halo, at a distance of 590 pc.
}{
The BH in the \gaia BH3 system is more massive than any other Galactic stellar-origin BH known thus far. The low metallicity of the star companion supports the scenario that metal-poor massive stars are progenitors of the high-mass BHs detected by gravitational-wave telescopes.
The Galactic orbit of the system and its metallicity indicate that it might belong to the Sequoia halo substructure. Alternatively, and more plausibly, it could belong to the ED-2 stream, which likely originated from a globular cluster that had been disrupted by the Milky Way.
}

\authorrunning{Gaia Collaboration et al.}
\begin{document}

\maketitle


\section{Introduction}

Since the first event detected in 2015 \citep{2016PhRvL.116f1102A} by the LIGO/Virgo collaboration,
the detection of black-hole (BH) mergers via gravitational waves has uncovered the existence of a population of BHs residing in short-period binaries
with masses higher than 30\Msun, ranging up to 85\Msun \citep{2020ApJ...900L..13A,2021ApJ...913L...7A}.

Stellar evolution models have difficulties in explaining such large masses for BHs of stellar origin: stars with an initial mass larger than 30\Msun are predicted to lose most of their mass during their evolution, due to the onset of strong winds, producing BHs with masses below 20\Msun \citep{2008NewAR..52..419V,2007ApJ...662..504B,2016ApJ...821...38S}.

The masses of the merging BHs detected via gravitational waves are also larger than any known stellar-origin BHs in our Galaxy: all confirmed or candidate BHs of stellar origin in the Milky Way have typical masses around or below 10\Msun, with Cyg X-1 
\citep[$\sim$20\Msun,][]{2021Sci...371.1046M} being the most massive one known thus far.
However, the known stellar-origin BHs, mainly limited to short-period X-ray binaries, are only a very tiny fraction of the 
expected number of BHs in our Galaxy \citep[$\sim 10^8$ e.g. ][]{2020A&A...638A..94O}. 
In fact, stellar-origin BHs are hard to detect because most of them do not interact with a companion. So, the lack of data on BHs with masses larger than 20\Msun could be due to an observational bias.
Indirect methods such as gravitational microlensing have also yielded   only one robust discovery of a single BH with mass, of about 7\Msun \citep{2022ApJ...933...83S}.

The detection of mergers of BHs with masses larger than 30\Msun can be reconciled with stellar evolution models if the progenitors of the high-mass BHs are low-metallicity stars \citep{2009MNRAS.395L..71M,2010ApJ...714.1217B,2016Natur.534..512B,2014MNRAS.441.3703Z,2012ApJ...749...91F}. The lack of metals  substantially decreases the mass loss during the stellar lifetime \citep{2008NewAR..52..419V} and reduces the radius of the evolving progenitors \citep{2000MNRAS.315..543H,2010ApJ...714.1217B}, the latter effect decreasing the probability of merging during the common-envelope phase \citep{2007ApJ...662..504B} in binary systems. Finally, the higher mass of the BHs produced by low-metallicity progenitors is expected to decrease substantially or eliminate the natal kick strength at the birth of the BHs, preserving the binary as a bound system \citep{2010ApJ...715L.138B}. The maximum metallicity for the formation of the high-mass BHs is a matter of active debate, with some models predicting the formation of 30\Msun BHs even at solar metallicities \citep{2023NatAs...7.1090B}.

Since its Data Release 3 \citep[DR3,][]{2023A&A...674A...1G}, the \gaia mission \citep{2016A&A...595A...1G} has increased the number of detected stellar binary systems by two orders of magnitude \citep[][Gosset et al. in prep.]{2023A&A...674A..34G,2023A&A...674A...9H}. This has opened up the possibility of detecting BHs in binary systems that do not interact with their companion.
Moreover, the ability of \gaia to measure the astrometric orbit of such systems allows the measurement of the inclination of the orbit, providing a robust estimate of the mass of the dark companion.
Two dormant BHs, \gaia BH1 and BH2 \citep{2023MNRAS.521.4323E,2023MNRAS.518.1057E,2023ApJ...946...79T,2023AJ....166....6C} were discovered in \gaia binaries of DR3.
\gaia Data Release 4 (DR4) is expected to contain a larger number of binary 
systems than \gaia DR3;  consequently, this will provide a greater number of BH-hosting systems, 
which will help to shed light on 
the BH population and the mechanisms in action in the BHs' formation.

In this Letter, we report the serendipitous discovery of a nearby ($\sim$590\,pc) binary system composed of an old, very metal-poor,\footnote{We use the nomenclature from \citet{BC2005} where very metal-poor are those stars having $[\mathrm{Fe/H}] < -2$.} giant star orbiting a BH in 11.6 yr. The estimated BH mass,
33\Msun, is substantially higher than all known Galactic BHs and is in the mass range of the extra-galactic BHs detected by gravitational waves.

The system was identified while validating the preliminary \gaia astrometric binary solutions produced in preparation for 
DR4 and subsequently confirmed by \gaia RVS radial-velocity data. 
We took the exceptional step of the publication of this paper based on preliminary data 
ahead of the official DR4 due to the unique nature of the discovery, 
which we believe should not be kept from the scientific community until the next release. An early disclosure will also enable an early and extensive follow-up by the community.


\section{Observations and analysis}\label{sec:obs}

\subsection{Properties of the source}\label{sec:propsource}

\begin{figure}
    \centering
    \columnImage{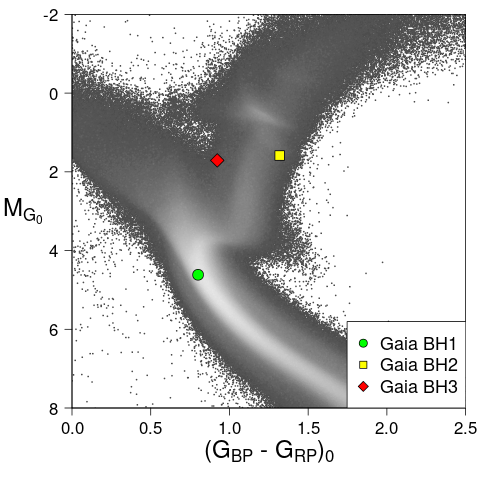}
    \caption{\gaia BH3 position in the \gaia color-magnitude diagram, compared with the position of Gaia BH1, BH2 and the low extinction ($A_0<0.05$\,mag) \gaia DR3 color-magnitude diagram. 
    All extinctions are estimated through the \citet{2022A&A...661A.147L} extinction map.}
    \label{fig:hr}
\end{figure}

\object{Gaia DR3 4318465066420528000} (also known as LS~II~+14~13 and
2MASS~J19391872+1455542), 
hereafter denoted as \gaia BH3, is a bright 
source in the constellation Aquila, known to be a high proper-motion star \citep{2005AJ....129.1483L}.
Its basic properties from the \gaia DR3 archive are reported in \tabref{tab:basic_params}.
Its absolute magnitude and color \citep{2021A&A...649A...3R,2023A&A...674A...6S} identify it as 
a star climbing the giant branch (see \figref{fig:hr}). 
The source was analysed by the Astrophysical Parameters Inference System \citep[Apsis,][]{2023A&A...674A..26C,2023A&A...674A..28F}.  It has been identified as a G spectral-type star by the ESP-ELS algorithm \citep[Sect 11.3.7 of the online documentation,][]{CU8-DR3-documentation} and the GSP-Spec ANN parameters \citep{2023A&A...674A..29R} indicate it as a metal-poor giant. 
No GSP-Phot result \citep{2023A&A...674A..27A} is published in the \gaia archive, while the parameters provided by GSP-Spec MatisseGauguin \citep{2023A&A...674A..29R} carry large uncertainties. 

\begin{table}[t]
    \caption{Basic properties of \gaia BH3 from the \gaia DR3 catalogue. Astrophysical parameters are from GSP-Spec ANN.}
    \label{tab:basic_params}
    \centering
    \begin{tabular}{lc}
    \hline
    \hline
    Parameter & Value\\
    \hline
    $\alpha$ [deg] & 294.8278625082\tablefootmark{a} \\
    $\sigma_{\alpha^*}$ [mas] & 0.051 \\
    $\delta$ [deg] & 14.9309796086\tablefootmark{a}\\
    $\sigma_\delta$ [mas] & 0.052 \\
    $\varpi$ [mas] & $1.679\pm 0.069$\tablefootmark{b} \\
    $\mu^*_\alpha$ [mas yr$^{-1}$] & $-22.235\pm 0.062$\\
    $\mu_\delta$ [mas yr$^{-1}$] & $-155.276\pm0.059$\\
    $G$ [mag]& $11.2311\pm 0.0028$\\
    \bprp\ [mag]& $1.2156\pm 0.0048$\\
    \grvs\ [mag]& $10.2289\pm 0.0122$\\
    \teff\ [K] & 5340$_{-198}^{+212}$  \\
    \logg\ & 3.08$_{-0.30}^{+0.36}$ \\
    \mh & $-2.76_{-0.09}^{+0.19}$ \\
    \aabun & $0.54\pm 0.06$ \\
    RV [\kms] & $-333.2\pm 3.4$ \\
    \hline
    \end{tabular}
    \tablefoot{\tablefoottext{a}{The reference epoch of DR3 coordinates is J2016.0 (JD 2457389.0).}
    \tablefoottext{b}{A zero-point correction \citep{2021A&A...649A...4L} of 35.4\muas\ has been applied to the parallax value given in the catalogue.}
}
\end{table}

The source is not known as a variable star in the literature, neither in the AAVSO International Database, nor in the ASAS-SN database. We inspected ASAS-SN, ZTF, and TESS photometry, finding that the source does not present any significant periodic variability. The source was not observed with XMM-Newton, Chandra nor GALEX, nor it is present in the RAVE, APOGEE, LAMOST, or GALAH spectroscopic surveys. No eROSITA data have been made available yet for \gaia BH3, which belongs to the eastern Galactic hemisphere.


\subsection{Astrometry and orbital solution}\label{sec:astrometry}

\begin{figure*}
    \centering
    \includegraphics[width=\textwidth]{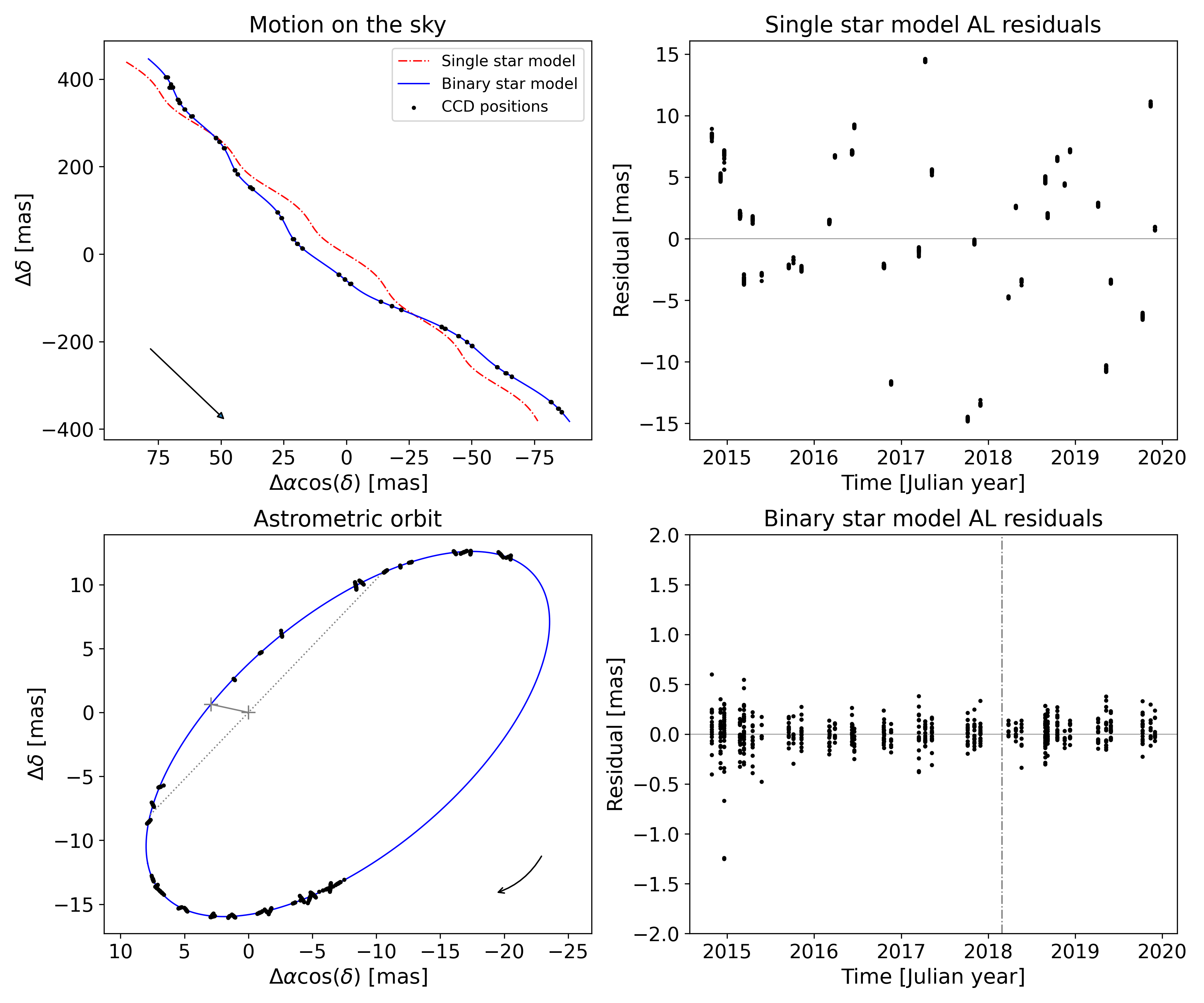}
    \caption{Astrometric data of \gaia BH3. Top-left panel:\ Motion on the sky of the photocentre of the source, as seen by \gaia in the different CCD transits (dots), compared with the best fitting single-star solution from AGIS and the astrometric-binary solution from the NSS pipeline; the arrow indicates the direction of the proper motion. Bottom-left panel:\ Derived astrometric orbit of the photocentre, after a subtraction of parallax and proper motion, compared with the astrometric measurements.
    We note that only one-dimensional (1D) along-scan (AL) astrometry was used by the NSS pipeline. The position of the photocentre on the sky corresponding to each measurement is derived combining the measured one-dimensional AL position and the assumed orbital solution.
    The + signs show the barycentre and the position of the  periastron, the dotted line shows the line of nodes, and the arrow indicates the direction of the motion along the orbit. In the top-right and bottom-right panels, we can see the residuals of the along-scan (AL) astrometric measurements for, respectively, the single-star solution and the binary-star solution. The vertical dot-dashed line in the bottom-right panel marks the time of the periastron passage.}
    \label{fig:astrometry}
\end{figure*}

The system was identified while validating astrometric binaries orbital solutions produced by the Non-Single Star (NSS) pipeline in a preliminary run (identified as NSS 4.1), done in preparation of \gaia DR4. The NSS pipeline used in the NSS 4.1 run is similar to the one used in the \gdrthree, which is described in \citet{2023A&A...674A...9H} and in Section 7.2.2 of the DR3 NSS documentation \citep{NSS-DR3-documentation}, with improvements in the filtering of spurious solutions.
The NSS pipeline processed astrometric data produced by preliminary runs of the Intermediate Data Update \citep[IDU; see][]{2016A&A...595A...3F} and the Astrometric Global Iterative Solution \citep[AGIS; see][]{2021A&A...649A...2L} pipelines,
covering the time range from JD 2456941.6218 to JD 2458869.4177 (TCB\footnote{TCB: Barycentric Coordinate Time, the time scale used here for all \gaia dates.}), for a total of about 64 months. 
The NSS 4.1 run was executed on an input list of 10.4 million sources, chosen to be brighter than $G_{\mathrm{RVS}}=14$\,mag, and produced almost 1.5 million orbital solutions. We note that the final NSS run for DR4 will extend to fainter magnitudes, and it is expected to produce a much larger number of binary solutions. Further details on the NSS 4.1 run can be found in \appref{sec:nssrun}.

For each orbital solution, we computed the astrometric mass function from the angular semi-major axis of the photocentre orbit ($a_0$), period ($P$), and parallax ($\varpi$) as: 
\begin{equation}
   f_\text{M}=(a_0/\varpi)^3(1\,\mathrm{yr}/P)^2\Msun~.  
\end{equation}

For an astrometric binary, the mass function depends on the masses of the components (${M}_1$, ${M}_2$) and on their flux ratio, ${\cal F}_2/{\cal F}_1$ \citep{2023A&A...674A...9H}. For an invisible companion (${\cal F}_2/{\cal F}_1=0$) the mass function simplifies to:
\begin{equation}\label{eq:fM}
f_\text{M} = {M}_2 \left(\frac{{M}_2}{{M}_1+{M}_2}\right)^2\, ,
\end{equation}
from which it follows that $M_2\ge f_\text{M}$. Given $f_\text{M}$ and an estimate of the mass of the visible component ($M_1$), Eq.~(\ref{eq:fM}) can be used to solve for the mass of the dark companion ($M_2$).

Among the 1.5 million orbital solutions, \gaia BH3 yielded the largest mass function, $32.03\pm0.64$\Msun, with a significance ($a_0/\sigma_{a_0}$) of 48.1; 
no other solution has a mass function larger than 20\Msun. In
\figref{fig:astrometry}, we show the orbital solution and the residuals, from which the strength of the 
astrometric signal of the orbit, along with the the robustness and quality of the solution can be appreciated.
The Campbell orbital elements of the source are reported in the central column of \tabref{tab:orbit_params}. We note that the NSS pipeline used in this preliminary run produces Thiele-Innes elements; the Campbell elements and their uncertainties were computed using the equations in Appendix A of \citet{2023A&A...674A...9H}. The astrometric mass function value and its uncertainty were computed using Monte Carlo resampling of the Thiele-Innes elements, the parallax and the period, in order to take into account the correlations between parameters; in particular, between $a_0$ and the period. To make sure this procedure would yield reliable results, we first checked that the correlations are sufficiently well behaved to allow for Monte Carlo resampling, following Section 6.1 of \citet{2023A&A...674A..32B}. 

A word of caution is necessary on the parallax value, and thus on the astrometric mass function: as in previous releases, the \gaia parallaxes are affected by a small bias \citep[see][]{2021A&A...649A...4L}, but we do not have enough information at this stage to quantify the bias for the preliminary NSS solutions. As a consequence, the uncertainty on the mass function reported in \tabref{tab:orbit_params} is underestimated.

\begin{table}[t]
    \caption{Campbell orbital elements of the \gaia BH3 system and the astrometric parameters of its barycentre.}
    \label{tab:orbit_params}
    \centering
    \begin{tabular}{lcc}
    \hline
    \hline
    Parameter & Astrometric & Combined\\
              & solution & solution \\
    \hline
    $\alpha$ [deg] & 294.8278502411\tablefootmark{a} & 294.8278502301\tablefootmark{a} \\
    $\sigma_{\alpha^*}$ [mas] & 0.060 & 0.054\\
    $\delta$ [deg] &  14.9309190720\tablefootmark{a} & 14.9309190869\tablefootmark{a}\\
    $\sigma_\delta$ [mas] & 0.086 & 0.074 \\
    $\varpi$ [mas] & $1.6747\pm 0.0094$ & $1.6933\pm 0.0164$\tablefootmark{b} \\
    $\mu^*_\alpha$ [mas yr$^{-1}$] & $-28.372\pm 0.077$ & $-28.317\pm 0.067$\\
    $\mu_\delta$ [mas yr$^{-1}$] & $-155.150\pm 0.129$ & $-155.221\pm 0.111$\\
    $P$ [days] & $4194.7\pm 112.3$ & $4253.1\pm 98.5$\\
    $e$ & $0.7262\pm 0.0056$ & $0.7291 \pm 0.0048$\\
    $a_0$ [mas] & $27.07\pm 0.56$ & $27.39\pm 0.49$\\
    $i$ [deg] & $110.659\pm 0.107$ & $110.580 \pm 0.095$\\
    $T_p$ [JD, TCB] & $2458177.28\pm 0.98$ & $2458177.39\pm 0.88$\\
    $\Omega$ [deg] & $136.200\pm 0.147$ & $136.236\pm 0.128$\\
    $\omega$ [deg] & $77.77\pm 0.66$ & $77.34 \pm 0.76$\\
    $a_1$ [AU] & ... & $16.17 \pm 0.27$\\  
    $\gamma$ [\kms] & ... & $-357.31\pm 0.44$\\
    $f_\text{M}$ [\Msun] & $32.03 \pm 0.64$ & $31.23 \pm 0.81$ \\
    GoF & 2.17 & $-0.53$ \\
    \hline
    \end{tabular}
    \tablefoot{\tablefoottext{a}{The reference epoch of DR4 coordinates is J2017.5 (JD 2457936.875).}
    \tablefoottext{b}{Derived as $a_0/a_1$, see text.}
    }
\end{table}


\subsection{Spectroscopy and combined orbital solution}\label{sec:comb_sol}

\begin{figure}
    \centering
    \columnImage{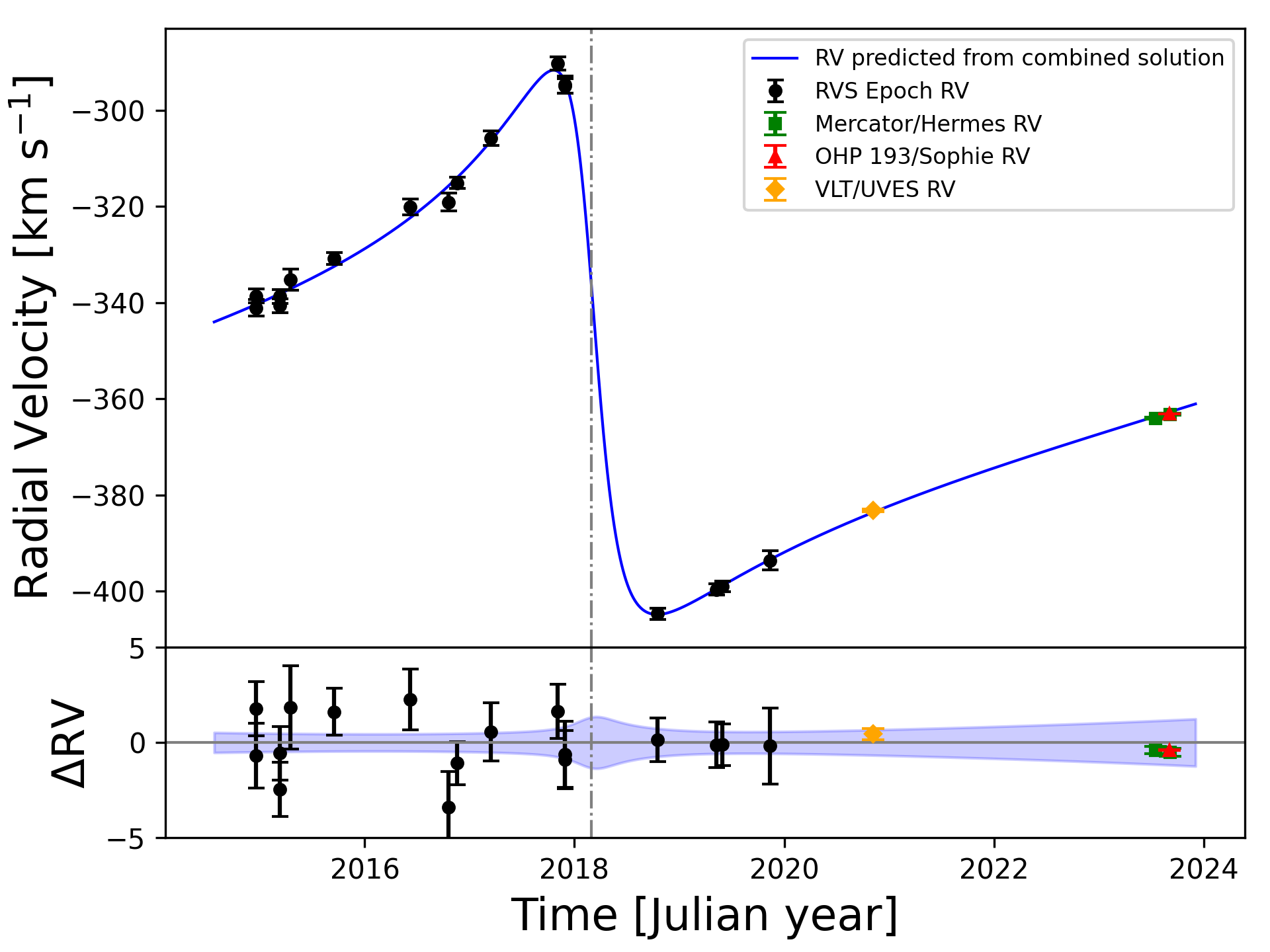}
    \caption{Radial-velocity evolution of \gaia BH3. Top panel: Comparison between the radial-velocity evolution predicted from the combined \gaia astrometric-spectroscopic binary model (blue solid line) and the epoch radial velocities measured with the \gaia RVS instrument (black filled circles), and ground-based measurements for \gaia BH3.  Bottom panel: Radial-velocity residuals with respect to the binary solution  compared with the 1-$\sigma$ uncertainty of the predicted radial-velocity evolution (blue shaded area). The vertical dot-dashed line in both panels marks the time of the periastron passage.}
    \label{fig:radvel}
\end{figure}

\begin{figure}
    \centering
    \columnImage{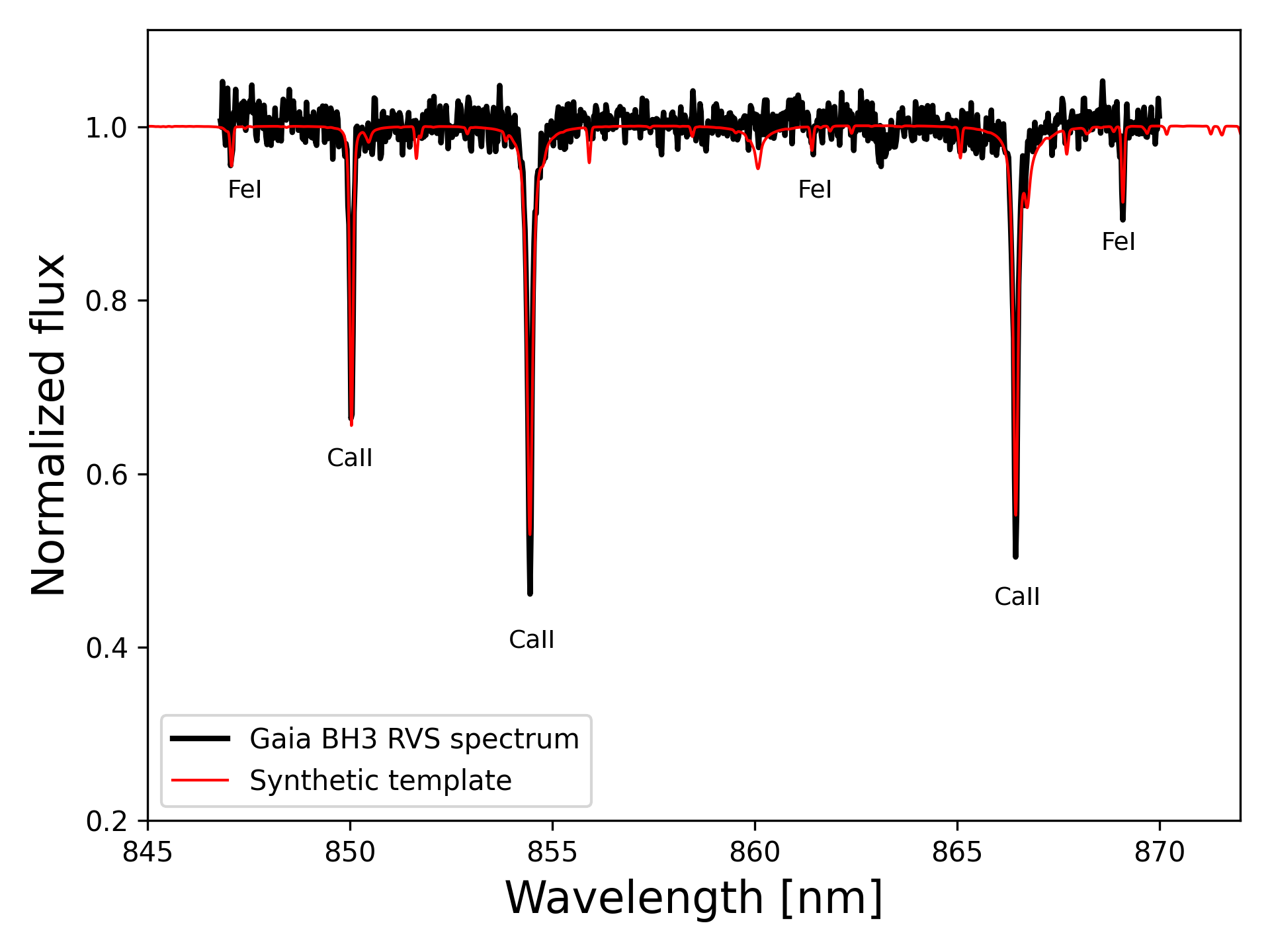}
    \caption{\gaia RVS combined spectrum of \gaia BH3, in restframe, compared with the template spectrum.}
    \label{fig:rvs_spec}
\end{figure}

The \gaia RVS \citep{2018A&A...616A...5C} data of sources with an orbital solution from the NSS 4.1 run were processed with the DR4 operational RVS pipeline. Improvements with respect to the DR3 version will be described in the \gaia DR4 documentation; the DR3 RVS pipeline is described in \citet{2018A&A...616A...6S}, \citet{2023A&A...674A...5K}, and \citet{CU6-DR3-documentation}. It is worth mentioning that the DR4 RVS pipeline includes the correction for the effect of the astrometric orbital motion, discussed in \citet{2023A&A...674A..25H}; in particular, Section 3.3.1.
The DR4 RVS data cover the time range from JD 2456863.9385 to JD 2458869.4177, namely, about 67 months.
The pipeline produced 17 valid epoch radial velocities for \gaia BH3, reported in \appref{sec:epoch_data}, using a template spectrum with $T_\mathrm{eff}=6000$\,K, $\log g =3.5$ and $[\mathrm{Fe/H}]=-1.5$ \citep[see][]{LL:RHB-005}. The template parameters were estimated by the pipeline from the RVS spectrum itself. 

We used the DR3 NSS pipeline code to compute an orbital solution combining astrometric data and \gaia RVS radial velocities (\texttt{nss\_solution\_type = AstroSpectroSB1}). The details of the adopted model are described in Section 7.7.3 of \citet{NSS-DR3-documentation}. 
We recall that in the combined solution model only the period, eccentricity, and periastron time are in common between the astrometric and the spectroscopic part of the solution. There is no constraint to impose that the semi-major axis of the photocentre orbit in AU ($=a_0/\varpi$) would be equal to the semi-major axis of the spectroscopic orbit $a_1$,
as expected in the case of a dark companion. 
The consistency between $a_0$ and $a_1$ allows us to check whether the flux ratio is indeed compatible with a value of zero.

As discussed in \secref{sec:astrometry}, the parallax derived by the NSS pipeline for the combined solution ($\varpi=1.6808\pm0.0086$\,mas) is affected by a bias which we cannot quantify. In order to avoid its effect on the mass function, we estimate the latter using $a_1$ instead of $a_0/\varpi$, namely: 
\begin{equation}
f_\text{M}=\left(\frac{a_1}{1\,\mathrm{AU}}\right)^3 \left(\frac{1\,\mathrm{yr}}{P}\right)^2\Msun~,    
\end{equation}
resulting in a value of $31.23\pm 0.81$\Msun. Assuming the equality between the photocentre and the spectroscopic orbit, we can also provide an alternative estimation of the parallax as 
 \begin{equation}
  \varpi=a_0/a_1~,
 \end{equation}
which results in a value of $1.6933\pm 0.0164$\,mas.

The Campbell orbital elements of the combined solution for \gaia BH3 are reported in  \tabref{tab:orbit_params}. 
The combined solution is very similar to the astrometric solution, with slightly smaller uncertainties, and a better goodness-of-fit \citep[GoF, in the Hipparcos sense, see][]{NSS-DR3-documentation}, as a result of a stronger filtering of outliers.
Radial velocities predicted by the combined solution are compared with the measurements in \figref{fig:radvel}. In \figref{fig:rvs_spec}, we show the combined and normalised 
RVS spectrum (see Seabroke et al. in prep.) compared with the template spectrum.

Given the extreme value of the mass function of the system and the importance of its detection, a confirmation with ground-based observations was indispensable to discard the possibility of a spurious solution. We thus observed \gaia BH3 with the HERMES spectrograph \citep{2011A&A...526A..69R} mounted on the 1.2-meter Mercator telescope  at the Roque de los Muchachos Observatory (Spain), and with the SOPHIE spectrograph \citep{2008SPIE.7014E..0JP} mounted on the 1.93-meter telescope of the Observatoire de Haute-Provence (France). 
A search in the ESO archive revealed that the source was observed with the UVES spectrograph \citep{2000SPIE.4008..534D} mounted on the VLT. Details on the data reduction of these observations can be found in \appref{sec:gb_spec}. The spectra do not show any sign of the presence of a second component, nor of continuum filling of absorption lines; furthermore, no emission line was detected.
Radial velocities were derived for each ground-based observation and their values  are reported in \tabref{tab:rv_ground}. Although these values were not used to derive the orbital solution described above, they are in agreement with the predicted radial velocity within 0.5\,\kms; this is less than the uncertainty of the orbital solution,
as can be seen in \figref{fig:radvel}. This result confirms the reality and accuracy of the orbital solution derived from \gaia data. 


\subsection{Stellar parameters, abundances and Galactic orbit}

We derived new stellar parameters of the luminous component, using the \gdrthree photometry ($G$ magnitude and \bprp\ colour), the parallax from the combined astrometric-spectroscopic solution, and the extinction ($A_0$) derived from the dust extinction maps of \citet{2022A&A...664A.174V}, with an iterative procedure described in \appref{sec:derstellar}.
The UVES spectrum was used to determine the metallicity and abundances (see \appref{sec:abundances}). We then compared the extinction-corrected absolute $G$ magnitude ($M_{\mathrm{G},0}$) and the dereddened colour $(\bprp)_0$ with the ones given by the isochrones libraries  
PARSEC \citep{2012MNRAS.427..127B} and BaSTI \citep{2021ApJ...908..102P}. Thus, we derived the mass ($M_\star$) of the visible component.
The parameters are reported in \tabref{tab:new_params}.

\begin{table}[t]
    \caption{Stellar parameters of \gaia BH3 derived in this work.}
    \label{tab:new_params}
    \centering
    \begin{tabular}{lc}
    \hline
    \hline
    Parameter & Value\\
    \hline
    \teff\ [K] & $5212\pm 80$  \\
    \logg & $2.929\pm 0.003$ \\
    \feh & $-2.56\pm 0.11$ \\
    \aabun & $0.43\pm 0.12$ \\
    $[\mathrm{M/H}]$ & $-2.21\pm 0.15$ \\
    $\log(L_\star/L_\sun)$ & $1.208\pm 0.030$ \\
    $M_\star$ [\Msun] & $0.76\pm 0.05$ \\
    $R_\star$ [\Rsun] & $4.936\pm 0.016$ \\
    $M_{\mathrm{G},0}$ [mag]& $1.778\pm 0.082$ \\
    $(\bprp)_0$ [mag]& $0.921\pm 0.031$ \\
    \hline
    \end{tabular}
\end{table}

\begin{figure}
    \centering
   \columnImage{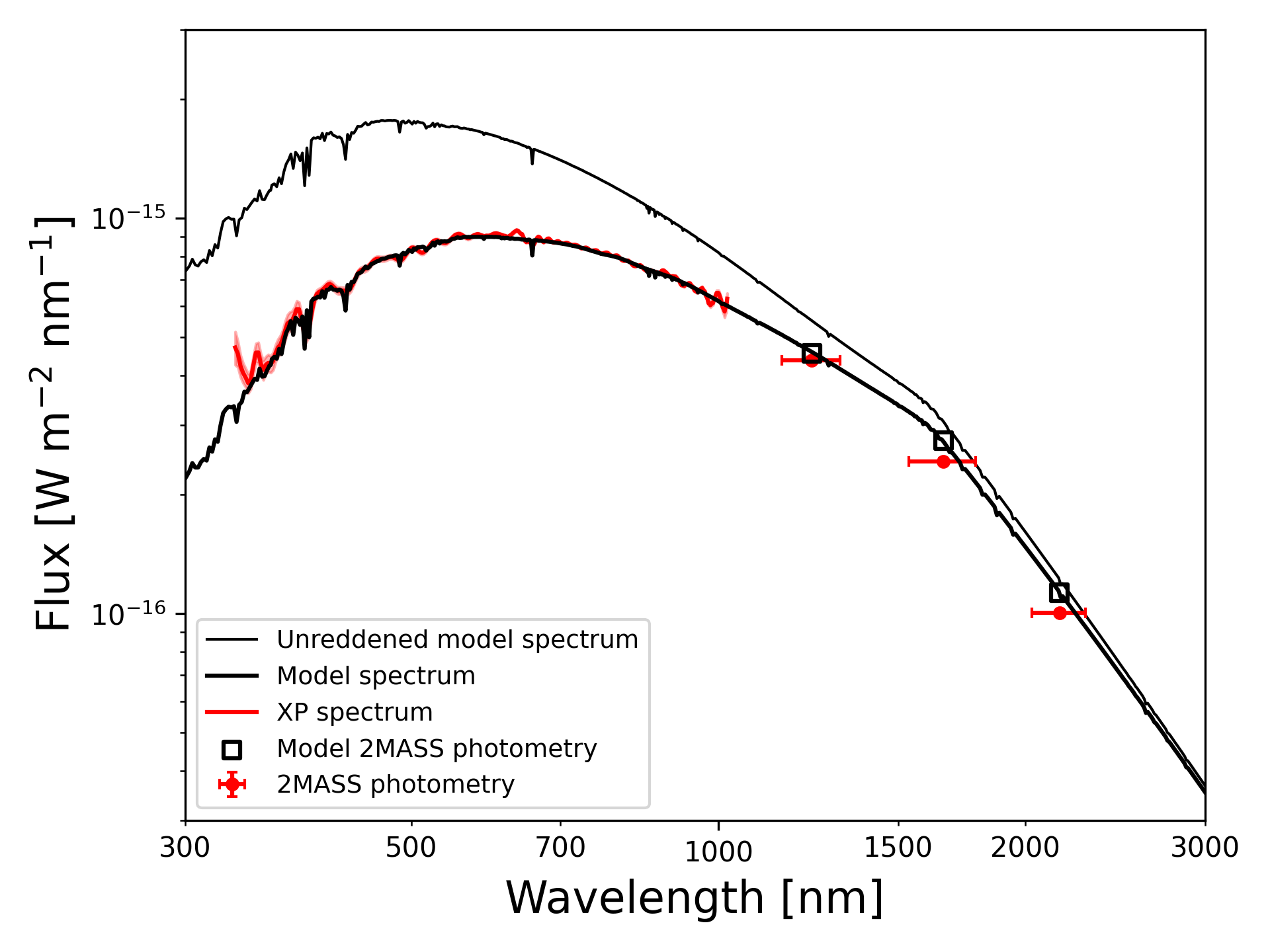}
    \caption{\gaia BH3 modelled SED, compared with the \gaia XP spectrum and 2MASS photometry. The thin black line shows the unreddened model, while the thick line shows the SED assuming $A_0=0.71$\,mag.}
    \label{fig:sed}
\end{figure}

In \figref{fig:sed}, we compare the expected spectral energy distribution (SED) with 
the \gaia XP spectrum \citep{2021A&A...652A..86C,2023A&A...674A...2D,2023A&A...674A...3M} and 2MASS photometry. The agreement between the predicted SED and the \gaia XP spectrum is very good, with the only exception of the blue edge, where the XP spectrum is noisier.

The abundances of \gaia BH3 (reported in \appref{sec:abundances}) show that the star is $\alpha$-enhanced, as expected for a very metal-poor star.
There is no trace of \element[][13]{C} in the spectrum and the $[\mathrm{Ba/Fe}]$ is nearly solar, indicating the star has not been enriched by material processed in the CNO cycle, as expected if it had, for instance, accreted material from a companion star in the AGB phase.
The star has no chemical peculiarity, except an enhancement of Eu ($[\mathrm{Eu/Fe}]=0.52$). Thus, it can be classified as an r-I neutron-capture-rich star, following the classification of \citet{BC2005}.

Using the systemic radial velocity, proper motion, 
position, and distance, and assuming the Milky Way gravitational potential from \citet{2017MNRAS.465...76M}, we may find that the source has a high-energy retrograde orbit\footnote{Here we use the same conventions as in \citet{2019MNRAS.488.1235M}.} ($E=-1.29\times 10^5$ km$^2$ s$^{-2}$, $L_z=-2.3\times 10^3$ kpc \kms, $L_{\perp} = 1.05\times 10^3$ kpc km s$^{-1}$) in the Galaxy.
Its kinematic characteristics are consistent with those of the halo substructure known as Sequoia \citep{2019MNRAS.488.1235M}, but are also in agreement to those of the recently discovered ED-2 stream \citep{2023A&A...670L...2D,2023A&A...678A.115B}, a likely remnant of a globular cluster that was disrupted by the Milky Way. The metallicity of \gaia BH3 is more consistent with ED-2 ($[\mathrm{Fe/H}]=-2.6\pm0.2$) than with the median metallicity of Sequoia ($[\mathrm{Fe/H}]\sim-1.7$). 


\section{Discussion}\label{sec:discussion}

With an estimated mass of $0.76\pm 0.05\Msun$ for the luminous companion, 
we derived a mass of: 
\begin{equation}
 M_{\mathrm{BH}}=32.70\pm 0.82\Msun~, 
\end{equation}
for the dark companion. 
The observed luminosity of \gaia BH3 is too low by several orders of magnitude to be compatible with the hypothesis that the companion is a main sequence star or even two main sequence stars in a close orbit.
The estimated mass is also too large for one neutron star or two neutron stars in a close orbit, so we are left with the possibility of: 1) a single BH; 2) an inner binary containing two BHs; or 3) a BH and another compact object. Although the single BH is the simplest explanation, the hypothesis of an inner binary of two BHs cannot be excluded. \citet{2023ApJ...958...26H} proposed a method to test this hypothesis by detecting radial-velocity perturbations at the periastron. Using the \citet{2023ApJ...958...26H} formulation, we estimated radial-velocity perturbations\footnote{We use $K_{\mathrm{short}}(1-e)^{-7/2}$ as perturbations level estimation,  \citep[see Sect. 2.1 in][]{2023ApJ...958...26H}, assuming an inner equal-mass binary with a circular coplanar orbit, and a period corresponding to the maximum allowed by dynamic stability \citep[126 days, according to eq. 6 in][]{2023ApJ...958...26H}.} with a maximum amplitude of the order of 0.2 \kms. Such perturbations are too small to be detected in the \gaia RVS data, but can be verified with ground-based instruments \citep[see][for an application to \gaia BH1]{2024PASP..136a4202N}.
For the purposes of the subsequent discussion, we have adopted the single BH hypothesis as the most likely explanation.

The estimated mass of the BH in \gaia BH3 makes it the most massive BH of stellar origin discovered in our Galaxy. It is striking that the only BH with a mass larger than 20\Msun found in the \gaia data so far is in orbit with a very metal-poor star, while such stars make up only a tiny fraction of the stars analysed in the NSS pipeline run (0.4\% of sources which produced a binary solution have  $[\mathrm{M/H}] < -2$ from DR3 GSP-Phot). Such stars also make up 
a small fraction of our Galactic halo \citep[less than 
5\%\ according to][]{bonifacio22}
where this star and the majority of metal-poor stars are located.
Although we can not exclude that this BH is the result of the merger of two less massive BHs, this discovery strongly supports the scenario where high-mass BHs are remnants of low-metallicity stars. 
The above considerations also raise the question of the maximum metallicity value for the formation of high-mass BHs, which in \citet{2016Natur.534..512B} is identified at $[\mathrm{M/H}] =-1$. The much lower metallicity of \gaia BH3 may be an indication that high-mass BHs form only at very low metallicities rather than at moderately low ones.

An in-depth discussion of the possible formation scenarios for this binary system is beyond the scope of the paper; nevertheless,   a few aspects ought to be highlighted. 
As discussed in \cite{2023MNRAS.518.1057E,2023MNRAS.521.4323E}, the formation of the \gaia BH1 and BH2 systems as isolated binaries is unlikely. This is also true for the recently discovered \gaia NS1 system \citep{2024arXiv240206722E}, composed of a high-mass neutron star and a low-metallicity star.  Given the size of their orbits, these systems should have experienced a common-envelope phase and then a mass transfer toward the light companion, which would then have resulted in much closer orbits than the observed ones. 
For \gaia BH2, the common-envelope phase could have been avoided if the BH progenitor was more massive than 65~\Msun.
In the case of \gaia BH3, the present-day minimum separation is of the order of 1000~\Rsun and the common-envelope phase could not have been avoided because models predict that the BH progenitor becomes a red supergiant even at 150~\Msun \citep{2015MNRAS.452.1068C}.
Similarly to \gaia BH1 and BH2, the chemical composition of the luminous component does not show any unusual abundance; in particular, the absence of \element[][13]{C} and the observed $[\mathrm{Ba/Fe}]$ point toward a lack of contamination by the BH progenitor during its evolution.
The observed enhanced Eu abundance could be due to the contamination from the SN at the birth of the BH, but also due to the medium in which the star formed.
An alternative formation scenario, proposed to explain the \gaia BH1 and BH2 systems, is that the BH acquired the low-mass companion via dynamical exchange in a dense environment \citep[see for example][]{2023MNRAS.526..740R,2024MNRAS.527.4031T,2024ApJ...965...22D}.  Such a scenario might be supported by the probable association of \gaia BH3 with the ED-2 stream, which could be a remnant of a globular cluster \citep{2023A&A...670L...2D,2023A&A...678A.115B}.


\section{Conclusions}\label{sec:conclusion}

In this Letter, we present the discovery of a wide binary composed of a very metal-poor giant orbiting a dark object of 33~\Msun, 
using \gaia preliminary DR4 astrometric data, corroborated by \gaia spectroscopy. Most probably, the massive dark object is a single black hole (BH).
The 33~\Msun\ of the BH mass makes it the
most massive BH of stellar origin discovered in our Galaxy. All Galactic BHs that reside in short-period X-ray binaries have masses generally below 10~\Msun \citep[e.g.][]{2016A&A...587A..61C}, except Cyg-X1 ($M_\mathrm{BH} \sim 20 \Msun$). Even the first two dormant BHs discovered by \gaia in wide astrometric orbits have masses of about 10~\Msun. The mass of \gaia BH3 puts it in the mass range of the BHs discovered by gravitational waves \citep[e.g.][]{2023PhRvX..13a1048A}, and, in fact, it is close to the peak of the observed mass distribution for the merging BHs \citep[e.g.][]{2024ApJ...962...69F}.
The metallicity of the system supports the scenario \citep{2016Natur.534..512B} that the high-mass BHs observed by LIGO/Virgo/KAGRA \citep{2020LRR....23....3A} are the remnants of metal-poor stars. 

The discovered system, with its extremely low-mass ratio, wide orbit, and specific chemical composition, can also provide constraints for stellar evolution and binary models. As in the case of the \gaia BH1 and BH2 systems, the formation scenario as an isolated binary appears unlikely and alternative scenarios should be considered. 
The BHs discovered by \gaia in wide binaries in our Galaxy and those detected by LIGO/Virgo/KAGRA in external galaxies (i.e. BH merger events of extremely short-period binaries) constitute two ends of the BH population. When studied together, they can help to formulate a comprehensive view of BH formation and the evolution of their progenitors.

Finally, the bright magnitude of the system and its relatively small distance makes it an easy target for further observations and detailed analyses by the astronomical community. This discovery should be also seen as a preliminary teaser for the content of \gaia DR4, which will undoubtedly reveal other binary systems hosting a BH.


\begin{acknowledgements} 
This work has made use of data from the European Space Agency (ESA) mission
{\it Gaia} (\url{https://www.cosmos.esa.int/gaia}), processed by the {\it Gaia}
Data Processing and Analysis Consortium (DPAC,
\url{https://www.cosmos.esa.int/web/gaia/dpac/consortium}). Funding for the DPAC
has been provided by national institutions, in particular the institutions
participating in the {\it Gaia} Multilateral Agreement.
Based on observations made with the Mercator Telescope, operated on the island of La Palma by the Flemish Community, at the Spanish Observatorio del Roque de los Muchachos of the Instituto de Astrofisica de Canarias. Based on observations obtained with the HERMES spectrograph, which is supported by the Research Foundation - Flanders (FWO), Belgium, the Research Council of KU Leuven, Belgium, the Fonds National de la Recherche Scientifique (F.R.S.-FNRS), Belgium, the Royal Observatory of Belgium, the Observatoire de Gen\`eve, Switzerland and the Th\"uringer Landessternwarte Tautenburg, Germany. 
This publication has also made use of observations collected with the SOPHIE spectrograph on the 1.93-m telescope at Observatoire de Haute-Provence (CNRS), France (program 23B.PNPS.AREN) using support by the French Programme National de Physique Stellaire (PNPS).
Based on observations collected at the European Southern Observatory under ESO programme 106.21JJ.001.
We warmly thank Piercarlo Bonifacio for help with the use of ATLAS9 models and helpful discussions, Hans Van Winckel, HERMES PI, for granting observational time, and Rosine Lallement for helpful discussions on the use of extinction maps. We thank the referee for comments that helped to improve the paper.

The full acknowledgements are available in \appref{sec:ack}.
\end{acknowledgements}

\bibliographystyle{aa}
\bibliography{biblio}


\begin{appendix}

\section{Astrometric processing in the NSS 4.1 run}\label{sec:nssrun}

Here, we provide the details of the Non-Single Star (NSS) pipeline used in the preliminary run NSS 4.1, underlining this is not the final version of the NSS pipeline for the generation of \gaia DR4 data. 

The pipeline is similar to the one used in \gdrthree \citep{2023A&A...674A...9H}, with a few updates -- the main differences are in the improvement of the filters to remove spurious solutions. In particular, two new filters were introduced to remove sources which are partially resolved: the first one filters out sources with significant scan-angle-dependent signals \citep{2023A&A...674A..25H} in the $G$ flux before attempting an astrometric solution; the second one removes solutions which have periods that correspond to combinations of the \gaia spacecraft precession frequency and the yearly frequency \citep[see][Sect. 4.1]{2023A&A...674A..25H}. These new filters allow us to relax  the filters based on the significance and on $\varpi/\sigma_\varpi$, described in \citet{2023A&A...674A...9H}, which are replaced by the following criteria: the goodness-of-fit (GoF) must be smaller than 15, the eccentricity error smaller than 0.2, semi-major axis significance $a_0/\sigma_{a_0} > 5$, and
\begin{equation}
\varpi/\sigma_\varpi > \mathrm{max}\left(15,-208.02\cdot\log(P/1\,\mathrm{day})+548.03\right)~.
\end{equation}

The NSS 4.1 run was executed using astrometric data produced with preliminary runs (namely IDU 4.1 and AGIS 4.1) of the Intermediate Data Update \citep[IDU; see][]{2016A&A...595A...3F} and the Astrometric Global Iterative Solution \citep[AGIS; see][]{2021A&A...649A...2L} pipelines. The preliminary astrometry provided by AGIS 4.1 covers the entire range of DR4, i.e. from JD 2456863.9385 to JD 2458869.4177, for a total of about 67 months, but only data after JD 2456941.6218 were used in the NSS 4.1 run. 
The local perspective effect \citep{2009MNRAS.394.1075H} was not included in the model for the run NSS 4.1 and the variability-induced mover (VIM) solutions were not attempted.

The NSS 4.1 run was executed on a list of 10\,450\,939 sources chosen with the following criteria: the source must be brighter than $G_{\mathrm{RVS}}=14$\,mag and either $G<18$\,mag, an astrometric renormalised unit weight error (RUWE) larger than $1.05$, or $\varpi > 5$\,mas and RUWE $> 0.9$, and a number of visibility periods used in AGIS solution larger than 11. In order to exclude partially resolved sources, sources with a percent of successful Image Parameter Determination (IPD) windows with more than one peak larger than 10\% or with amplitudes of the IPD GoF versus the scan angle larger than 0.2, were excluded (see \gdrthree archive documentation for details on the above quantities).
The run produced a total of 1\,469\,196 orbital solutions. 

The selection function of the NSS 4.1 run is not trivial to characterise, thus, it is not possible to estimate how common or rare are systems like \gaia BH3. If we push Gaia BH3 to the distance corresponding to the cut in magnitude ($G_\mathrm{RVS} = 14$ mag would correspond to a distance of 3.3 kpc, ignoring the extinction), the semi-major axis of the orbit would be 4.8 mas, which is still very large with respect to the precision of Gaia epoch measurements at that magnitude. However, given that the DR4 time range covers only half of the orbital period, the resulting significance could be below the acceptance thresholds. 
We note that during the \gdrthree preparation, \gaia BH3 produced an astrometric acceleration solution \citep{2023A&A...674A...9H}
and an SB1 solution (Gosset et al. in prep.), which were both discarded from the release due to a low significance, because the orbital period was much longer than the \gaia DR3 time span and the periastron passage was not covered by the DR3 time range.

If we consider a Gaia BH3-like system but with an orbital period similar or shorter of the DR4 time range (i.e. below 2000 days), the significance would be always higher than the acceptance threshold solution, with the exception of very short ($<$20 day) periods. We note that the NSS 4.1 is limited to periods larger than 10 days.  For binaries with shorter periods, the giant would almost fill its Roche lobe and the source would probably be detected as a X-ray source. 

Although it has not yet been finalised, the input list for DR4 will be significantly larger than for NSS 4.1, probably built as the sum of a volume-limited sample and a $G<18$ magnitude-limited sample, as indicated above, though without the $G_{\mathrm{RVS}}<14$\,mag criterion. The motivation for the dedicated NSS 4.1 run and the reason for the latter criterion was the analysis of the effect of a deviation of the astrometry from the assumed single-star model on the calibration of the spectroscopic instrument.


\section{\gaia epoch data}\label{sec:epoch_data}

Here, we describe the \gaia epoch astrometric data and epoch radial velocities used to produce the binary solution of \gaia BH3.

The astrometric measurements of \gaia BH3 are provided in \tabref{tab:epoch_astrometry}. They were produced from preliminary pipelines, provisional instrument models and calibrations; as a consequence they will not be identical (but still similar) to the corresponding data to be produced and published for this star with DR4. Furthermore, the final epoch astrometry table in DR4 will contain many additional details and quality diagnostics on the individual measurements. A full explanation of the epoch astrometry is beyond the scope of the present short appendix. We refer the reader to the Gaia Technical Document \citet{LL:LL-061} and to \citet{2012A&A...538A..78L}. 

Each \gaia epoch astrometry record in \tabref{tab:epoch_astrometry} corresponds to a transit of the source on one of the CCDs of the AF instrument. 
The Table is arranged as follows. Col. 1:\ \fieldName{transit_id}\footnote{A decoder for the \fieldName{transit_id} is available on-line at
\url{https://gaia.esac.esa.int/decoder/transitidDecoder.jsp}}, a unique identifier assigned to each detected celestial light source as its image transits the \gaia focal plane; Col. 2:\ AF CCD strip;
Col. 3:\ Barycentric time, in JD, corresponding to the middle of the 4.41-second CCD exposure time; 
Col. 4:\ Along-scan position of the photocentre, with its associated uncertainty, which corresponds to the longitude of the observed photocentre in a 2D tangential coordinate system having its origin at a reference equatorial position, and having the axis of its longitude coordinate oriented corresponding to the scanning direction; 
Col. 5:\ Parallax factor, namely, the quantity by which it is necessary to multiply the parallax in order to obtain the contribution to the along-scan position due to the orbit of the spacecraft with respect to the Solar System barycentre;
Col. 6:\ Scan angle, which is the angle of the scanning direction with respect to local ICRS North;
Col. 7:\ Outlier flag, indicating whether the measurement was considered as an outlier (flag = 1) by the NSS pipeline and filtered out, or not (flag = 0), when solving for the astrometric-spectroscopic combined solution.

The reference position ($\alpha_0$, $\delta_0$), in the sense of \citet{LL:LL-061}, for \gaia BH3 is
\begin{equation}
\alpha_0=294\fdg82784900557243~,
\delta_0=14\fdg930918410309376~,
\end{equation}
while the reference time is J2017.5 (JD 2457936.875).

\begin{table*}[t]
    \caption{\gaia BH3 epoch astrometry.}
    \label{tab:epoch_astrometry}
    \centering
    \begin{tabular}{ccccccc}
    \hline
    \hline
        \fieldName{transit\_id} & AF    & Time      & Along-scan position & Parallax factor & Scan angle & Outlier\\
                                & strip & [BJD, TCB] & [mas] &  & [deg] & flag\\
        \hline
 20114916805338633 & 1 & 2456958.110978 & $ 147.066 \pm  0.370$ &  0.70827985 &  $-59.04672662$ & 0\\
 20114916805338633 & 2 & 2456958.111034 & $ 146.696 \pm  0.231$ &  0.70827991 &  $-59.04677105$ & 0\\
 20114916805338633 & 3 & 2456958.111091 & $ 146.685 \pm  0.183$ &  0.70828003 &  $-59.04681553$ & 0\\
 20114916805338633 & 4 & 2456958.111156 & $ 146.557 \pm  0.151$ &  0.70828015 &  $-59.04686711$ & 0\\
 20114916805338633 & 5 & 2456958.111203 & $ 146.396 \pm  0.097$ &  0.70828021 &  $-59.04690435$ & 0\\
 20114916805338633 & 6 & 2456958.111259 & $ 146.374 \pm  0.088$ &  0.70828027 &  $-59.04694869$ & 0\\
 20114916805338633 & 7 & 2456958.111325 & $ 146.436 \pm  0.100$ &  0.70828038 &  $-59.04700014$ & 0\\
 20114916805338633 & 8 & 2456958.111372 & $ 146.060 \pm  0.100$ &  0.70828044 &  $-59.04703723$ & 1\\
 20114916805338633 & 9 & 2456958.111437 & $ 146.256 \pm  0.130$ &  0.70828056 &  $-59.04708855$ & 0\\
 20119009095238725 & 1 & 2456958.184993 & $ 146.244 \pm  0.364$ &  0.70847327 &  $-59.11055165$ & 0\\
 20119009095238725 & 2 & 2456958.185043 & $ 146.148 \pm  0.296$ &  0.70847344 &  $-59.11059886$ & 0\\
 20119009095238725 & 3 & 2456958.185105 & $ 146.192 \pm  0.267$ &  0.70847368 &  $-59.11065699$ & 0\\
 20119009095238725 & 4 & 2456958.185162 & $ 146.271 \pm  0.225$ &  0.70847386 &  $-59.11070963$ & 0\\
 20119009095238725 & 5 & 2456958.185218 & $ 146.047 \pm  0.273$ &  0.70847410 &  $-59.11076223$ & 0\\
 20119009095238725 & 6 & 2456958.185274 & $ 146.068 \pm  0.158$ &  0.70847428 &  $-59.11081482$ & 0\\
 20119009095238725 & 7 & 2456958.185330 & $ 146.023 \pm  0.177$ &  0.70847452 &  $-59.11086742$ & 0\\
 20119009095238725 & 8 & 2456958.185386 & $ 146.095 \pm  0.191$ &  0.70847470 &  $-59.11092003$ & 0\\
 ... & ... & ... & ... & ... & ... & ... \\
        \hline
    \end{tabular}
    \tablefoot{The full table is available at the CDS.}
\end{table*}

Epoch RVS radial velocities, reported in \tabref{tab:epoch_rvs}, were produced with the final pipeline, but not finalised with the post-processing; their values or uncertainties may slightly differ from the final DR4 values. As for the astrometry, the final epoch radial velocity table in DR4 will contain 
additional details and quality diagnostics on the individual measurements. 
Each epoch radial velocity record in \tabref{tab:epoch_rvs} corresponds to a transit of the source on the RVS CCDs. The provided observation time of the radial velocity corresponds to the mean of the observation times of the three CCDs used to collect spectra in the RVS during the transit. 

\begin{table}[t]
    \caption{\gaia BH3 epoch radial velocities from \gaia RVS.}
    \label{tab:epoch_rvs}
    \centering
    \begin{tabular}{ccccccc}
    \hline
    \hline
        \fieldName{transit\_id}  & Time & Radial velocity \\
                   & [BJD, TCB] & [\kms] \\
        \hline
22989619581449144  & 2457010.096166  & $-338.62 \pm 1.42$ \\
22993711812243969  & 2457010.170169  & $-341.08 \pm 1.70$\\
27565894819405940  & 2457092.856649  & $-338.73 \pm 1.41$\\
27569987068420645  & 2457092.930660  & $-340.63 \pm 1.43$\\
29624665178404300  & 2457130.091203  & $-335.28 \pm 2.20$\\
37997278005744779  & 2457281.508798  & $-330.88 \pm 1.24$\\
52687653750370648  & 2457547.177378  & $-320.11 \pm 1.61$\\
60072325242450363  & 2457680.722202  & $-319.09 \pm 1.87$\\
61702852954182073  & 2457710.207118  & $-315.11 \pm 1.12$\\
68202800816349412  & 2457827.754668  & $-305.82 \pm 1.53$\\
81052816639971536  & 2458060.142534  & $-290.29 \pm 1.42$\\
82486181452083612  & 2458086.062277  & $-294.93 \pm 1.53$\\
82490273693622584  & 2458086.136278  & $-294.67 \pm 1.75$\\
100312802889225719 & 2458408.450948  & $-404.70 \pm 1.14$\\
111669627072959442 & 2458613.834346  & $-399.65 \pm 1.21$\\
112765863248582085 & 2458633.660613  & $-398.97 \pm 1.10$\\
121996985218154856 & 2458800.596176  & $-393.59 \pm 2.00$\\
        \hline
    \end{tabular}
    \tablefoot{The table is also available at the CDS.}
\end{table}

A public code, illustrating the use of epoch astrometric and radial velocity data to produce an orbital solution for \gaia BH3 is available online\footnote{\url{https://www.cosmos.esa.int/web/gaia/gaia-bhthree}}.


\section{Ground-based spectroscopy}\label{sec:gb_spec}

We observed \gaia BH3 with the HERMES spectrograph \citep{2011A&A...526A..69R} mounted on the 1.2-meter Mercator telescope  at the Roque de los Muchachos Observatory (Spain), at two dates (17 July 2023 and 7 September 2023), taking two consecutive exposures of 2700\,s each night. The spectra have a spectral coverage from 377 to 900\,nm, a resolving power of 85\,000, and a S/N $\sim 43$ at 520\,nm.

We observed \gaia BH3 also with the SOPHIE spectrograph \citep{2008SPIE.7014E..0JP} mounted on the 1.93-meter telescope of the Observatoire de Haute-Provence (France) on 4 September 2023. The source was observed with a single exposure of 6000\,s; the spectra cover the range 387 to 694\,nm with a resolving power of 40\,000, and have a S/N $\sim 66$ at 520\,nm.
HERMES and SOPHIE spectra of \gaia BH3 are available on request from the corresponding author.

A search in the ESO archive revealed that the source was observed with the UVES spectrograph \citep{2000SPIE.4008..534D} mounted on the VLT, on 5 November 2020, in the program 106.21JJ.001 proposed by T. Matsuno and collaborators. The aim of the program was to derive a complete chemical inventory of stars belonging to Galactic accretion events.
The exposure time of the UVES spectrum was 900\,s in the 390+580 setting (spectral coverage 326 to 454\,nm and 476 to 684\,nm) with a slit of 0.7\arcsec, producing a resolving power of 58\,000 in the blue and 62\,000 in the red  and a S/N $\sim 100$  at 520\,nm. 
 
\begin{figure*}
    \centering
    \includegraphics[width=\textwidth]{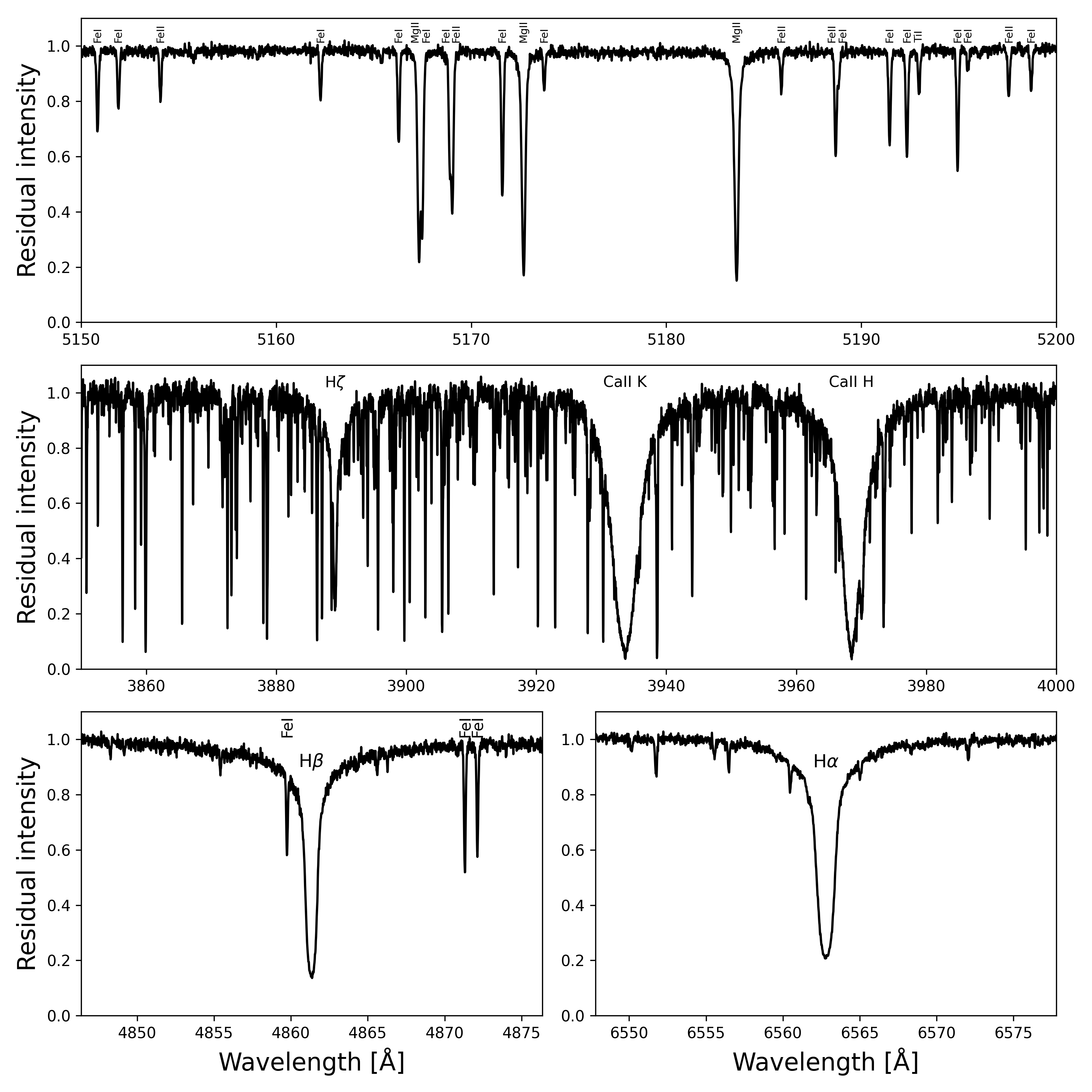}
    \caption{Residual intensity of the \gaia BH3 UVES spectrum in the magnesium triplet region (top panel), in the \ion{Ca}{ii} H+K region (middle panel), H$\beta$ (bottom-left) and H$\alpha$ (bottom-right).}
    \label{fig:uves_mg}
\end{figure*}

For HERMES and SOPHIE observations, radial velocities were derived by computing the cross-correlation functions with a G2 mask, while for the UVES spectrum the radial velocity was derived via template matching. 
The barycentric radial-velocity values are reported in  \tabref{tab:rv_ground}, and they are in good agreement with the orbit derived from \gaia data. The spectra do not show any sign of the presence of a second component, nor any emission line. The UVES normalised spectrum of the \gaia BH3 in selected spectral regions is shown in \figref{fig:uves_mg}.

\begin{table}[t]
    \caption{\gaia BH3 epoch radial velocities from ground-based observations.}
    \label{tab:rv_ground}
    \centering
    \begin{tabular}{lcc}
    \hline
    \hline
        Instrument & Time & Radial velocity \\
                   & [BJD, UTC] & [\kms] \\
        \hline
        UVES   & 2459158.5333 & $-383.20\pm 0.30$ \\
        HERMES & 2460143.4948 & $-364.05\pm 0.20$ \\
        SOPHIE & 2460192.4359 & $-363.15\pm 0.02$ \\
        HERMES & 2460193.4833 & $-363.24\pm 0.20$\\
        \hline
    \end{tabular}
\end{table}

\section{Derivation of stellar parameters}\label{sec:derstellar}

Here we describe the iterative procedure used to derive the stellar parameters of the luminous component in \gaia BH3. The procedure is similar to the one adopted in  \citet{lombardo21}.

We computed the emerging flux for a grid in \teff, \logg\ and \feh, of 1D plane-parallel model atmospheres, using ATLAS 9 \citep{2003IAUS..210P.A20C,2005MSAIS...8...14K}. The fluxes were then converted to spectral energy distributions 
by multiplying them by a factor $4\pi(R_\odot/10\,\mathrm{pc})^2$, where $R_\odot$ is the solar radius, as done in \citet{2014MNRAS.444..392C}. All models were computed assuming $[\alpha/\mathrm{Fe}]=0.4$. For each model, we computed theoretical values of the color \bprp, the bolometric correction (BC$_\mathrm{G}$), and the extinction coefficients $A_\mathrm{G}/A_0$ and $\ebpminrp/A_0$, using the average Milky Way reddening law from \citet{2019ApJ...886..108F}.

The iterative procedure starts with first-guess metallicity, temperature and gravity from \gaia DR3 values reported in \tabref{tab:basic_params}, a first-guess mass ($M_\star$) of 0.8\Msun, and a given reddening of $A_0$.
We use the above parameters to obtain \ebpminrp\ from the grid, which is then used to obtain a dereddened \bprp\ colour, $(\bprp)_0$ , the extinction $A_\mathrm{G}$, and the bolometric correction BC$_\mathrm{G}$.
We then compare the $(\bprp)_0$ value to theoretical colours in the grid to derive a new effective temperature, and we use the Stefan-Boltzmann equation  to derive a new surface gravity:
\begin{eqnarray}
    \log g &=& \log\frac{M_\star}{\Msun}+4\log\frac{T_\mathrm{eff}}{T_\odot}+\log g_\odot+ \\
    \nonumber
    &+& 0.4(G-A_\mathrm{G}+\mathrm{BC}_\mathrm{G}-m_{\mathrm{bol},\odot})+2\log\frac{\varpi}{1000\,\mathrm{mas}}~.
\end{eqnarray}
The procedure is repeated to convergence in \teff\ and \logg\, which is achieved after a few iterations.
The parameters \teff\ and \logg\ are then used to derive the metallicity \feh\ as described in \secref{sec:abundances} and the process repeated.

\begin{figure}
    \centering
   \columnImage{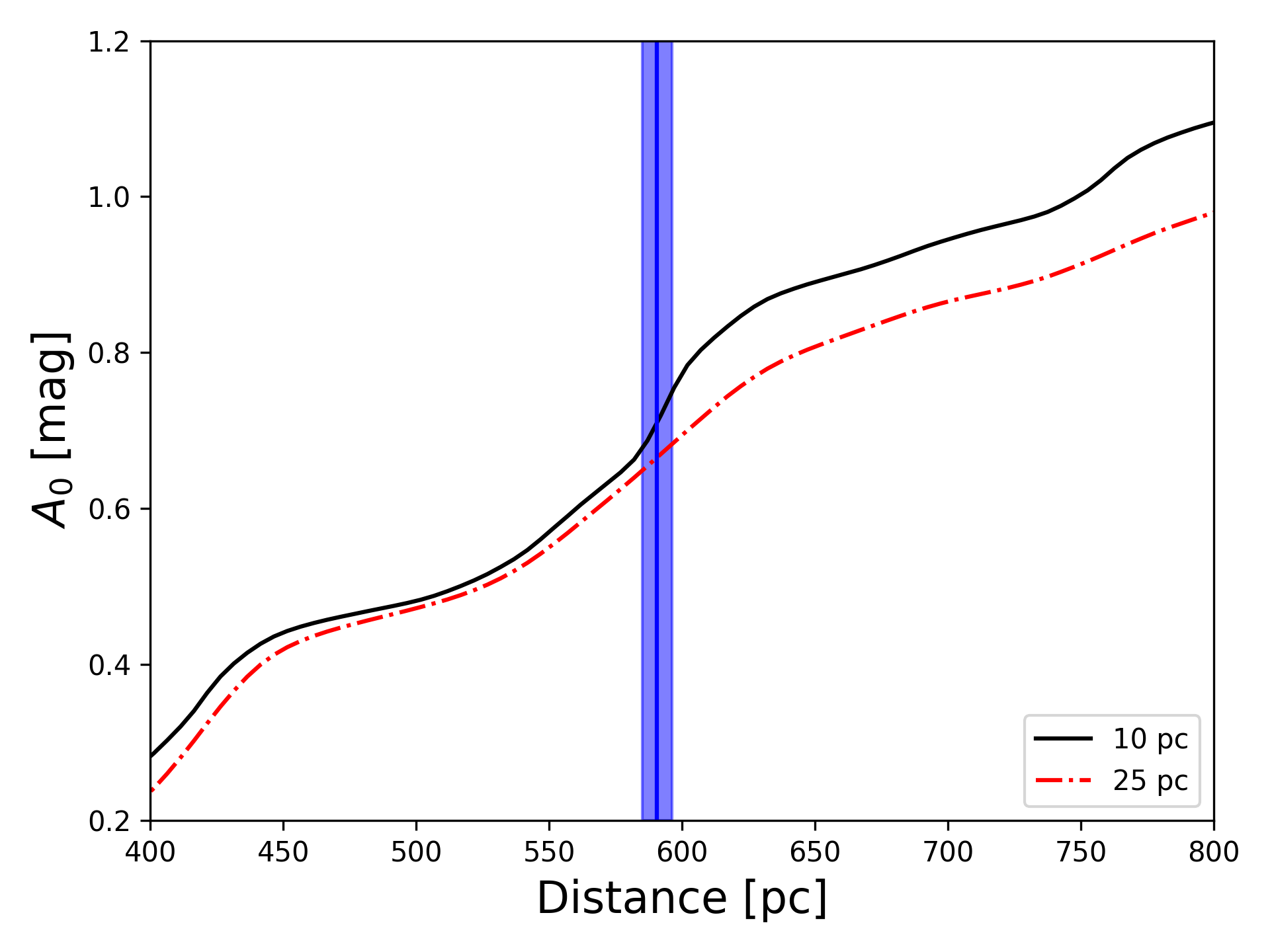}
    \caption{Extinction in the direction of \gaia BH3, as function of the distance. The extinction was derived with the maps with correlation length of 10\,pc (black solid line) and 25\,pc (red dot-dashed line) from \citet{2022A&A...664A.174V}; the vertical shaded region shows the distance range of \gaia BH3.}
    \label{fig:extinction}
\end{figure}

The \teff, \logg,\ and \feh\ derived with the above procedure depend mainly on the choice of  $A_0$. \gaia BH3 has a low Galactic latitude ($b=-3.49\degr$), located in a zone with a relatively high gradient of $A_0$, according to extinction maps of \citet{2022A&A...664A.174V}, as can be seen in \figref{fig:extinction}.
Adopting a distance of $590.6\pm5.8$\,pc, we obtain $A_0=0.710^{+0.041}_{-0.034}$\,mag for a correlation length of 10\,pc and $A_0=0.666\pm 0.017$\,mag for a correlation length of 25\,pc. We thus adopted $A_0=0.71\pm0.07$\,mag.

An updated value for $M_\star$ can be then estimated by comparing the absolute $G$ magnitude, $M_{\mathrm{G,0}}$ and the colour $(\bprp)_0$ with the ones given by theoretical isochrones. 
The procedure for the determination of stellar parameters is finally repeated with the updated value of $M_\star$.

\begin{figure}
    \centering
   \columnImage{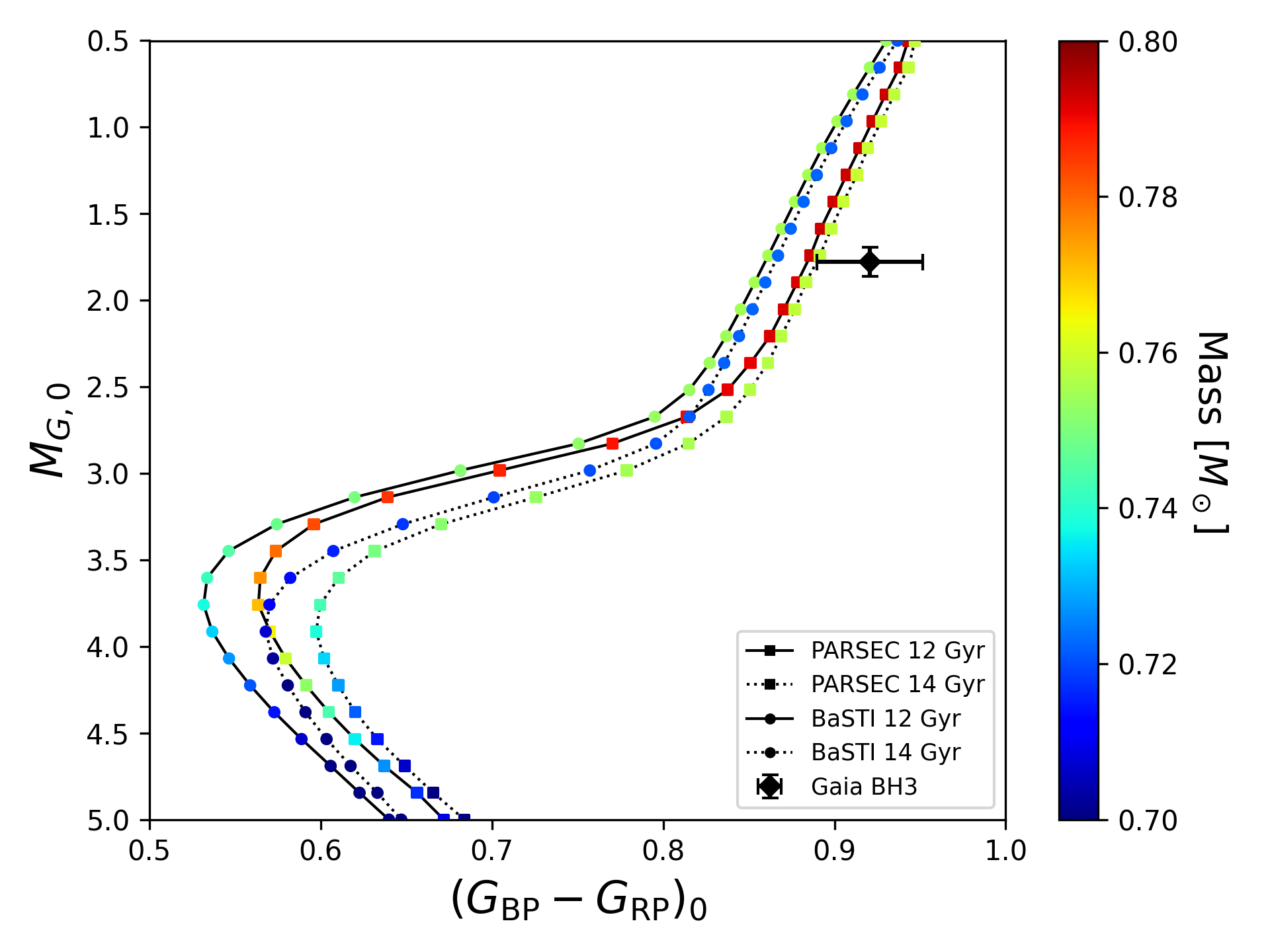}
    \caption{Comparison between the position of \gaia BH3 and isochrones in the colour-magnitude diagram. The colours of the symbols (filled circles for BaSTI and filled squares for PARSEC), on the two isochrones sets (12 and 14\,Gyr), correspond to the stellar masses.}
    \label{fig:isochrones}
\end{figure}

The value of $M_\star$ depends mainly on the assumed age and isochrone set, while it has a very small dependency on the assumed $A_0$. 
We used isochrones from both PARSEC\footnote{\url{http://stev.oapd.inaf.it/cgi-bin/cmd}} \citep{2012MNRAS.427..127B} and BaSTI\footnote{\url{http://basti-iac.oa-abruzzo.inaf.it/isocs.html}} \citep{2021ApJ...908..102P} libraries. 
For BaSTI, we adopt $[\mathrm{Fe/H}]=-2.5$, $[\alpha/\mathrm{Fe}]=0.4$ and $[\mathrm{M/H}]=-2.18$, while for PARSEC isochrones, which are only available with no $\alpha$-enhancement, we use metallicity $[\mathrm{Fe/H}]=-2.18$, i.e. we scale the \feh\ to match the $[\mathrm{M/H}]$ of BaSTI isochrones and take into account the contribution of $\alpha$-elements to the total metallicity \citep[e.g.][]{1993ApJ...414..580S}.
The comparison between isochrones and \gaia BH3 in the $M_{\mathrm{G,0}}$ versus $(\bprp)_0$ diagram is shown in \figref{fig:isochrones}. 

The value of $M_\star$ goes from 0.758 to 0.793\Msun for PARSEC and 0.723 to 0.755 \Msun for BaSTI, for isochrone of ages 12 and 14 Gyr, respectively. Younger ages would result in higher masses, but also bluer colours. We then estimate a mass $M_\star = 0.76 \pm 0.05$\Msun as the mass for the visible companion. 

With the procedure described above, we obtain the following parameters: $T_\mathrm{eff}=5211\pm 80$\,K, $\log g= 2.929\pm 0.003$, $[\mathrm{Fe/H}]=-2.56\pm 0.12$. From the \ion{Mg}{i} and \ion{Ca}{i} abundances, we derived an $\alpha$-enhancement of $[\alpha/\mathrm{Fe}]=0.43\pm 0.12$. Using the relation between iron content, enhancement and metallicity from \citet{1998A&A...333..419T}, we obtain $[\mathrm{M/H}]=-2.21\pm 0.15$.

It can be seen that the source is slightly redder and cooler than what is predicted by the models, albeit not significantly. In order to check that the \teff\ that we determined above is not underestimated, we derived an alternative temperature estimation from the excitation equilibrium of the \ion{Fe}{i} lines, including the NLTE corrections by \citet{2013ApJ...769...57F}. With this method, we obtained $T_\mathrm{eff}\sim 5100$\,K, which is even cooler, confirming that 
our estimation is not underestimated. A more detailed analysis of the stellar parameters is outside the scope of this work.


\section{Abundances}\label{sec:abundances}

We estimated the metallicity and abundances of the luminous component of \gaia\ BH3 from the UVES spectrum, which is of higher quality than our HERMES and SOPHIE spectra.

The abundances were derived with the code \textsc{MyGIsFOS} \citep{2014A&A...564A.109S}, with the exception of those for C, Ba and Eu; the carbon abundance was derived by fitting the Fraunhofer $G$-band of the CH molecule, while 
line-profile fitting was used to determine Ba and Eu abundances.
The resulting values are listed in \tabref{tab:abundances}. In the table, we provide the line-to-line scatter and the variation of the abundance corresponding to the uncertainty on the effective temperature. We adopted the solar abundances for C, Fe and Eu from \citet{caffausun}, and those from \citet{lodders09} for the other elements. We derived a value of 1.19\kms\ for the micro-turbulence, by forcing the same Fe abundance from \ion{Fe}{i} lines of different equivalent width. 

We sought the non-local thermal equilibrium (NLTE) correction for \ion{Mg}{i} and \ion{Ca}{i} in \citet{NLTE_MPIA}.
For the four \ion{Mg}{i} lines available, we derived a NLTE correction of 0.06.
Eleven lines of \ion{Ca}{i} provided a NLTE correction of 0.15.
The five \ion{Cr}{i} lines available in the database provide a large NLTE correction: 0.46.
For the \ion{Mn}{i} features, the NLTE correction is 0.54.
The 67 \ion{Fe}{i} lines available provide a mean NLTE correction of 0.1.
The NLTE effect on the \ion{Zn}{i} line at 481\,nm in \citet{sitnova_zn} and the Ba in \citet{korotin15} are small (see \tabref{tab:abundances}).

The \feh\ ratio we derived from the UVES spectrum ($[\mathrm{Fe/H}]=-2.56\pm 0.12$ from 223 \ion{Fe}{i} lines)
is in perfect agreement with the value derived from the SOPHIE spectrum (\feh $=-2.57\pm 0.12$ from 121 \ion{Fe}{i} lines)
and from the HERMES spectrum (\feh $=-2.54\pm 0.14$ from 156 \ion{Fe}{i} lines).
A good agreement was also obtained for the other elements, whose abundance is based on several lines.
The star, as expected for the metal-poor regime, is enhanced in $\alpha$ elements.

There is no trace of \element[][13]{C} in the spectrum, so the star has not been enriched by material processed in the CNO cycle, as it would if it had, for instance, accreted material from a companion star in the AGB phase.

\begin{figure}
    \centering
   \columnImage{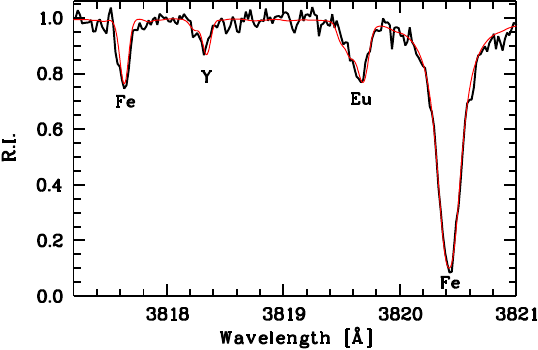}
    \caption{Residual intensity of the \gaia BH3 UVES spectrum (black line) compared with the modelled spectrum (thin red line) in the Eu region.}
    \label{fig:uves}
\end{figure}

The star has no chemical peculiarity, except a slight enhancement in Eu: $[\mathrm{Eu/Fe}]=0.52$.
When coupled with $[\mathrm{Ba/Fe}]=0.11$, this classifies this star as an r-I neutron-capture-rich star, following the classification of \citet{BC2005}. The UVES spectrum in the region of the Eu is shown in \figref{fig:uves}.

According to the stellar parameters, the star is expected to have a Li abundance on the Mucciarelli plateau \citep[see][]{mucciarelli12}.
At the wavelength of the Li feature at 607\,nm, we see an absorption line compatible with an abundance of A(Li)\,=\,1.2,
but the shape is not comparable with the synthetic spectrum, showing an absorption on the blue side of the feature. The Li feature is also visible in the SOPHIE spectrum, but with a lower S/N than in the UVES spectrum, while the S/N of the HERMES spectrum at 760\,nm is too low.  

\begin{table}[t]
    \caption{Abundances of \gaia BH3 from UVES spectrum. The values A(X) are expressed in the form A(X) = log(X/H) + 12, while $[\mathrm{X/H}]= \log\left(\mathrm{A(X)/A(X)}_\sun\right)$.}
    \label{tab:abundances}
    \centering
    \begin{tabular}{lcccccccc}
    \hline
    \hline
    Ion      & $N_\mathrm{lines}$ &A(X) & [X/H] & \multicolumn{2}{c}{$\sigma_\mathrm{A(X)}$} & NLTE \\
    & & & & (1) & (2) & corr.\\
    \hline
\ion{C}{i}   &  ...    & 6.11 & -2.39 & 0.10 & 0.17 &    ... \\
\ion{Mg}{i}  &    5    & 5.46 & -2.08 & 0.05 & 0.04 &   0.06\tablefootmark{a} \\
\ion{Ca}{i}  &   20    & 4.13 & -2.20 & 0.10 & 0.05 &   0.14\tablefootmark{b} \\
\ion{Sc}{ii} &    6    & 0.76 & -2.34 & 0.11 & 0.03 &    ... \\
\ion{Ti}{i}  &   17    & 2.67 & -2.23 & 0.06 & 0.11 &    ... \\
\ion{Ti}{ii} &   26    & 2.80 & -2.10 & 0.08 & 0.04 &    ... \\
\ion{V}{i}   &    3    & 1.34 & -2.66 & 0.13 & 0.10 &    ... \\
\ion{V}{ii}  &    6    & 1.61 & -2.39 & 0.07 & 0.03 &    ... \\
\ion{Cr}{i}  &    5    & 2.86 & -2.78 & 0.01 & 0.08 &    ... \\
\ion{Cr}{ii} &    3    & 3.23 & -2.41 & 0.06 & 0.03 &    ... \\
\ion{Mn}{i}  &    7    & 2.45 & -2.92 & 0.04 & 0.04 &    ... \\
\ion{Mn}{ii} &    1    & 2.79 & -2.58 & ...  & 0.08 &    ... \\
\ion{Fe}{i}  &  223    & 4.96 & -2.56 & 0.12 & 0.08 &   0.10\tablefootmark{c} \\
\ion{Fe}{ii} &   15    & 5.03 & -2.49 & 0.13 & 0.01 &    ... \\
\ion{Co}{i}  &    7    & 2.62 & -2.30 & 0.09 & 0.11 &    ... \\
\ion{Ni}{i}  &   18    & 3.68 & -2.55 & 0.12 & 0.06 &    ... \\
\ion{Zn}{i}  &    1    & 2.22 & -2.40 & ...  & 0.02 &   0.12\tablefootmark{d} \\
\ion{Sr}{ii} &    2    & 0.50 & -2.42 & 0.05 & 0.08 &    ... \\
\ion{Y}{ii}  &   10    &-0.44 & -2.65 & 0.16 & 0.05 &    ... \\
\ion{Zr}{ii} &    6    & 0.34 & -2.28 & 0.12 & 0.05 &    ... \\
\ion{Ba}{ii} &    4    &-0.28 & -2.45 & 0.13 & 0.06 &  -0.10\tablefootmark{e} \\
\ion{La}{ii} &    4    &-1.22 & -2.36 & 0.07 & 0.05 &    ... \\
\ion{Eu}{ii} &    1    &-1.45 & -1.97 & ...  & 0.05 &    ... \\
\hline
    \end{tabular}
    \tablefoot{The uncertainties on the abundances ($\sigma_\mathrm{A(X)}$) are reported as (1) line-to-line dispersion or (2) effect of \teff\ uncertainty. NLTE corrections are from \tablefoottext{a}{\citet{2017ApJ...847...15B}},
    \tablefoottext{b}{\citet{2007A&A...461..261M}},
    \tablefoottext{c}{\citet{2012MNRAS.427...27B}},
    \tablefoottext{d}{\citet{sitnova_zn}},
    \tablefoottext{e}{\citet{korotin15}}.
    }
\end{table}

\section{Acknowledgements}\label{sec:ack}

This work presents results from the European Space Agency (ESA) space mission \gaia. \gaia\ data are being processed by the \gaia\ Data Processing and Analysis Consortium (DPAC). Funding for the DPAC is provided by national institutions, in particular the institutions participating in the \gaia\ MultiLateral Agreement (MLA). The \gaia\ mission website is \url{https://www.cosmos.esa.int/gaia}. The \gaia\ archive website is \url{https://archives.esac.esa.int/gaia}.

The \gaia\ mission and data processing have financially been supported by, in alphabetical order by country:
\begin{itemize}
\item the Algerian Centre de Recherche en Astronomie, Astrophysique et G\'{e}ophysique of Bouzareah Observatory;
\item the Australian Research Council (ARC) through an Australian Laureate Fellowship (awarded to Prof. Joss Bland-Hawthorn);
\item the Austrian Fonds zur F\"{o}rderung der wissenschaftlichen Forschung (FWF) Hertha Firnberg Programme through grants T359, P20046, and P23737;
\item the BELgian federal Science Policy Office (BELSPO) for the provision of financial support in the framework of the PRODEX Programme of the European Space Agency (ESA), the Research Foundation Flanders (Fonds Wetenschappelijk Onderzoek) through grant VS.091.16N, the Fonds de la Recherche Scientifique (FNRS), and the Research Council of Katholieke Universiteit (KU) Leuven through grant C16/18/005 (Pushing AsteRoseismology to the next level with TESS, GaiA, and the Sloan DIgital Sky SurvEy -- PARADISE);
\item the Brazil-France exchange programmes Funda\c{c}\~{a}o de Amparo \`{a} Pesquisa do Estado de S\~{a}o Paulo (FAPESP) and Coordena\c{c}\~{a}o de Aperfeicoamento de Pessoal de N\'{\i}vel Superior (CAPES) - Comit\'{e} Fran\c{c}ais d'Evaluation de la Coop\'{e}ration Universitaire et Scientifique avec le Br\'{e}sil (COFECUB);
\item the Chilean Agencia Nacional de Investigaci\'{o}n y Desarrollo (ANID) through Fondo Nacional de Desarrollo Cient\'{\i}fico y Tecnol\'{o}gico (FONDECYT) Regular Project 1210992 (L.~Chemin);
\item the National Natural Science Foundation of China (NSFC) through grants 11573054, 11703065, and 12173069, the China Scholarship Council through grant 201806040200, and the Natural Science Foundation of Shanghai through grant 21ZR1474100;  
\item the Tenure Track Pilot Programme of the Croatian Science Foundation and the \'{E}cole Polytechnique F\'{e}d\'{e}rale de Lausanne and the project TTP-2018-07-1171 `Mining the Variable Sky', with the funds of the Croatian-Swiss Research Programme;
\item the Czech-Republic Ministry of Education, Youth, and Sports through grant LG 15010 and INTER-EXCELLENCE grant LTAUSA18093, and the Czech Space Office through ESA PECS contract 98058;
\item the Danish Ministry of Science;
\item the Estonian Ministry of Education and Research through grant IUT40-1;
\item the European Commission's Sixth Framework Programme through the European Leadership in Space Astrometry (\href{https://www.cosmos.esa.int/web/gaia/elsa-rtn-programme}{ELSA}) Marie Curie Research Training Network (MRTN-CT-2006-033481), through Marie Curie project PIOF-GA-2009-255267 (Space AsteroSeismology \& RR Lyrae stars, SAS-RRL), and through a Marie Curie Transfer-of-Knowledge (ToK) fellowship (MTKD-CT-2004-014188); the European Commission's Seventh Framework Programme through grant FP7-606740 (FP7-SPACE-2013-1) for the \gaia\ European Network for Improved data User Services (\href{https://gaia.ub.edu/twiki/do/view/GENIUS/}{GENIUS}) and through grant 264895 for the \gaia\ Research for European Astronomy Training (\href{https://www.cosmos.esa.int/web/gaia/great-programme}{GREAT-ITN}) network;
\item the European Cooperation in Science and Technology (COST) through COST Action CA18104 `Revealing the Milky Way with \gaia\ (MW-\gaia)';
\item the European Research Council (ERC) through grants 320360 (The Gaia-ESO Milky Way Survey), 647208 (Do intermediate-mass black holes exist?),  687378 (Small Bodies: Near and Far), 682115 (Using the Magellanic Clouds to Understand the Interaction of Galaxies), 695099 (A sub-percent distance scale from binaries and Cepheids -- CepBin), 745617 (Our Galaxy at full HD -- Gal-HD), 834148 (Accelerating Galactic Archeology), 895174 (The build-up and fate of self-gravitating systems in the Universe), 947660 (Measuring Hubble's Constant to 1\% with Pulsating Stars -- H1PStars), 951549 (Sub-percent calibration of the extragalactic distance scale in the era of big surveys -- UniverScale), 101004214 (Innovative Scientific Data Exploration and Exploitation Applications for Space Sciences -- EXPLORE), 101004719 (OPTICON-RadioNET Pilot), 101055318 (The 3D motion of the Interstellar Medium with ESO and ESA telescopes -- ISM-FLOW), 101063193 (Evolutionary Mechanisms in the Milky waY: the Gaia Data Release 3 revolution -- EMMY), 101093572 (StarDance: the non-canonical evolution of stars in clusters) and 101135205 (HORIZON-CL4-2023-SPACE-01-71 SPACIOUS project);
\item the European Science Foundation (ESF), in the framework of the \gaia\ Research for European Astronomy Training Research Network Programme (\href{https://www.cosmos.esa.int/web/gaia/great-programme}{GREAT-ESF});
\item the European Space Agency (ESA) in the framework of the \gaia\ project, through the Plan for European Cooperating States (PECS) programme through contracts C98090 and 4000106398/12/NL/KML for Hungary, through contract 4000115263/15/NL/IB for Germany, through PROgramme \break de D\'{e}veloppement d'Exp\'{e}riences scientifiques (PRODEX) Experiment Arrangement grants  4000132054 for Hungary, 4000142234 (Inference of radial velocities from astrometric stellar data - ASTRO2RV) and 4000138941 (Gaia Astrometric Microlensing Events - GAME) for Slovenia and through contract 4000132226/20/ES/CM;
\item the Research Council of Finland through grants 336546 and 345115 and Waldemar von Frenckells stiftelse;
\item the French Centre National d'\'{E}tudes Spatiales (CNES), the Agence Nationale de la Recherche (ANR) through grant ANR-10-IDEX-0001-02 for the `Investissements d'avenir' programme, through grant ANR-15-CE31-0007 for project `Modelling the Milky Way in the \gaia\ era' (MOD4\gaia), through grant ANR-14-CE33-0014-01 for project `The Milky Way disc formation in the \gaia\ era' (ARCHEOGAL), through grant ANR-15-CE31-0012-01 for project `Unlocking the potential of Cepheids as primary distance calibrators' (UnlockCepheids), through grant ANR-19-CE31-0017 for project `Secular evolution of galaxies' (SEGAL), and through grant ANR-18-CE31-0006 for project `Galactic Dark Matter' (GaDaMa), the Centre National de la Recherche Scientifique (CNRS) and its SNO \gaia\ of the Institut des Sciences de l'Univers (INSU), its Programmes Nationaux: Cosmologie et Galaxies (PNCG), Gravitation R\'{e}f\'{e}rences Astronomie M\'{e}trologie (PNGRAM), Plan\'{e}tologie (PNP), Physique et Chimie du Milieu Interstellaire (PCMI), and Physique Stellaire (PNPS), supported by INSU along with the Institut National de Physique  (INP) and the Institut National de Physique nucl\'{e}aire et de Physique des Particules (IN2P3), and co-funded by CNES; the `Action F\'{e}d\'{e}ratrice \gaia' of the Observatoire de Paris, and the R\'{e}gion de Franche-Comt\'{e};
\item the German Aerospace Agency (Deutsches Zentrum f\"{u}r Luft- und Raumfahrt e.V., DLR) through grants 50QG0501, 50QG0601, 50QG0602, 50QG0701, 50QG0901, 50QG1001, 50QG1101, 50\-QG1401, 50QG1402, 50QG1403, 50QG1404, 50QG1904, 50QG2101, 50QG2102, and 50QG2202, and the Centre for Information Services and High Performance Computing (ZIH) at the Technische Universit\"{a}t Dresden for generous allocations of computer time;
\item the Hungarian Academy of Sciences through the J\'anos Bolyai Research Scholarship (G. Marton and Z. Nagy) and the Hungarian National Research, Development, and Innovation Office (NKFIH) through grants KKP-137523 (`SeismoLab'), OTKA FK 146023 and TKP2021-NKTA-64;
\item the Science Foundation Ireland (SFI) through a Royal Society - SFI University Research Fellowship (M.~Fraser);
\item the Israel Ministry of Science and Technology through grant 3-18143 and the Israel Science Foundation (ISF) through grant 1404/22;
\item the Agenzia Spaziale Italiana (ASI) through contracts I/037/08/0, I/058/10/0, 2014-025-R.0, 2014-025-R.1.2015, and 2018-24-HH.0 and its addendum 2018-24-HH.1-2022 to the Italian Istituto Nazionale di Astrofisica (INAF), contract 2014-049-R.0/1/2, 2022-14-HH.0 to INAF for the Space Science Data Centre (SSDC, formerly known as the ASI Science Data Center, ASDC), contracts I/008/10/0, 2013/030/I.0, 2013-030-I.0.1-2015, and 2016-17-I.0 to the Aerospace Logistics Technology Engineering Company (ALTEC S.p.A.), INAF, and the Italian Ministry of Education, University, and Research (Ministero dell'Istruzione, dell'Universit\`{a} e della Ricerca) through the Premiale project `MIning The Cosmos Big Data and Innovative Italian Technology for Frontier Astrophysics and Cosmology' (MITiC);
\item the Netherlands Organisation for Scientific Research (NWO) through grant NWO-M-614.061.414, through a VICI grant (A.~Helmi), and through a Spinoza prize (A.~Helmi), and the Netherlands Research School for Astronomy (NOVA);
\item the Polish National Science Centre through HARMONIA grant 2018/30/M/ST9/00311 and DAINA grant 2017/27/L/ST9/03221; the Ministry of Science and Higher Education (MNiSW) through grant DIR/WK/2018/12; the Polish National Agency for Academic Exchange through BEKKER fellowship BPN/BEK/2022/1/00106;
\item the Portuguese Funda\c{c}\~{a}o para a Ci\^{e}ncia e a Tecnologia (FCT) through national funds, grants 2022.06962.PTDC and 2022.03993.PTDC, and work contract DL 57/2016/CP1364/CT0006, grants UIDB/04434/2020 and 
UIDP/04434/2020 for the Instituto de Astrof\'{\i}sica e Ci\^{e}ncias do Espa\c{c}o (IA), grants UIDB/00408/2020 and UIDP/00408/2020 for LASIGE, and grants UIDB/00099/2020 and UIDP/00099/2020 for the Centro de Astrof\'{\i}sica e Gravita\c{c}\~{a}o (CENTRA);  
\item the Slovenian Research Agency through grants P1-0188, P1-0031, I0-0033, J1-8136, J1-2460 and N1-0344;
\item the Spanish Ministry of Economy (MINECO/FEDER, UE), the Spanish Ministry of Science and Innovation (MCIN), the Spanish Ministry of Education, Culture, and Sports, and the Spanish Government through grants BES-2016-078499, BES-2017-083126, BES-C-2017-0085, ESP2016-80079-C2-1-R, FPU16/03827, RTI2018-095076-B-C22, PID2021-122842OB-C22, PDC2021-121059-C22,  and TIN2015-65316-P (`Computaci\'{o}n de Altas Prestaciones VII'), the Juan de la Cierva Incorporaci\'{o}n Programme (FJCI-2015-2671 and IJC2019-04862-I for F.~Anders), the Severo Ochoa Centre of Excellence Programme (SEV2015-0493) and MCIN/AEI/10.13039/501100011033/ EU FEDER and Next Generation EU/PRTR (PRTR-C17.I1 and CNS2022-135232); the European Union through European Regional Development Fund `A way of making Europe' through grants PID2021-122842OB-C21 and PID2021-125451NA-I00, the Institute of Cosmos Sciences University of Barcelona (ICCUB, Unidad de Excelencia `Mar\'{\i}a de Maeztu') through grant CEX2019-000918-M, the University of Barcelona's official doctoral programme for the development of an R+D+i project through an Ajuts de Personal Investigador en Formaci\'{o} (APIF) grant, the \href{https://svo.cab.inta-csic.es/}{Spanish Virtual Observatory} project funded by MCIN/AEI/10.13039/501100011033/ through grant PID2020-112949GB-I00; the Centro de Investigaci\'{o}n en Tecnolog\'{\i}as de la Informaci\'{o}n y las Comunicaciones (CITIC), funded by the Xunta de Galicia through the collaboration agreement to reinforce CIGUS research centers, research consolidation grant ED431B 2021/36 and scholarships from Xunta de Galicia and the EU - ESF ED481A-2019/155 and ED481A 2021/296; the Red Espa\~{n}ola de Supercomputaci\'{o}n (RES) computer resources at MareNostrum, the Barcelona Supercomputing Centre - Centro Nacional de Supercomputaci\'{o}n (BSC-CNS) through activities AECT-2017-2-0002, AECT-2017-3-0006, AECT-2018-1-0017, AECT-2018-2-0013, AECT-2018-3-0011, AECT-2019-1-0010, AECT-2019-2-0014, AECT-2019-3-0003, AECT-2020-1-0004, and DATA-2020-1-0010, the Departament d'Innovaci\'{o}, Universitats i Empresa de la Generalitat de Catalunya through grant 2014-SGR-1051 for project `Models de Programaci\'{o} i Entorns d'Execuci\'{o} Parallels' (MPEXPAR), and Ramon y Cajal Fellowships RYC2018-025968-I,  RYC2021-031683-I and RYC2021-033762-I, funded by MICIN/AEI/10.13039/501100011033 and by the European Union NextGenerationEU/PRTR and the European Science Foundation (`Investing in your future'); the Port d'Informaci\'{o} Cient\'{i}fica (PIC), through a collaboration between the Centro de Investigaciones Energ\'{e}ticas, Medioambientales y Tecnol\'{o}gicas (CIEMAT) and the Institut de F\'{i}sica d’Altes Energies (IFAE), supported by the grant EQC2021-007479-P funded by MCIN/AEI/ 10.13039/501100011033 and by the "European Union NextGenerationEU/PRTR), and also by MICIIN with funding from European Union NextGenerationEU(PRTR-C17.I1) and by Generalitat de Catalunya;
\item the Swedish National Space Agency (SNSA/Rymdstyrelsen);
\item the Swiss State Secretariat for Education, Research, and Innovation through the Swiss Activit\'{e}s Nationales Compl\'{e}mentaires and the Swiss National Science Foundation through an Eccellenza Professorial Fellowship (award PCEFP2\_194638 for R.I.~Anderson) and in the framework of the National Centre of Competence in Research PlanetS under grants 51NF40\_182901 and 51NF40\_205606;
\item the United Kingdom Particle Physics and Astronomy Research Council (PPARC), the United Kingdom Science and Technology Facilities Council (STFC), and the United Kingdom Space Agency (UKSA) through the following grants to the University of Bristol, Brunel University London, the Open University, the University of Cambridge, the University of Edinburgh, the University of Hertfordshire, the University of Leicester, the Mullard Space Sciences Laboratory of University College London, and the United Kingdom Rutherford Appleton Laboratory (RAL): PP/D006503/1, PP/D006511/1, PP/D006546/1, PP/D006570/1, PP/D006791/1, ST/I000852/1, ST/J005045/1, ST/K00056X/1, ST/K000209/1, ST/K000756/1, ST/K000578/1, ST/L002388/1, ST/L006553/1, ST/L006561/1, ST/N000595/1, ST/N000641/1, ST/N000978/1, ST/N001117/1, ST/S000089/1, ST/S000976/1, ST/S000984/1, ST/S001123/1, ST/S001948/1, ST/S001980/1, ST/S002103/1, ST/V000624/1, ST/V000969/1, EP/V520342/1, ST/W002469/1, ST/W002493/1, ST/W002671/1, ST/W002809/1, ST/W507490/1, ST/X00158X/1, ST/X001601/1, ST/X001636/1, ST/X001687/1, ST/X002667/1, ST/X002683/1 and ST/X002969/1.
\end{itemize}

The \gaia\ project and data processing have made use of:
\begin{itemize}
\item the Set of Identifications, Measurements, and Bibliography for Astronomical Data \citep[SIMBAD,][]{2000AAS..143....9W}, the `Aladin sky atlas' \citep{2000A&AS..143...33B,2014ASPC..485..277B}, and the VizieR catalogue access tool \citep{2000A&AS..143...23O}, all operated at the Centre de Donn\'{e}es astronomiques de Strasbourg (\href{http://cds.u-strasbg.fr/}{CDS});
\item the National Aeronautics and Space Administration (NASA) Astrophysics Data System (\href{http://adsabs.harvard.edu/abstract_service.html}{ADS});
\item the SPace ENVironment Information System (SPENVIS), initiated by the Space Environment and Effects Section (TEC-EES) of ESA and developed by the Belgian Institute for Space Aeronomy (BIRA-IASB) under ESA contract through ESA’s General Support Technologies Programme (GSTP), administered by the BELgian federal Science Policy Office (BELSPO);
\item the software products \href{http://www.starlink.ac.uk/topcat/}{TOPCAT}, \href{http://www.starlink.ac.uk/stil}{STIL}, and \href{http://www.starlink.ac.uk/stilts}{STILTS} \citep{2005ASPC..347...29T,2006ASPC..351..666T};
\item Matplotlib \citep{Hunter:2007};
\item IPython \citep{PER-GRA:2007};  
\item Astropy, a community-developed core Python package for Astronomy \citep{2018AJ....156..123A};
\item R \citep{RManual};
\item the HEALPix package \citep[][\url{http://healpix.sourceforge.net/}]{2005ApJ...622..759G};
\item Vaex \citep{2018A&A...618A..13B};
\item the \hip-2 catalogue \citep{2007A&A...474..653V}. The \hip and \tyc catalogues were constructed under the responsibility of large scientific teams collaborating with ESA. The Consortia Leaders were Lennart Lindegren (Lund, Sweden: NDAC) and Jean Kovalevsky (Grasse, France: FAST), together responsible for the \hip Catalogue; Erik H{\o}g (Copenhagen, Denmark: TDAC) responsible for the \tyc Catalogue; and Catherine Turon (Meudon, France: INCA) responsible for the \hip Input Catalogue (HIC);  
\item the \tyctwo catalogue \citep{2000A&A...355L..27H}, the construction of which was supported by the Velux Foundation of 1981 and the Danish Space Board;
\item the Tycho double star catalogue \citep[TDSC,][]{2002A&A...384..180F}, based on observations made with the ESA \hip astrometry satellite, as supported by the Danish Space Board and the United States Naval Observatory through their double-star programme;
\item data products from the Two Micron All Sky Survey \citep[2MASS,][]{2006AJ....131.1163S}, which is a joint project of the University of Massachusetts and the Infrared Processing and Analysis Center (IPAC) / California Institute of Technology, funded by the National Aeronautics and Space Administration (NASA) and the National Science Foundation (NSF) of the USA;
\item the ninth data release of the AAVSO Photometric All-Sky Survey (\href{https://www.aavso.org/apass}{APASS}, \citealt{apass9}), funded by the Robert Martin Ayers Sciences Fund;
\item the first data release of the Pan-STARRS survey \citep{panstarrs1,panstarrs1b,panstarrs1c,panstarrs1d,panstarrs1e,panstarrs1f}. The Pan-STARRS1 Surveys (PS1) and the PS1 public science archive have been made possible through contributions by the Institute for Astronomy, the University of Hawaii, the Pan-STARRS Project Office, the Max-Planck Society and its participating institutes, the Max Planck Institute for Astronomy, Heidelberg and the Max Planck Institute for Extraterrestrial Physics, Garching, The Johns Hopkins University, Durham University, the University of Edinburgh, the Queen's University Belfast, the Harvard-Smithsonian Center for Astrophysics, the Las Cumbres Observatory Global Telescope Network Incorporated, the National Central University of Taiwan, the Space Telescope Science Institute, the National Aeronautics and Space Administration (NASA) through grant NNX08AR22G issued through the Planetary Science Division of the NASA Science Mission Directorate, the National Science Foundation through grant AST-1238877, the University of Maryland, Eotvos Lorand University (ELTE), the Los Alamos National Laboratory, and the Gordon and Betty Moore Foundation;
\item the second release of the Guide Star Catalogue \citep[GSC2.3,][]{2008AJ....136..735L}. The Guide Star Catalogue II is a joint project of the Space Telescope Science Institute (STScI) and the Osservatorio Astrofisico di Torino (OATo). STScI is operated by the Association of Universities for Research in Astronomy (AURA), for the National Aeronautics and Space Administration (NASA) under contract NAS5-26555. OATo is operated by the Italian National Institute for Astrophysics (INAF). Additional support was provided by the European Southern Observatory (ESO), the Space Telescope European Coordinating Facility (STECF), the International GEMINI project, and the European Space Agency (ESA) Astrophysics Division (nowadays SCI-S);
\item the eXtended, Large (XL) version of the catalogue of Positions and Proper Motions \citep[PPM-XL,][]{2010AJ....139.2440R};
\item data products from the Wide-field Infrared Survey Explorer (WISE), which is a joint project of the University of California, Los Angeles, and the Jet Propulsion Laboratory/California Institute of Technology, and NEOWISE, which is a project of the Jet Propulsion Laboratory/California Institute of Technology. WISE and NEOWISE are funded by the National Aeronautics and Space Administration (NASA);
\item the first data release of the United States Naval Observatory (USNO) Robotic Astrometric Telescope \citep[URAT-1,][]{urat1};
\item the fourth data release of the United States Naval Observatory (USNO) CCD Astrograph Catalogue \citep[UCAC-4,][]{2013AJ....145...44Z};
\item the sixth and final data release of the Radial Velocity Experiment \citep[RAVE DR6,][]{2020AJ....160...83S,rave6a}. Funding for RAVE has been provided by the Leibniz Institute for Astrophysics Potsdam (AIP), the Australian Astronomical Observatory, the Australian National University, the Australian Research Council, the French National Research Agency, the German Research Foundation (SPP 1177 and SFB 881), the European Research Council (ERC-StG 240271 Galactica), the Istituto Nazionale di Astrofisica at Padova, the Johns Hopkins University, the National Science Foundation of the USA (AST-0908326), the W.M.\ Keck foundation, the Macquarie University, the Netherlands Research School for Astronomy, the Natural Sciences and Engineering Research Council of Canada, the Slovenian Research Agency, the Swiss National Science Foundation, the Science \& Technology Facilities Council of the UK, Opticon, Strasbourg Observatory, and the Universities of Basel, Groningen, Heidelberg, and Sydney. The RAVE website is at \url{https://www.rave-survey.org/};
\item the first data release of the Large sky Area Multi-Object Fibre Spectroscopic Telescope \citep[LAMOST DR1,][]{LamostDR1};
\item the K2 Ecliptic Plane Input Catalogue \citep[EPIC,][]{epic-2016ApJS..224....2H};
\item the ninth data release of the Sloan Digitial Sky Survey \citep[SDSS DR9,][]{SDSS9}. Funding for SDSS-III has been provided by the Alfred P. Sloan Foundation, the Participating Institutions, the National Science Foundation, and the United States Department of Energy Office of Science. The SDSS-III website is \url{http://www.sdss3.org/}. SDSS-III is managed by the Astrophysical Research Consortium for the Participating Institutions of the SDSS-III Collaboration including the University of Arizona, the Brazilian Participation Group, Brookhaven National Laboratory, Carnegie Mellon University, University of Florida, the French Participation Group, the German Participation Group, Harvard University, the Instituto de Astrof\'{\i}sica de Canarias, the Michigan State/Notre Dame/JINA Participation Group, Johns Hopkins University, Lawrence Berkeley National Laboratory, Max Planck Institute for Astrophysics, Max Planck Institute for Extraterrestrial Physics, New Mexico State University, New York University, Ohio State University, Pennsylvania State University, University of Portsmouth, Princeton University, the Spanish Participation Group, University of Tokyo, University of Utah, Vanderbilt University, University of Virginia, University of Washington, and Yale University;
\item the thirteenth release of the Sloan Digital Sky Survey \citep[SDSS DR13,][]{2017ApJS..233...25A}. Funding for SDSS-IV has been provided by the Alfred P. Sloan Foundation, the United States Department of Energy Office of Science, and the Participating Institutions. SDSS-IV acknowledges support and resources from the Center for High-Performance Computing at the University of Utah. The SDSS web site is \url{https://www.sdss.org/}. SDSS-IV is managed by the Astrophysical Research Consortium for the Participating Institutions of the SDSS Collaboration including the Brazilian Participation Group, the Carnegie Institution for Science, Carnegie Mellon University, the Chilean Participation Group, the French Participation Group, Harvard-Smithsonian Center for Astrophysics, Instituto de Astrof\'isica de Canarias, The Johns Hopkins University, Kavli Institute for the Physics and Mathematics of the Universe (IPMU) / University of Tokyo, the Korean Participation Group, Lawrence Berkeley National Laboratory, Leibniz Institut f\"ur Astrophysik Potsdam (AIP),  Max-Planck-Institut f\"ur Astronomie (MPIA Heidelberg), Max-Planck-Institut f\"ur Astrophysik (MPA Garching), Max-Planck-Institut f\"ur Extraterrestrische Physik (MPE), National Astronomical Observatories of China, New Mexico State University, New York University, University of Notre Dame, Observat\'ario Nacional / MCTI, The Ohio State University, Pennsylvania State University, Shanghai Astronomical Observatory, United Kingdom Participation Group, Universidad Nacional Aut\'onoma de M\'{e}xico, University of Arizona, University of Colorado Boulder, University of Oxford, University of Portsmouth, University of Utah, University of Virginia, University of Washington, University of Wisconsin, Vanderbilt University, and Yale University;
\item the second release of the SkyMapper catalogue \citep[SkyMapper DR2,][Digital Object Identifier 10.25914/5ce60d31ce759]{2019PASA...36...33O}. The national facility capability for SkyMapper has been funded through grant LE130100104 from the Australian Research Council (ARC) Linkage Infrastructure, Equipment, and Facilities (LIEF) programme, awarded to the University of Sydney, the Australian National University, Swinburne University of Technology, the University of Queensland, the University of Western Australia, the University of Melbourne, Curtin University of Technology, Monash University, and the Australian Astronomical Observatory. SkyMapper is owned and operated by The Australian National University's Research School of Astronomy and Astrophysics. The survey data were processed and provided by the SkyMapper Team at the Australian National University. The SkyMapper node of the All-Sky Virtual Observatory (ASVO) is hosted at the National Computational Infrastructure (NCI). Development and support the SkyMapper node of the ASVO has been funded in part by Astronomy Australia Limited (AAL) and the Australian Government through the Commonwealth's Education Investment Fund (EIF) and National Collaborative Research Infrastructure Strategy (NCRIS), particularly the National eResearch Collaboration Tools and Resources (NeCTAR) and the Australian National Data Service Projects (ANDS);
\item the \gaia-ESO Public Spectroscopic Survey \citep[GES,][]{GES_final_release_paper_1,GES_final_release_paper_2}. The \gaia-ESO Survey is based on data products from observations made with ESO Telescopes at the La Silla Paranal Observatory under programme ID 188.B-3002. Public data releases are available through the \href{https://www.gaia-eso.eu/data-products/public-data-releases}{ESO Science Portal}. The project has received funding from the Leverhulme Trust (project RPG-2012-541), the European Research Council (project ERC-2012-AdG 320360-\gaia-ESO-MW), and the Istituto Nazionale di Astrofisica, INAF (2012: CRA 1.05.01.09.16; 2013: CRA 1.05.06.02.07).
\end{itemize}

The GBOT programme (\href{https://gea.esac.esa.int/archive/documentation/GDR3/Data_processing/chap_cu3ast/sec_cu3ast_prop/ssec_cu3ast_prop_gbot.html}{GBOT}) uses observations collected at (i) the European Organisation for Astronomical Research in the Southern Hemisphere (ESO) with the VLT Survey Telescope (VST), under ESO programmes
092.B-0165,
093.B-0236,
094.B-0181,
095.B-0046,
096.B-0162,
097.B-0304,
098.B-0030,
099.B-0034,
0100.B-0131,
0101.B-0156,
0102.B-0174,
0103.B-0165,
0104.B-0081,
0106.20ZA.001 (OmegaCam),
0106.20ZA.002 (FORS2),
0108.21YF;
and under INAF programs 110.256C,
112.266Q;
%
%
and (ii) the Liverpool Telescope, which is operated on the island of La Palma by Liverpool John Moores University in the Spanish Observatorio del Roque de los Muchachos of the Instituto de Astrof\'{\i}sica de Canarias with financial support from the United Kingdom Science and Technology Facilities Council, and (iii) telescopes of the Las Cumbres Observatory Global Telescope Network.

In addition to the currently active DPAC (and ESA science) authors of the peer-reviewed papers accompanying \gdr{3}, there are large numbers of former DPAC members who made significant contributions to the (preparations of the) data processing. Among those are, in alphabetical order:
Stephanie Accart, 
Christopher Agard, 
Juan Jos\'{e} Aguado, 
Micha\"{e}l Ajaj, 
Fernando Aldea-Montero, 
Alexandra Alecu, 
Bruno Alessi, 
Peter Allan, 
France Allard, 
Walter Allasia, 
Carlos Allende Prieto, 
Javier \'{A}lvarez Cid-Fuentes, 
Marco Antonio \'{A}lvarez, 
Jo\~{a}o Alves, 
Antonio Amorim, 
Kader Amsif, 
Alexandre Andrei, 
Antonino Angi, 
Guillem Anglada-Escud\'{e}, 
Erika Antiche, 
Sonia Ant\'{o}n, 
Bernardino Arcay, 
Clément Arnaudeau, 
Borja Arroyo Galende, 
Vladan Arsenijevic, 
Tri Astraatmadja, 
Rajesh Kumar Bachchan, 
Adrien Bangma, 
Carlos Barata, 
Domenico Barbato, 
Fabio Barblan, 
Paul Barklem, 
Mickael Batailler, 
Duncan Bates, 
Alexandre Baudesson-Stella, 
Mathias Beck, 
Luigi Bedin, 
Dan Beilis, 
Antonio Bello Garc\'{\i}a, 
Vasily Belokurov, 
Philippe Bendjoya, 
Ángel Berihuete, 
Hans Bernstein$^\dagger$, 
Olivier Bienaym\'{e}, 
Lionel Bigot, 
Albert Bijaoui, 
Louis Bil, 
Fran\c{c}oise Billebaud, 
Nadejda Blagorodnova, 
Thierry Bloch, 
Klaas de Boer$^\dagger$, 
Marco Bonfigli, 
Giuseppe Bono, 
Simon Borgniet, 
Raul Borrachero-Sanchez, 
Fran\c{c}ois Bouchy, 
Steve Boudreault, 
Geraldine Bourda, 
Guy Boutonnet, 
Lorenzo Bramante, 
Pascal Branet, 
Maarten Breddels, 
Scott Brown, 
Pierre-Marie Brunet, 
Thomas Br\"{u}semeister, 
Peter Bunclark$^\dagger$, 
Roberto Buonanno, 
Alexandru Burlacu, 
Robert Butorafuchs, 
Joan Cambras, 
Heather Campbell, 
Sylvain Cannizzo, 
Christophe Carret, 
Manuel Carrillo, 
C\'{e}sar Carri\'{o}n, 
Pau Castro Sampol, 
Francisco Javier Casquero, 
Laurence Chaoul, 
Jonathan Charnas, 
Fabien Ch\'{e}reau, 
Vincenzo Chiaramida 
Mathurin Chritin, 
Maria-Rosa Cioni, 
Uma Cladellas Sanjuan, 
Marcial Clotet, 
Gabriele Cocozza, 
Ross Collins, 
Gabriele Comoretto, 
Gabriele Contursi, 
Leonardo Corcione, 
Gr\'{a}inne Costigan, 
Fran\c{c}oise  Crifo,  
Alessandro Crisafi, 
Nick Cross, 
Jan Cuypers$^\dagger$, 
Jean-Charles Damery, 
Anastasios Dapergolas, 
Eric Darmigny, 
Pedro David, 
Jonas Debosscher, 
Peter De Cat, 
Domitilla De Martino, 
Rafael De Souza, 
Enrique Del Pozo, 
H\'{e}ctor Delgado, 
David Delhoume, 
C\'{e}line Delle Luche, 
Markus Demleitner, 
L\'{e}o Denglos, 
S\'{e}kou Diakite, 
Paola Di Matteo, 
Carla Domingues, 
Sandra Dos Anjos, 
Laurent Douchy, 
Petros Drazinos, 
Pierre Dubath, 
Javier Dur\'{a}n, 
Yifat Dzigan, 
Bengt Edvardsson, 
Deepak Eappachen, 
Sebastian Els, 
Arjen van Elteren, 
Kjell Eriksson, 
Pilar Esquej, 
Carolina von Essen, 
Wyn Evans, 
Guillaume Eynard Bontemps, 
Antonio Falc\~{a}o, 
Mart\'{\i} Farr\`{a}s Casas, 
Jacopo Federici, 
Luciana Federici, 
Fernando de Felice, 
Agn\`{e}s Fienga, 
Krzysztof Findeisen, 
Christian Fischer, 
Florin Fodor, 
Yori Fournier, 
Fr\'{e}d\'{e}ric Franke, 
Benoit Frezouls, 
Aidan Fries, 
Jan Fuchs, 
Flavio Fusi Pecci, 
Diego Fustes, 
Duncan Fyfe, 
Eva Gallardo, 
Silvia Galleti, 
Fernando Garc\'{i}a, 
Alberto Garc\'{i}a Guti\'{e}rrez, 
Mar\'{i}a Garc\'{i}a-Reinaldos, 
Daniele Gardiol, 
Nora Garralda Torres, 
Emilien Gaudin, 
Alvin Gavel, 
Marwan Gebran, 
Yoann G\'{e}rard, 
Nathalie Gerbier, 
Joris Gerssen, 
Miguel Gomes, 
Roy Gomel, 
Anita G\'{o}mez, 
Ana Gonz\'{a}lez-Marcos, 
Juan Gonz\'{a}lez-N\'{u}\~{n}ez, 
Juan Jos\'{e} Gonz\'{a}lez-Vidal, 
Eva Grebel, 
Michel Grenon, 
Bj\"{o}rn Grieger, 
Eric Grux, 
Alain Gueguen, 
Pierre Guillout, 
Julie Guiraud, 
Andr\'{e}s G\'{u}rpide, 
Leanne Guy, 
Jean-Louis Halbwachs, 
Marcus Hauser, 
Aurelien Hees, 
Kevin Henares, 
Julien Heu, 
Albert Heyrovsky, 
Thomas Hilger, 
Nathan Himpens, 
Natalia H\l{}adczuk, 
Wilfried Hofmann, 
Erik H{\o}g, 
David Hogg, 
Andrew Holland, 
Greg Holland, 
Gordon Hopkinson$^\dagger$, 
Claude Huc, 
Pablo Huijse, 
Jason Hunt, 
Brigitte Huynh, 
Arkadiusz Hypki, 
Giacinto Iannicola, 
Sergio Ibarmia, 
Vilma Icardi, 
Laura Inno, 
Mike Irwin, 
Yago Isasi Parache, 
Javier Izquierdo, 
Maja Jab\l{}o\'{n}ska, 
Thierry Jacq, 
Asif Jan, 
Anne-Marie Janotto, 
Kevin Jardine, 
G\'{e}rard Jasniewicz, 
Anne Jean-Antoine Piccolo, 
Laurent Jean-Rigaud, 
Isabelle J{\'{e}}gouzo-Giroux, 
Christian Jezequel, 
Fran\c{c}ois Jocteur-Monrozier, 
Paula Jofr\'{e}, 
Anthony Jonckheere, 
Peter Jonker, 
\'{A}ron Juh\'{a}sz, 
Francesc Julbe, 
Antonios Karampelas, 
Lea Karbevska, 
Ralf Keil, 
Adam Kewley, 
Dae-Won Kim, 
Peter Klagyivik, 
Jochen Klar, 
Jonas Kl\"{u}ter, 
Jens Knude, 
Angela Kochoska, 
Oleg Kochukhov, 
Katrien Kolenberg, 
Indrek Kolka, 
Pavel Koubsky, 
Janez Kos, 
Irina Kovalenko, 
Daniel Krefl, 
Maria Kudryashova, 
Ilya Kull, 
Alex Kutka, 
Fr\'{e}d\'{e}ric Lacoste-Seris, 
Sylvain Lafosse, 
Val\'{e}ry Lainey, 
Pascal Laizeau, 
Yannick Lasne, 
Antoni Latorre, 
Felix Lauwaert, 
Claudia Lavalley, 
Jean-Baptiste Lavigne, 
David Le Bouquin, 
Jean-Fran\c{c}ois Le Campion, 
Isabelle Lecoeur-Taibi, 
Yann Le Fustec, 
Vassili Lemaitre, 
Helmut Lenhardt, 
Christophe Le Poncin-Lafitte, 
Fr\'{e}d\'{e}ric Leroux, 
Thierry Levoir, 
Hans Lindstr{\o}m, 
Tim Lister, 
Chao Liu, 
Mauro L\'{o}pez Del Fresno, 
Davide Loreggia, 
Denise Lorenz, 
Cristina Luengo, 
Ian MacDonald, 
Marc Madaule, 
Pau Madrero Pardo, 
Tiago Magalh\~{a}es Fernandes, 
Arrate Magdaleno Romeo, 
Kirill Makan, 
Valeri Makarov, 
Jean-Christophe Malapert, 
Sandrine Managau, 
Herv\'{e} Manche, 
Carmelo Manetta, 
Gregory Mantelet, 
Jos\'{e} Marcos, 
Miguel Marcos Santos, 
Federico Marocco, 
Gabor Marschalko, 
Mathieu Marseille, 
Christophe Martayan, 
\'{O}scar Mart\'{i}nez-Rubi, 
Michele Martino, 
Paul Marty, 
Nicolas Mary, 
Davide Massari, 
Benjamin Massart, 
Gal Matijevi\v{c}, 
Mohamed Meharga, 
Emmanuel Mercier, 
Maria Messineo, 
Fr\'{e}d\'{e}ric Meynadier, 
Daniel Michalik, 
Anthony Michon, 
Shan Mignot, 
Hadi Minbashian, 
Bruno Miranda, 
L\'{a}szl\'{o} Moln\'{a}r, 
Marco Molinaro, 
Giacomo Monari, 
Marc Moniez, 
\'{A}ngel Montero, 
Alain Montmory, 
Roger Mor, 
Thierry Morel, 
Stephan Morgenthaler, 
Angelo Mulone, 
Ulisse Munari, 
Daniel Mu\~{n}oz, 
Cillian Murphy, 
J\'{e}r\^{o}me Narbonne, 
Gijs Nelemans, 
Anne-Th\'{e}r\`{e}se Nguyen, 
Luciano Nicastro, 
Sara Nieto, 
Thomas Nordlander, 
Alexandre Nouvel, 
Louis Noval, 
Markus Nullmeier, 
Derek O'Callaghan, 
Francisco Oca\~{n}a, 
Pierre Ocvirk, 
Alex Ogden, 
Joaqu\'{\i}n Ordieres-Mer\'{e}, 
Diego Ordonez, 
Giuseppe Orr\`{u}, 
Patricio Ortiz, 
Jos\'{e} Osinde, 
Jose Osorio, 
Dagmara Oszkiewicz, 
Alex Ouzounis, 
Hugo Palacin, 
Max Palmer, 
Aviad Panahi, 
Chantal Panem, 
Vincent Papy, 
Peregrine Park, 
Ester Pasquato, 
Xavier Passot, 
Stefan Payne-Wardenaar, 
Louis Pegoraro, 
Roselyne Pedrosa, 
Christian Peltzer, 
Hanna Pentik\"{a}inen, 
Xavier Pe\~{n}alosa Esteller, 
Jordi Peralta, 
Rub\'{e}n P\'{e}rez, 
Jean-Marc Petit, 
Fabien P\'{e}turaud, 
Bernard Pichon, 
Tuomo Pieniluoma, 
Anna Marina Piersimoni, 
Fran\c{c}ois-Xavier Pineau, 
Enrico Pigozzi, 
Federic Pireddu, 
Bertrand Plez, 
Joel Poels$^\dagger$, 
Aurelian Polidoro, 
Eric Poujoulet, 
Arnaud Poulain, 
Guylaine Prat, 
Thibaut Prod'homme, 
Andrej Pr\v{s}a, 
Elena Racero, 
Adrien Raffy, 
Silvia Ragaini, 
Serena Rago, 
Nicolas Rambaux, 
Piero Ranalli, 
Gregor Rauw, 
Andrew Read, 
Jos\'{e} Rebordao, 
Philippe Redon, 
Rita Ribeiro, 
Ariadna Ribes Metidieri, 
Pascal Richard, 
Phil Richards, 
Carlos R\'{i}os D\'{i}az, 
Daniel Risquez, 
Adrien Rivard, 
Clement Robin, 
Brigitte Rocca-Volmerange, 
Maroussia Roelens, 
Herv\'{e} Rogues, 
Laurent Rohrbasser, 
Nicolas de Roll, 
Julia Roquette, 
Siv Ros\'{e}n, 
Frederic Royer, 
Stefano Rubele, 
Laura Ruiz Dern, 
Idoia Ruiz-Fuertes, 
Federico Russo, 
Jan Rybizki, 
Albert S\'{a}ez N\'{u}\~{n}ez, 
Jes\'{u}s Salgado, 
Eugenio Salguero, 
Nik Samaras, 
Paula S\'{a}nchez Gayet, 
V\'{i}ctor S\'{a}nchez Gim\'{e}nez, 
Toni Santana, 
Helder Savietto, 
Maud Segol, 
Juan Carlos Segovia, 
Damien Segransan, 
L\'{e}a Sellahannadi, 
Didier Semeux, 
I-Chun Shih, 
Hassan Siddiqui, 
Lauri Siltala, 
Andr\'{e} Silva, 
Helder Silva, 
Arturo Silvelo, 
Dimitris Sinachopoulos, 
Christos Siopis, 
Riccardo Smareglia, 
Kester Smith, 
Michael Soffel, 
Sergio Soria Nieto, 
Danuta Sosnowska, 
Alessandro Spagna, 
Maxime Spano, 
Lorenzo Spina, 
Ulrike Stampa, 
Craig Stephenson, 
Hristo Stoev, 
Vytautas Strai\v{z}ys, 
Frank Suess, 
Maria S\"{u}veges, 
Elza Szegedi-Elek, 
Francis T\^{a}che, 
Jeff Tambouez, 
Guy Tauran, 
Dirk Terrell, 
David Terrett, 
Pierre Teyssandier, 
Stephan Theil, 
William Thuillot, 
Carola Tiede, 
Brandon Tingley, 
Kre\v{s}imir Tisani\'{c}, 
Anastasia Titarenko, 
Jordi Torra$^\dagger$, 
Scott Trager, 
Licia Troisi, 
Paraskevi Tsalmantza, 
David Tur, 
Stefano Uzzi, 
Mattia Vaccari, 
Fr\'{e}d\'{e}ric Vachier, 
Emmanouil Vachlas, 
Marc Vaillant, 
Gaetano Valentini, 
Pau Vall\`{e}s, 
Veronique Valette, 
Emmanuel van Dillen, 
Walter Van Hamme, 
Eric Van Hemelryck, 
Wouter van Reeven, 
Mihaly Varadi, 
Marco Vaschetto, 
Jovan Veljanoski, 
Lionel Veltz, 
Sjoert van Velzen, 
Teresa Via, 
Yves Viala, 
Jenni Virtanen, 
Antonio Volpicelli, 
Holger Voss, 
Viktor Votruba, 
Stelios Voutsinas, 
Jean-Marie Wallut, 
Gavin Walmsley, 
Olivier Wertz, 
Thomas Wevers, 
Rainer Wichmann, 
Mark Wilkinson, 
Abdullah Yoldas, 
Patrick Yvard, 
Petar Ze\v{c}evi\'{c}, 
Tim de Zeeuw, 
Maruska Zerjal, 
Houri Ziaeepour, 
Claude Zurbach, 
and Sven Zschocke. 

In addition to the DPAC consortium, past and present, there are numerous people, mostly in ESA and in industry, who have made or continue to make essential contributions to \gaia, for instance those employed in science and mission operations or in the design, manufacturing, integration, and testing of the spacecraft and its modules, subsystems, and units. Many of those will remain unnamed yet spent countless hours, occasionally during nights, weekends, and public holidays, in cold offices and dark clean rooms. At the risk of being incomplete, we specifically acknowledge, in alphabetical order,
from Airbus DS (Toulouse):
Alexandre Affre,
Marie-Th\'{e}r\`{e}se Aim\'{e},
Audrey Albert,
Aur\'{e}lien Albert-Aguilar,
Jeanine Alloun-Etcheberry,
Hania Arsalane,
Arnaud Aurousseau,
Denis Bassi,
Franck Bayle,
Bernard Bayol,
Pierre-Luc Bazin,
Emmanuelle Benninger,
Philippe Bertrand,
Jean-Bernard Biau,
Fran\c{c}ois Binter,
C\'{e}dric Blanc,
Eric Blonde,
Patrick Bonzom,
Bernard Bories,
Jean-Jacques Bouisset,
Jo\"el Boyadjian, 
Isabelle Brault,
Corinne Buge,
Bertrand Calvel, 
Jean-Michel Camus,
France Canton,
Lionel Carminati, 
Michel Carrie,
Didier Castel,
Philippe Charvet, 
Fran\c{c}ois Chassat, 
Fabrice Cherouat,
Ludovic Chirouze,
Michel Choquet,
Claude Coatantiec, 
Emmanuel Collados,
Philippe Corberand,
Christelle Dauga,
Robert Davancens, 
Catherine Deblock,
Eric Decourbey,
Charles Dekhtiar,
Michel Delannoy,
Michel Delgado,
Damien Delmas,
Emilie Demange, 
Victor Depeyre,
Isabelle Desenclos,
Christian Dio,
Kevin Downes,
Marie-Ange Duro,
Eric Ecale, 
Omar Emam,
Elizabeth Estrada,
Coralie Falgayrac,
Benjamin Farcot,
Claude Faubert,
Fr\'{e}d\'{e}ric Faye, 
S\'{e}bastien Finana,
Gr\'{e}gory Flandin, 
Loic Floury,
Gilles Fongy,
Michel Fruit, 
Florence Fusero, 
Christophe Gabilan,
J\'{e}r\'{e}mie Gaboriaud,
Cyril Gallard,
Damien Galy,
Benjamin Gandon,
Patrick Gareth,
Eric Gelis,
Andr\'{e} Gellon,
Laurent Georges, 
Philippe-Marie Gomez,
Jos\'{e} Goncalves,
Fr\'{e}d\'{e}ric Guedes,
Vincent Guillemier,
Thomas Guilpain,
St\'{e}phane Halbout,
Marie Hanne,
Gr\'{e}gory Hazera,
Daniel Herbin,
Tommy Hercher,
Claude Hoarau le Papillon,
Matthias Holz,
Philippe Humbert, 
Sophie Jallade, 
Gr\'{e}gory Jonniaux, 
Fr\'{e}d\'{e}ric Juillard,
Philippe Jung,
Charles Koeck,
Marc Labaysse, 
R\'{e}n\'{e} Laborde,
Anouk Laborie, 
J\'{e}r\^{o}me Lacoste-Barutel,
Baptiste Laynet,
Virginie Le Gall, 
Julien L'Hermitte,
Marc Le Roy, 
Christian Lebranchu, 
Didier Lebreton,
Patrick Lelong, 
Jean-Luc Leon,
Stephan Leppke,
Franck Levallois,
Philippe Lingot,
Laurant Lobo,
Didier Loche,
C\'{e}line Lopez,
Jean-Michel Loupias,
Carlos Luque,
S\'{e}bastien Maes,
Bruno Mamdy, 
Denis Marchais,
Alexandre Marson,
Benjamin Massart, 
R\'{e}mi Mauriac,
Philippe Mayo,
Caroline Meisse, 
Herv\'{e} Mercereau,
Olivier Michel,
Florent Minaire,
Xavier Moisson, 
David Monteiro, 
Denis Montperrus,
Boris Niel,
C\'{e}dric Papot,
Jean-Fran\c{c}ois Pasquier, 
Gareth Patrick,
Pascal Paulet, 
Martin Peccia,
Sylvie Peden,
Sonia Penalva, 
Michel Pendaries,
Philippe Peres,
Gr\'{e}gory Personne, 
Dominique Pierot,
Jean-Marc Pillot,
Lydie Pinel, 
Fabien Piquemal,
Vincent Poinsignon, 
Maxime Pomelec,
Andr\'{e} Porras,
Pierre Pouny, 
Severin Provost, 
S\'{e}bastien Ramos,
Fabienne Raux,
Audrey Rehby,
Florian Reuscher,
Xavier Richard,
Nicolas Riguet,
Mickael Roche,
Gilles Rougier, 
Bruno Rouzier, 
Stephane Roy,
Jean-Paul Ruffie,
Fr\'{e}d\'{e}ric Safa, 
Heloise Scheer, 
Claudie Serris,
Andr\'{e} Sobeczko, 
Jean-Fran\c{c}ois Soucaille,
Romain Suze,
Philippe Tatry, 
Th\'{e}o Thomas,
Pierre Thoral,
Dominique Torcheux,
Vincent Tortel,
Damien Tourbez,
Stephane Touzeau, 
Didier Trantoul,
Cyril V\'{e}tel, 
Jean-Axel Vatinel,
Jean-Paul Vormus, and 
Marc Zanoni;
from Airbus DS (Friedrichshafen):
Jan Beck,
Frank Blender,
Volker Hashagen,
Armin Hauser,
Bastian Hell,
Cosmas Heller,
Matthias Holz,
Heinz-Dieter Junginger,
Klaus-Peter Koeble,
Karin Pietroboni,
Ulrich Rauscher,
Rebekka Reichle,
Florian Reuscher,
Ariane Stephan,
Christian Stierle,
Riccardo Vascotto,
Christian Hehr,
Markus Schelkle,
Rudi Kerner,
Udo Schuhmacher,
Peter Moeller,
Rene Stritter,
J\"{u}rgen Frank,
Wolfram Beckert,
Evelyn Walser,
Steffen Roetzer,
Fritz Vogel, and
Friedbert Zilly;
from Airbus DS (Stevenage):
Mohammed Ali,
Bill Bental,
David Bibby,
Leisha Carratt,
Veronica Carroll,
Clive Catley,
Patrick Chapman,
Christoper Chetwood,
Alison Colegrove,
Tom Colegrove,
Andrew Davies,
Denis Di Filippantonio,
Andy Dyne,
Alex Elliot,
Omar Emam,
Colin Farmer,
Steve Farrington,
Nick Francis,
Albert Gilchrist,
Brian Grainger,
Yann Le Hiress,
Vicky Hodges,
Jonathan Holroyd,
Haroon Hussain,
Roger Jarvis,
Lewis Jenner,
Steve King,
Chris Lloyd,
Neil Kimbrey,
Alessandro Martis,
Bal Matharu,
Karen May,
Florent Minaire,
Katherine Mills,
James Myatt,
Chris Nicholas,
Paul Norridge,
David Perkins,
Michael Pieri,
Matthew Pigg,
Angelo Povoleri,
Robert Purvinskis,
Phil Robson,
Julien Saliege,
Satti Sangha,
Paramijt Singh,
John Standing,
Dongyao Tan,
Keith Thomas,
Rosalind Warren,
Andy Whitehouse,
Robert Wilson,
Hazel Wood,
Steven Danes,
Scott Englefield,
Juan Flores-Watson,
Chris Lord,
Allan Parry,
Juliet Morris,
Nick Gregory, and
Ian Mansell.

From ESA, in alphabetical order:
Ricard Abello, 
Asier Abreu, 
Ivan Aksenov, 
Matthew Allen,  
Salim Ansari,  
Philippe Armbruster,  
Mari-Liis Aru, 
Alessandro Atzei,  
Liesse Ayache,  
Samy Azaz,  
Nana Bach, 
Jean-Pierre Balley,  
Paul Balm, 
Manuela Baroni,  
Rainer Bauske, 
Thomas Beck, 
Gabriele Bellei, 
Carlos Bielsa, 
Gerhard Billig, 
Carmen Blasco,  
Andreas Boosz, 
Bruno Bras,  
Julia Braun, 
Thierry Bru, 
Frank Budnik,  
Joe Bush, 
Marco Butkovic, 
Jacques Cande\'{e}, 
David Cano, 
Carlos Casas, 
Francesco Castellini, 
David Chapmann, 
Nebil Cinar, 
Mark Clements, 
Giovanni Colangelo,  
Peter Collins,  
Ana Colorado McEvoy, 
Gabriele Comoretto, 
Vincente Companys, 
Federico Cordero,
Yannis Croizat, 
Sylvain Damiani, 
Fabienne Delhaise, 
Gianpiero Di Girolamo, 
Yannis Diamantidis, 
John Dodsworth, 
Ernesto D\"olling, 
Jane Douglas,  
Jean Doutreleau,  
Dominic Doyle,  
Mark Drapes, 
Frank Dreger,  
Peter Droll, 
Gerhard Drolshagen,  
Michal \v{D}urovi\v{c}, 
Bret Durrett, 
Christina Eilers,  
Yannick Enginger, 
Alessandro Ercolani,  
Matthias Erdmann,  
Orcun Ergincan,  
Robert Ernst,  
Daniel Escolar,  
Maria Espina,  
Hugh Evans,  
Fabio Favata,  
Stefano Ferreri, 
Daniel Firre, 
Michael Flegel, 
Melanie Flentge, 
Alan Flowers, 
Steve Foley,  
Julia Fortuno-Benavent,  
Jens Freih\"ofer, 
Rob Furnell,  
Julio Gallegos,  
Maria De La Cruz Garcia Gonzalez,  
Philippe Gar\'{e},  
Wahida Gasti,  
Jos\'{e} Gavira,  
Frank Geerling,  
Franck Germes,  
Gottlob Gienger, 
B\'{e}n\'{e}dicte Girouart,  
Bernard Godard, 
Nick Godfrey, 
C\'{e}sar G\'omez Hern\'andez,  
Roy Gouka,  
Cosimo Greco, 
Robert Guilanya, 
Kester Habermann, 
Manfred Hadwiger, 
Ian Harrison,  
Angela Head, 
Martin Hechler,  
Javier Hernando Bravo,  
Kjeld Hjortnaes,  
John Hoar,  
Jacolien Hoek,  
Frank Hoffmann, 
Justin Howard, 
Fredrik H\"{u}lphers, 
Arjan Hulsbosch,  
Christopher Hunter,  
Premysl Janik,  
Jos\'{e} Jim\'{e}nez, 
Emmanuel Joliet,  
Helma van de Kamp-Glasbergen,  
Simon Kellett, 
Andrea Kerruish, 
Kevin Kewin, 
Oliver Kiddle, 
Sabine Kielbassa, 
Volker Kirschner,  
Kees van 't Klooster,  
Ralf Kohley, 
Jan Kolmas, 
Oliver El Korashy,  
Arek Kowalczyk, 
Holger Krag, 
Beno\^{\i}t Lain\'{e},  
Markus Landgraf,  
Sven Landstr\"om,  
Mathias Lauer, 
Robert Launer, 
Laurence Tu-Mai Levan,  
Mark ter Linden,  
Santiago Llorente, 
Tim Lock$^\dagger$,  
Alejandro Lopez-Lozano, 
Guillermo Lorenzo, 
Tiago Loureiro,  
James Madison, 
Juan Manuel Garcia, 
Federico di Marco,  
Jonas Marie,  
Filip Marinic, 
Pier Mario Besso, 
Arturo Mart\'{\i}n Polegre,  
Ander Mart\'{\i}nez, 
Monica Mart\'{\i}nez Fern\'{a}ndez,  
Marco Massaro, 
Paolo de Meo, 
Ana Mestre, 
Claudio Mevi, 
Luca Michienzi, 
David Milligan, 
Ali Mohammadzadeh,  
David Monteiro,  
Richard Morgan-Owen,  
Trevor Morley,  
Prisca M\"uhlmann,  
Jana Mulacova,  
Michael M\"uller, 
Pablo Mu\~{n}oz, 
Petteri Nieminen,  
Alfred Nillies, 
Wilfried Nzoubou, 
Alistair O'Connell, 
Karen O'Flaherty,  
Alfonso Olias Sanz,  
William O'Mullane, 
Jos\'{e} Osinde, 
Oscar Pace,  
Mohini Parameswaran, 
Ramon Pardo, 
Taniya Parikh,  
Paul Parsons,  
Panos Partheniou, 
Torgeir Paulsen,  
Dario Pellegrinetti, 
Jos\'{e}-Louis Pellon-Bailon,  
Joe Pereira,  
Michael Perryman,  
Christian Philippe,  
Alex Popescu,  
Fr\'{e}d\'{e}ric Raison,  
Riccardo Rampini,  
Florian Renk,  
Alfonso Rivero, 
Andrew Robson, 
Gerd R\"ossling, 
Martina Rossmann, 
Markus R\"uckert, 
Andreas Rudolph,  
Fr\'{e}d\'{e}ric Safa,  
Johannes Sahlmann, 
Eugenio Salguero, 
Jamie Salt,  
Giovanni Santin,  
Fabio de Santis, 
Rui Santos, 
Giuseppe Sarri,  
Stefano Scaglioni,  
Melanie Schabe, 
Dominic Sch\"afer, 
Micha Schmidt, 
Rudolf Schmidt,  
Ared Schnorhk,  
Klaus-J\"urgen Schulz, 
Jean Sch\"utz, 
Julia Schwartz,  
Andreas Scior, 
J\"org Seifert, 
Christopher Semprimoschnig$^\dagger$,  
Ed Serpell,  
I\~{n}aki Serraller Vizcaino,  
Gunther Sessler, 
Felicity Sheasby,  
Alex Short,  
Hassan Siddiqui, 
Heike Sillack, 
Swamy Siram,
Chloe Sivac,  
Christopher Smith,  
Claudio Sollazzo,  
Steven Straw,
Daniel Tapiador,  
Pilar de Teodoro,  
Mark Thompson, 
Giulio Tonelloto,  
Felice Torelli,  
Raffaele Tosellini,  
Cecil Tranquille,  
Irren Tsu-Silva,  
Livio Tucci, 
Aileen Urwin,  
Jean-Baptiste Valet, 
Martin Vannier,  
Enrico Vassallo, 
David Verrier, 
Sam Verstaen,  
R\"udiger Vetter, 
Jos\'{e} Villalvilla, 
Raffaele Vitulli,  
Mildred V\"ogele, 
Sandra Vogt, 
Sergio Volont\'{e}, 
Catherine Watson, 
Karsten Weber, 
Daniel Werner, 
Gary Whitehead$^\dagger$,  
Gavin Williams, 
Alistair Winton,  
Michael Witting,  
Peter Wright, 
Karlie Yeung, 
Marco Zambianchi, and  
Igor Zayer,  
and finally Vincenzo~Innocente from the Conseil Europ\'{e}en pour la Recherche Nucl\'{e}aire (CERN).

In case of errors or omissions, please contact the \href{https://www.cosmos.esa.int/web/gaia/gaia-helpdesk}{\gaia\ Helpdesk}.

A.J. acknowledges support from the Fonds de la Recherche Fondamentale Collective (FNRS, F.R.F.C.) of Belgium through grant PDR T.0115.23 and from Belspo/PRODEX/ESA under grant PEA nr. 4000119826. This research made use of \texttt{pystrometry}, an open source Python package for astrometry timeseries analysis \citep{johannes_sahlmann_2019_3515526}, and \texttt{galpy}, a Python library for Galactic dynamics \citep{2015ApJS..216...29B}.

\end{appendix}

\end{document}